%% file: main.tex
\UseRawInputEncoding

\documentclass[aps,prl,amsmath,amssymb,floatfix,reprint,citeautoscript,noeprint,superscriptaddress,twocolumn]{revtex4-2}
\usepackage{xr}

\usepackage{dsfont}
\usepackage{lipsum} 
\usepackage{bibentry}
\usepackage[english]{babel}
\usepackage[dvipsnames]{xcolor}
\usepackage{graphicx}
\usepackage[caption=false]{subfig} 
\usepackage{amsmath,amssymb,bm}
\usepackage[version=3]{mhchem}
\usepackage{verbatim}
\usepackage{multirow}
\usepackage{dcolumn}
\usepackage{float}
\usepackage{ragged2e}
\usepackage{nicefrac}
\usepackage{siunitx}
\usepackage{booktabs}
\usepackage{chemformula}
\usepackage{wrapfig}
\usepackage{enumitem}  
\usepackage{transparent}
\usepackage[colorlinks,allcolors=black,citecolor=blue,urlcolor=blue]{hyperref}
\usepackage{tcolorbox}
\usepackage{relsize}

\newcommand{\etal}{\emph{et al.}}

\makeatletter
\long\def\@makecaption#1#2{%
  \vskip\abovecaptionskip
  \sffamily %
  \small    %
  \begingroup
    \leftskip=0pt plus 1fil
    \rightskip=0pt plus 1fil
    \parfillskip=0pt
    \justifying %
    \noindent
    \bfseries #1: \normalfont #2\par
  \endgroup
  \vskip\belowcaptionskip
}
\makeatother

\makeatletter
\input{si-labels.aux}
\makeatother

\usepackage{sansmath}
\sansmath

\begin{document}

\def\mytitle{From Accurate Quantum Chemistry to Converged Thermodynamics for Ion Pairing in Solution}
\title{\mytitle}

\author{Niamh O'Neill}%
\affiliation{Yusuf Hamied Department of Chemistry, University of Cambridge, Lensfield Road, Cambridge, CB2 1EW, UK}
\affiliation{Cavendish Laboratory, Department of Physics, University of Cambridge, Cambridge, CB3 0HE, UK}
\affiliation{Lennard-Jones Centre, University of Cambridge, Trinity Ln, Cambridge, CB2 1TN, UK}
\affiliation{Max Planck Institute for Polymer Research, Mainz 55128, Germany}

\author{Benjamin X. Shi}%
\affiliation{Initiative for Computational Catalysis, Flatiron Institute, 160 5th Avenue, New York, NY 10010, USA}

\author{William C. Witt}%
\affiliation{Harvard John A. Paulson School of Engineering and Applied Sciences, Harvard University, Cambridge, MA, USA}

\author{Blake I. Armstrong}%
\affiliation{School of Molecular and Life Sciences, Curtin University, PO Box U1987, Perth, Western Australia 6845, Australia}
\author{Asmita Niyogi}%
\affiliation{Yusuf Hamied Department of Chemistry, University of Cambridge, Lensfield Road, Cambridge, CB2 1EW, UK}
\affiliation{Lennard-Jones Centre, University of Cambridge, Trinity Ln, Cambridge, CB2 1TN, UK}

\author{William J. Baldwin}%
\affiliation{Lennard-Jones Centre, University of Cambridge, Trinity Ln, Cambridge, CB2 1TN, UK}
\affiliation{%
Engineering Department, Trumpington St, Cambridge CB2 1PZ, UK
}

\author{Paolo Raiteri}%
\affiliation{School of Molecular and Life Sciences, Curtin University, PO Box U1987, Perth, Western Australia 6845, Australia}

\author{Christoph Schran}%
\email{cs2121@cam.ac.uk}
\affiliation{Cavendish Laboratory, Department of Physics, University of Cambridge, Cambridge, CB3 0HE, UK}
\affiliation{Lennard-Jones Centre, University of Cambridge, Trinity Ln, Cambridge, CB2 1TN, UK}

\author{Angelos Michaelides}%
\email{am452@cam.ac.uk}
\affiliation{Yusuf Hamied Department of Chemistry, University of Cambridge, Lensfield Road, Cambridge, CB2 1EW, UK}
\affiliation{Lennard-Jones Centre, University of Cambridge, Trinity Ln, Cambridge, CB2 1TN, UK}

\author{Julian D. Gale}%
\email{J.Gale@curtin.edu.au}
\affiliation{School of Molecular and Life Sciences, Curtin University, PO Box U1987, Perth, Western Australia 6845, Australia}

\newcommand*\mycommand[1]{\texttt{\emph{#1}}}

\begin{abstract}
Quantitative prediction of thermodynamic properties in solution is essential for translating atomistic simulations into reliable chemical insight.
As an exemplar system, the behaviour of CaCO\textsubscript{3} in water has been widely studied to understand its mineralization in seawater, with potential implications for carbon-capture strategies.
However, making accurate computational predictions has been a long-standing challenge, requiring both highly accurate electronic structure methods and extensive statistical sampling.
Here, we combine advances in machine learning and electronic structure theory to fully resolve the ion pairing free energy of CaCO\textsubscript{3} with explicit solvation.
We show that achieving quantitative agreement with experiment requires going beyond density functional theory up to the ``gold-standard'' coupled cluster theory with single, double, and perturbative triple excitations [CCSD(T)].
We generate a set of systematically improvable models, enabling reliable insights into the initial association mechanism of Ca and CO\textsubscript{3} ions prior to nucleation while fully quantifying enthalpic and entropic effects.
\textcolor{black}{These quantitatively accurate models further resolve a long-standing lack of consensus surrounding the hydration mechanism of the calcium ion in solution.}
Our results demonstrate that CCSD(T)-level thermodynamic predictions of complex aqueous systems can now be routinely achieved.
\end{abstract}

{\maketitle}

\section{Introduction}
Ions in solution play central roles in processes ranging from biomolecular function and mineralization \cite{luRoleWaterCaCO32021} to energy storage and conversion technologies \cite{caoSolvationStructureDesign2020}.
As an archetypal example, calcium carbonate (\ch{CaCO3}) is a crucial ionic species in seawater, which -- upon mineralisation -- forms the primary structural component of geological structures such as coral reefs \cite{noonanProgressiveChangesCoral2025} and serves as a promising route for long-term carbon sequestration \cite{huangBridgingMechanismsMaterials2026a, hepburnTechnologicalEconomicProspects2019}.
More fundamentally, it is often used as a model system for nucleation theories \cite{smeetsCalciumCarbonateNucleation2015}.
However the molecular mechanism of ion pairing and its relationship to nucleation remains poorly understood, despite sustained experimental and computational investigation \cite{raiteriWaterKeyNonclassical2010, gebauerStablePrenucleationCalcium2008, smeetsClassicalViewNonclassical2017,
luRoleWaterCaCO32021, henzlerSupersaturatedCalciumCarbonate2018,
gebauerPrenucleationClustersNonclassical2011, demichelisStablePrenucleationMineral2011, gebauerPrenucleationClustersSolute2014, raiteriIonPairingMultiple2020, armstrongSolubilityconsistentForceField2023, jinFormationChemicalEvolution2025}.
A major reason for this is that the macroscopic phenomena of mineralisation and dissolution are ultimately governed by the highly delicate nature of the interactions between individual 
ions and their aqueous solvation environment.
\textcolor{black}{Remarkably, even the seemingly simple question of the structure of water around an individual calcium ion remains poorly resolved, with experimental and computational studies reporting between 5 and 10 water molecules coordinating the ion (See Supplementary Information (SI), Fig.~\ref{si-fig:literature} and Tables~\ref{si-tab:aimd_mlip_calcium_cn}, \ref{si-tab:force_field_calcium_cn} and \ref{si-tab:experimental_calcium_cn}).}
Since these interactions are difficult to probe directly in experiments, atomistic simulations can provide a powerful framework to link microscopic structure to macroscopic behaviour \cite{tuckermanNatureTransportMechanism2002, neeseChemistryQuantumMechanics2019}.

\begin{figure*}[t!]
  \centering
   \includegraphics[width=\textwidth]{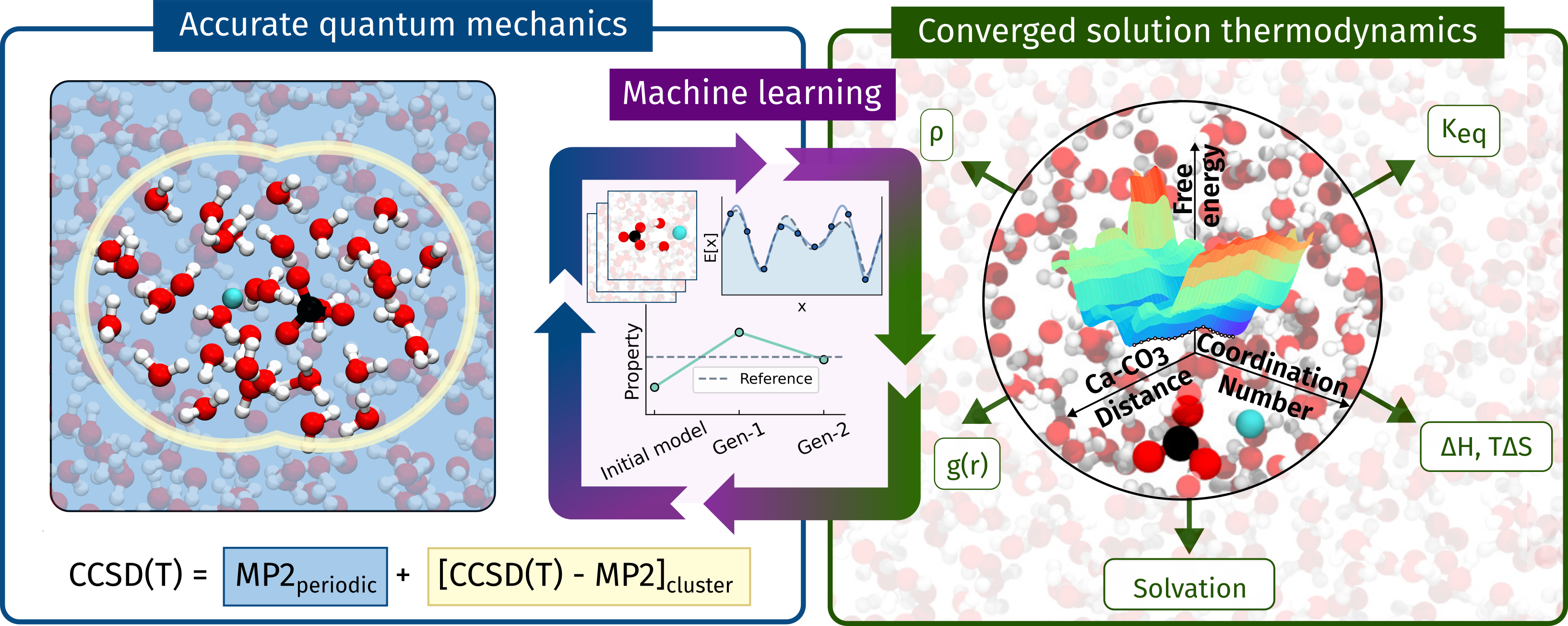}
    \caption{\textbf{Workflow combining accurate electronic structure, MLPs and enhanced sampling:} Schematic of the framework developed in this work, using machine learning potentials (MLPs) to access converged thermodynamics in solution at the level of coupled cluster theory with single, double and perturbative triple excitations [CCSD(T)]. We start from a baseline MLP trained to second-order M{\o}ller-Plesset perturbation theory (MP2) data under periodic boundary conditions. We employ a $\Delta$-learning framework to develop a second MLP that corrects the MP2 MLP up to the CCSD(T) level, using data from gas phase clusters extracted from periodic snapshots. In general, the models are `self-consistently' converged to ensure reliable predictions of the density and potential of mean force (PMF) (generically labelled `Property' in the middle panel) and adequate sampling of configuration space. Finally efficient implementations of the MLPs are used in tandem with enhanced sampling simulations to resolve the thermodynamics of complex potential (free) energy surfaces.} 
   \label{fig:schematic}
\end{figure*}

The fundamental property of interest for CaCO$_3$, and most ions, is the ion pair association free energy in solution.
It can be measured experimentally and serves as a stringent benchmark for assessing the reliability of atomistic insights obtained from simulations.
However, accurately predicting this quantity is challenging \cite{mcdonoghRedefinedIonAssociation2024}.
First, it requires a faithful description of the underlying potential energy surface (PES), governed by a delicate balance of ion--ion, ion--water, and water--water interactions \cite{tuckermanInitioMolecularDynamics1995, zhangDissociationSaltsWater2020}.
Second, reliable estimates demand thorough sampling of this PES to capture finite-temperature and entropic contributions \cite{aufortComputationalInsightsMg22022}.
In practice, enhanced sampling methods are needed to efficiently explore multiple ion pairing minima within accessible simulation timescales.
Finally, the solvent must be modeled explicitly to account for its effects both on the PES and the thermodynamic sampling.
{\color{black} We show in the SI, Sec.~\ref{si-sec:lit_comp_ip} via comparison to previous literature, the challenges in obtaining accurate ion pair association free energies when these considerations are not fully accounted for.}

These stringent requirements, illustrated here for \ch{CaCO3}, form the central challenge for any atomistic simulation in solution \cite{a.youngTransferableActivelearningStrategy2021}.
Classical force fields are computationally efficient but often insufficiently accurate, whereas first-principles (\textit{ab initio}) methods may be more accurate but the concurrent sampling demands make it too costly for well-converged \textit{ab initio} molecular dynamics (AIMD) simulations.
Machine learning potentials (MLPs) have emerged as a promising strategy to bridge this gap, training surrogate models to reproduce first-principles methods at significantly reduced cost.
It has become routine to train MLPs to density functional theory (DFT), the workhorse first-principles method, with notable success achieved across diverse systems, including \ch{CaCO3} \cite{piaggiInitioMachinelearningSimulation2025}.
The efficiency of computing DFT energies and gradients, together with mature implementations of periodic boundary conditions to describe the condensed phase, has enabled relatively standardized protocols for fitting DFT-level MLPs.
However, the accuracy of the resulting MLP is ultimately limited by the underlying density functional approximation (DFA).
In particular, delocalization error~\cite{bryentonDelocalizationErrorGreatest2023} -- especially problematic for charged anions~\cite{jensenDescribingAnionsDensity2010, otero-de-la-rozaAnalysisDensityFunctionalErrors2020, vydrovTestsFunctionalsSystems2007} -- can lead to qualitatively incorrect densities and solvation structures \cite{oneillPairNotPair2024, monterodehijesDensityIsobarWater2024}.
For example, a recent MLP trained to the SCAN functional surpasses many force-fields in describing \ch{CaCO3} from aqueous solution to the solid state \cite{piaggiInitioMachinelearningSimulation2025}, but its predicted ion pair association free energy remains several kJ/mol from experiment.

Correlated wavefunction theory (cWFT) on the other hand, provides a systematically improvable hierarchy of methods.
The random phase approximation (RPA) and second-order M{\o}ller--Plesset perturbation theory (MP2) form the lower tiers of this hierarchy, and often improve the description of ions over DFT as they naturally capture van der Waals interactions \cite{oneillPairNotPair2024, duignanQuantifyingHydrationStructure2020} while mitigating (albeit only partially in the case of RPA) delocalization error.
Periodic implementations of RPA \cite{delbenEnablingSimulationFifth2015} and MP2 \cite{delbenElectronCorrelationCondensed2013, delbenForcesStressSecond2015} that also compute gradients are now mature enough to train MLPs using strategies similar to those developed for DFT \cite{oneillPairNotPair2024}.
However, achieving quantitative accuracy with these methods is still challenging.
For example, MP2 significantly overstructures liquid water, resulting in a substantially lower diffusion coefficient~\cite{beckMoreConvergedLess2026,willowWhyMP2WaterCooler2016, delbenBulkLiquidWater2013}, while RPA tends to underestimate lattice energies of ice \cite{zenFastAccurateQuantum2018}.
The method of choice -- the ``gold standard'' -- is coupled cluster theory with single, double, and perturbative triple excitations [CCSD(T)], which has demonstrated reliable accuracy across diverse material and molecular systems \cite{kartonW4TheoryComputational2006, shiAccurateEfficientFramework2025, shiManyBodyMethodsSurface2023, schaferLocalEmbeddingCoupled2021, carboneCOAdsorptionPt1112024, yangInitioDeterminationCrystalline2014, yeAdsorptionVibrationalSpectroscopy2024, yePeriodicLocalCoupledCluster2024}.
However, its steep computational cost -- formally scaling as $N^7$ with $N$ electrons -- along with the lack of mature periodic and gradient implementations, makes CCSD(T) impractical for routine molecular dynamics simulations and incompatible with standard DFT-based MLP training strategies.

Nevertheless, there have been approaches that aim to achieve CCSD(T)-level accuracy in finite-temperature simulations of condensed-phase aqueous systems.
One strategy uses a multi-level embedding approach: simulations are performed at a lower level of theory, with CCSD(T) corrections applied to localized regions \cite{yuDensityPotentialFunctional2017,martirezSolventDynamicsAre2023,boynProbingPHDependentDehydration2023}.
The most relevant study to date explored Ca (and Mg) ion pairing with \ch{CO3}, combining DFT-based biased AIMD simulations with quantum embedding of selected fragments \cite{boynCharacterizingMechanismsCa2023}.
Although this approach can attain high accuracy at key points along the ion pairing profile, its computational cost limits exploration of the full profile. 
Recent work \cite{bianTransferLearningMeets2026} has shown that transfer-learned MLPs trained on the embedded PES can alleviate this sampling issue.
However, these approaches currently emphasise local corrections around the ions, and extending them to obtain a fully resolved ion pairing free energy for direct comparison with experiment -- along with a complete description of the solvent at CCSD(T) level -- remains an open question.

Alternative approaches aim to directly parameterise the global CCSD(T) PES with cost-effective surrogate models to be directly sampled during simulations.
One highly successful family of approaches is based on the many-body expansion, with the MB-pol/ MB-nrg formalism achieving excellent agreement with experiment for hydration properties of simple alkali and halide ions in water \cite{agnewMolecularInsightsInfluence2025, savojMolecularInsightsInfluence2024, sahaHydrationFreeEnergies2026, agnewManybodyInteractionsResolve2026} while q-AQUA-pol also accurately reproduces bulk water properties \cite{quInterfacingQAQUAPolarizable2023}. 
Most recent is the development of CCSD(T)-level MLPs to directly model the condensed phase PES \cite{daruCoupledClusterMolecular2022, chenDataEfficientMachineLearning2023, oneillRoutineCondensedPhase2025}.
In particular, $\Delta$-learning targets the CCSD(T) correction to DFT in the condensed phase by extrapolating from gas-phase clusters, alleviating the aforementioned limitations with periodic CCSD(T).

Although successful for liquid water \cite{daruCoupledClusterMolecular2022, meszarosShortRangeDMachineLearning2025, oneillRoutineCondensedPhase2025} and gas phase molecular systems \cite{ramakrishnanBigDataMeets2015}, extending these strategies to more complex multicomponent systems in the condensed phase remains a major challenge.
{\color{black}
On the electronic structure side, long-range interactions between ions necessarily require larger clusters, significantly increasing the cost. Furthermore, the aforementioned issues with DFT for charged systems require more advanced methods for a baseline model, making generating a reliable baseline already a formidable challenge.} 

In this work, we generalize our recently proposed $\Delta$-learning framework for liquid water \cite{oneillRoutineCondensedPhase2025} to study ions in solution with CCSD(T)-level MLPs, using the ion pairing of \ch{CaCO3} as a case study system.
As illustrated in Figure \ref{fig:schematic}, we leverage recent advances in efficient periodic MP2 implementations \cite{delbenForcesStressSecond2015} to construct a baseline MLP trained at the MP2 level {\color{black}(thus avoiding the issues for ionic systems associated with a DFT baseline). This provides a high-level baseline description of the full periodic system, while}
a separate $\Delta$-MLP is then trained to learn the MP2-to-CCSD(T) correction using gas-phase clusters (centred on both ions) extracted from the periodic configurations.
While trained on gas-phase clusters, we previously~\cite{oneillRoutineCondensedPhase2025} show that the resulting $\Delta$-MLP can accurately treat the condensed phase.
The final CCSD(T) MLP is the sum of (energies and forces) predicted by the baseline and $\Delta$-MLPs as illustrated in the left-hand panel of Figure \ref{fig:schematic}.
The underlying periodic dataset and derived $\Delta$-MLP dataset are improved iteratively to converge the potential of mean force (PMF) used to compute the ion pairing free energy, ensuring reliable prediction of the thermodynamic properties (illustrated in the centre panel of Figure \ref{fig:schematic}).
Further details on the model training and validation are given in the SI, Section \ref{si-sec:caco3si_models}.

Combining our CCSD(T) MLP model 
with state-of-the-art enhanced sampling approaches \cite{invernizziRethinkingMetadynamicsBias2020}, we achieve excellent agreement with experimental values for the ion pair association free energy of \ch{CaCO3} in water.
It must be emphasised here that this approach enables the entire (global) PES and thermodynamic sampling to be explicitly treated at the CCSD(T) level.
We decompose the ion pairing free energy into its enthalpic and entropic contributions, and compare to MLP models trained to several DFT functionals, as well as RPA and MP2.
We show that the CCSD(T) model is the only model that simultaneously agrees with experiment for the free energy, enthalpy and entropy, {\color{black}while also quantitatively reproducing experimental properties of the bulk water solvent.}
With the CCSD(T) model validated {\color{black} on experimentally realisable benchmarks},
\textcolor{black}{we are now able to explore and resolve long-standing questions about the hydration mechanism of the calcium ion in water. As previously mentioned, this property has no consensus from either experiment or simulation. Our model enables us to resolve the full mechanism of ion hydration, where we reveal a qualitatively different mechanism to that predicted by previous simulations at lower levels of theory as well as enabling critical analysis of experimental results. Such insight has vital consequences for example in the design of new membrane technologies for separations, as well as our understanding of ion transport in biological channels~\cite{zhangCa2PermeationMechanism2020, moRegulatingIonAffinity2024}.}
Overall, this work shows that thermodynamics of complex aqueous systems at cWFT accuracy -- and in particular at the CCSD(T) level -- is now accessible, opening a route towards routine, high-accuracy simulations of experimentally relevant processes in solution.

\section{Agreement with experiment on \ch{CaCO3} ion pairing}\label{sec:ionpairing}

\begin{figure}[h!]
  \centering
   \includegraphics[width=0.5\textwidth]{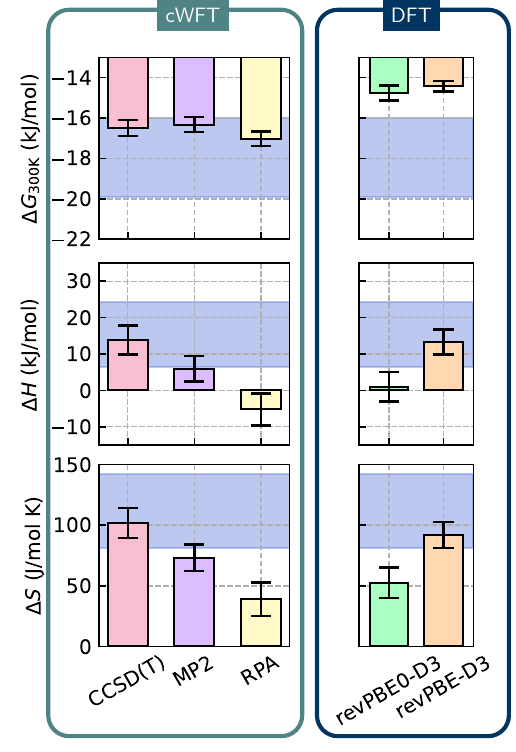}

    \caption{
	    \textbf{Ion pair association free energy, enthalpy and entropy from OPES simulations:} Comparison of cWFT MLPs (CCSD(T), MP2 and RPA) and DFT MLPs (revPBE0-D3, revPBE-D3) for (top) the standard ion pair association free energy at 300 K (middle) enthalpy and (bottom) entropy of ion pair association. The experimental range from literature \cite{kellermeierEntropyDrivesCalcium2016, plummerSolubilitiesCalciteAragonite1982} is shaded in blue for each quantity. Error bars on the cWFT and DFT results are the standard error on the mean of 6 independent OPES simulations. All results have also been corrected for finite size effects, as shown in Figure \ref{si-fig:finitesize} of the SI. Convergence of the OPES simulations is also shown in Figure \ref{si-fig:opes_convergence} of the SI, as well as the full temperature dependence of the ion pair association free energy used to obtain the enthalpy and entropy in Figure \ref{si-fig:dgdt} of the SI.}
   \label{fig:pmf}
\end{figure}

We begin by considering ion pairing free energies at 300 K ($\Delta$G$_\text{300K}$), as shown in the top panel of Figure~\ref{fig:pmf}.
Reported in the figure are experimental measurements and values obtained from MLPs trained at increasing levels of electronic structure theory and computed with on-the-fly probability enhanced sampling (OPES) \cite {invernizziRethinkingMetadynamicsBias2020} simulations.
{\color{black}Complete simulation details are given in the Methods section below, as well as the SI, Secs.~\ref{si-sec:simulations} and \ref{si-sec:ip_opes}, however we highlight a few salient details here.
Firstly, it is important to perform the simulations in a sufficiently large box to allow for correct alignment of the potential of mean force (PMF) with the tail of the analytical solution for two point charges interacting in a dielectric medium~\cite{aufortComputationalInsightsMg22022} (see SI, Fig.\ref{si-fig:alignment_cutoff} for the importance of correctly aligning the PMF).
Therefore our simulation box (containing one ion pair and 197 water molecules) is $\sim$ 18~\AA{} in length, (depending on the equilibrium density of the given model first computed from NPT simulations), and we quantify residual finite-size effects to the free energy in the SI, Table~\ref{si-tab:finite_size}.
Additionally, to fully converge the enthalpic and entropic contributions, we perform 6 independent replicas each between 20-30 ns in length.
We also emphasise the importance of fully sampling the orthogonal solvent degrees of freedom to the ion pair separation.
A full description of the molecular dynamics setup and free energy calculations is given in the SI, Secs.~\ref{si-sec:simulations} and \ref{si-sec:ip_opes}, respectively.}

Two experimental values for the ion pairing free energy are available, depending on the speciation model used to interpret the data \cite{plummerSolubilitiesCalciteAragonite1982,kellermeierEntropyDrivesCalcium2016} and cover a range of $4\,$kJ/mol, coincidentally corresponding to chemical accuracy.
We compare models trained in this work at the DFT level (revPBE-D3 \cite{zhangCommentGeneralizedGradient1998, grimmeConsistentAccurateInitio2010} and revPBE0-D3 \cite{adamoReliableDensityFunctional1999, goerigkThoroughBenchmarkDensity2011}), as well as RPA, MP2, and CCSD(T).
The predicted binding free energies span a range of roughly $3\,$kJ/mol.
A clear trend emerges: agreement with experiment improves 
with increasing levels of theory.
In particular, the RPA-, MP2-, and CCSD(T)-based models achieve quantitative agreement with experiment for $\Delta$G$_\text{300K}$.
In contrast, the widely used GGA functional revPBE-D3 underestimates the binding free energy.
Although hybrid DFT has been suggested to improve charge localization and the relative stability of contact and solvent-shared ion pairs \cite{raiteriIonPairingMultiple2020, aufortComputationalInsightsMg22022}, revPBE0-D3 shows little improvement over revPBE-D3 and still underestimates the binding strength.

The real challenge emerges in the middle and bottom plots of Figure~\ref{fig:pmf}, which highlight the difficulty of accurately capturing the delicate balance of enthalpy and entropy underlying the ion pairing.
These contributions differentiate the models, showing that while some can reproduce the free energy at $300\,$K, this is only through fortuitous error cancellation.
For example, although RPA yields the correct ion pairing free energy at $300\,$K, it predicts a negative enthalpy and a significantly underestimated entropy compared to experiments.
Conversely, revPBE-D3 reproduces the experimental enthalpy and entropy but underestimates the $300\,$K free energy.
Only CCSD(T) \textcolor{black}{and MP2} consistently agree with experiment for the free energy, enthalpy, and entropy.
These results emphasize the twofold requirement of an accurate cWFT-level PES and exhaustive sampling to properly capture entropic effects.

\section{Reliable chemical insights into ion pairing mechanisms}
\begin{figure*}[h!]
  \centering
   \includegraphics[width=\textwidth]{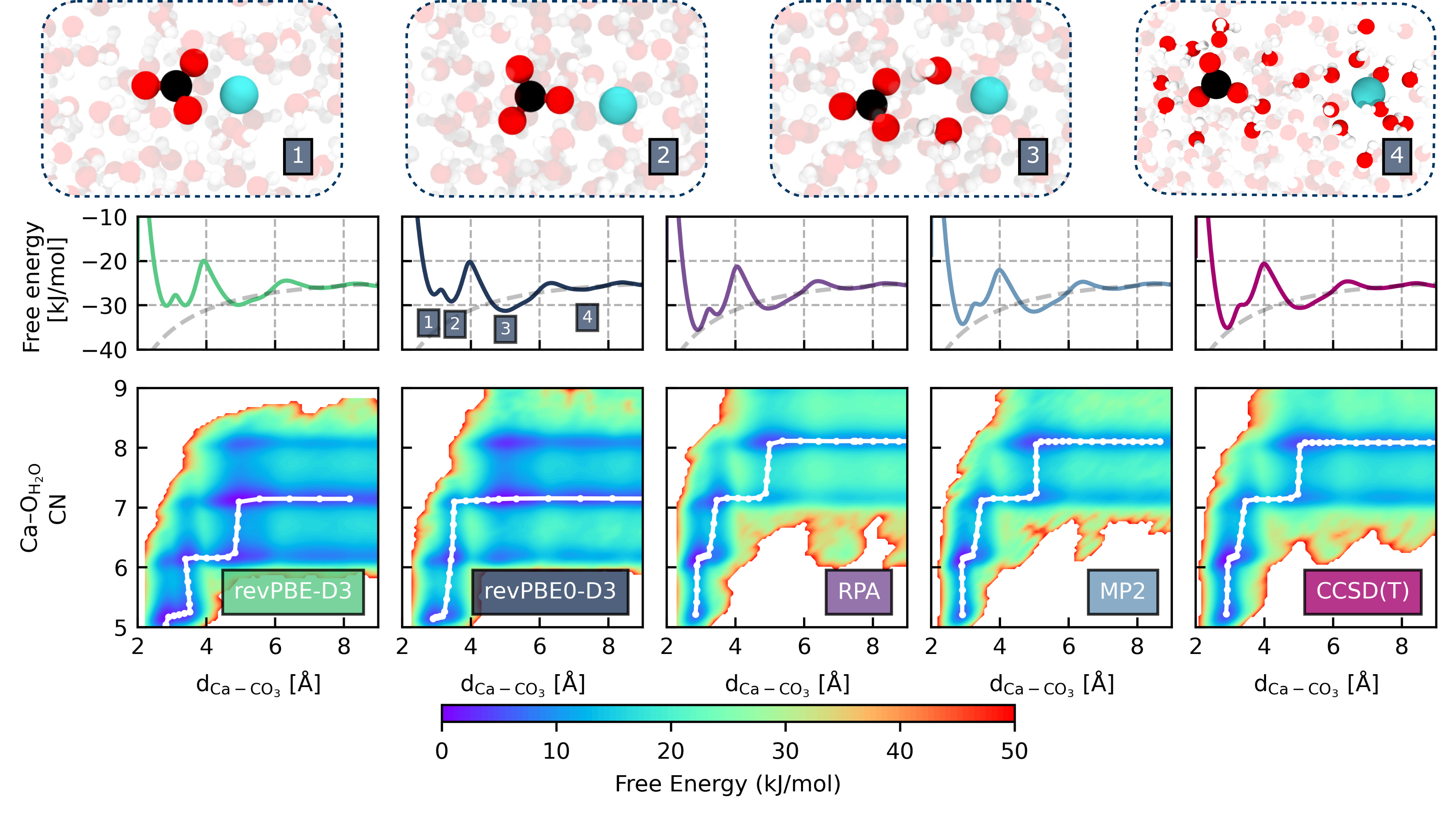}

    \caption{\textbf{\ch{CaCO3} ion pair dissociation pathways:}
    Ion pairing pathways for the models at 300 K: revPBE-D3, revPBE0-D3, RPA, MP2 and CCSD(T).  The minimum free energy path in each case is plotted in white. Above is the potential of mean force, aligned to the analytic solution for point charges in a dielectric medium (grey dashed line). The bidentate contact ion pair (1), monodentate contact ion pair (2), solvent-shared ion pair (3) and solvent separated  ion pair (4) are labelled for the revPBE0-D3 case, along with representative snapshots above (where only a small region of the full simulation cell is shown). In the snapshots, calcium, carbon, oxygen and hydrogen are coloured blue, black, red and white, respectively.
    }
   \label{fig:mfep}
\end{figure*}

Having established the quantitative accuracy of our CCSD(T) MLP, we now use it to examine the atomistic ion pairing pathways.
Figure \ref{fig:mfep} shows the two-dimensional (2D) free energy surface as a function of the Ca--\ch{CO_3} separation and Ca--O$_{\mathrm{H_2O}}$ coordination number (CN) for all of the models developed in this work.
The corresponding one-dimensional (1D) PMF, obtained by integrating over the Ca--O$_{\mathrm{H_2O}}$ coordination number, is shown above each surface. 
In order of increasing ion-ion separation, the key states are the bidentate (1) and monodentate (2) contact ion pairs (CIPs), the solvent-shared ion pair (SShIP) (3), and the solvent-separated ion pair (SSIP) (4), illustrated above the 1D PMFs.

The 1D PMF now clarifies why revPBE-D3 and revPBE0-D3 -- both widely used DFAs for aqueous systems -- underestimate the $300\,$K binding free energy.
Relative to CCSD(T) and MP2, the two DFAs predict incorrect stabilities of the CIPs, with the bidentate and monodentate CIPs nearly isoenergetic.
In contrast, CCSD(T) stabilizes the bidentate CIP significantly over the monodentate form.
Both DFT methods also predict the opposite stability of the CIP states and SShIP relative to CCSD(T).
The overstabilization of more separated states relative to the CIPs is consistent with delocalization error in DFT, which favors weaker ionic character.
\textcolor{black}{This is further manifested in the difference plots shown in the SI, Fig~\ref{si-fig:difference_fes}.}
We also analyse the charges on the ions predicted by the different levels of theory (up to MP2) via Bader charge partitioning in the SI, Sec.~\ref{si-sec:bader}, which confirms increased localization at higher levels of theory.
Physically, this aligns with prior force-field studies showing that larger ionic charges are necessary to achieve quantitative agreement with experiment \cite{armstrongSolubilityconsistentForceField2023}.

Among the first-principles methods studied in this work, only MP2 reproduces the CCSD(T) trend, although it slightly underestimates the barrier between CIP and SShIP (also shown in Figure \ref{si-fig:all_pmfs_wf} of the SI).
{\color{black}
The embedded correlated-wavefunction approach, including more recently its combination with transfer-learning, has also been applied to CaCO$_3$ ion pairing~\cite{boynCharacterizingMechanismsCa2023,bianTransferLearningMeets2026}. 
While our results agree closely with these studies in the contact ion pair region, appreciable differences emerge along the dissociation pathway. 
In particular, the reported CIP--SShIP barrier differs by approximately 5~\(\mathrm{kJ\,mol^{-1}}\), while the SShIP stability differs by up to $10-15\,$~\(\mathrm{kJ\,mol^{-1}}\). 
We investigate these discrepancies quantitatively in the SI, Sec. \ref{si-sec:lit_comp_ip}, where we show that they arise from a combination of finite-size (see SI Fig.\ref{si-fig:ccsdt_comparison}) and sampling effects, sensitivity of the association free energy to the long-range PMF alignment, and differences in the underlying potential-energy surface. 
Notably, reproducing the $12\,$\AA{} simulation cell used in Ref.~\citenum{bianTransferLearningMeets2026} largely reconciles the CIP--SShIP barrier, whereas substantial differences in the SShIP stability remain.
 We attribute the latter, at least in part, to the fact that the embedded CCSD(T) model retains a DFT-level description of the solvent.
 By contrast, our MLIP describes the complete ion--water system at CCSD(T) accuracy, allowing both the energetic and entropic contributions to the ion pairing free energy landscape to be sampled consistently at this level of theory (See SI, Figs~\ref{si-fig:density_caco3} and~\ref{si-fig:mp2ccsdt_valid})}.

The 2D free energy surface further highlights distinct differences in the minimum free energy path for ion pairing/dissociation between the different levels of theory.
While CCSD(T), MP2, RPA and revPBE0-D3 remain in the CIP state as the Ca--O$_{\mathrm{H_2O}}$ CN increases from 5 to 6 to 7, revPBE-D3 follows a different mechanism at each step.
It first increases the ion pair separation before adding further water molecules to its coordination shell, possibly arising from delocalisation error favouring more diffuse spread of charge.
Furthermore, both revPBE-D3 and revPBE0-D3 show increased preference for lower coordination with water at large ion separations compared to CCSD(T), which more easily accommodates an additional water in the Ca solvation shell (and will be discussed in more detail in the next Section).

\section{Resolving ion solvation structure}
\begin{figure*}[h!]
    \centering
    \includegraphics[width=1.0\linewidth]{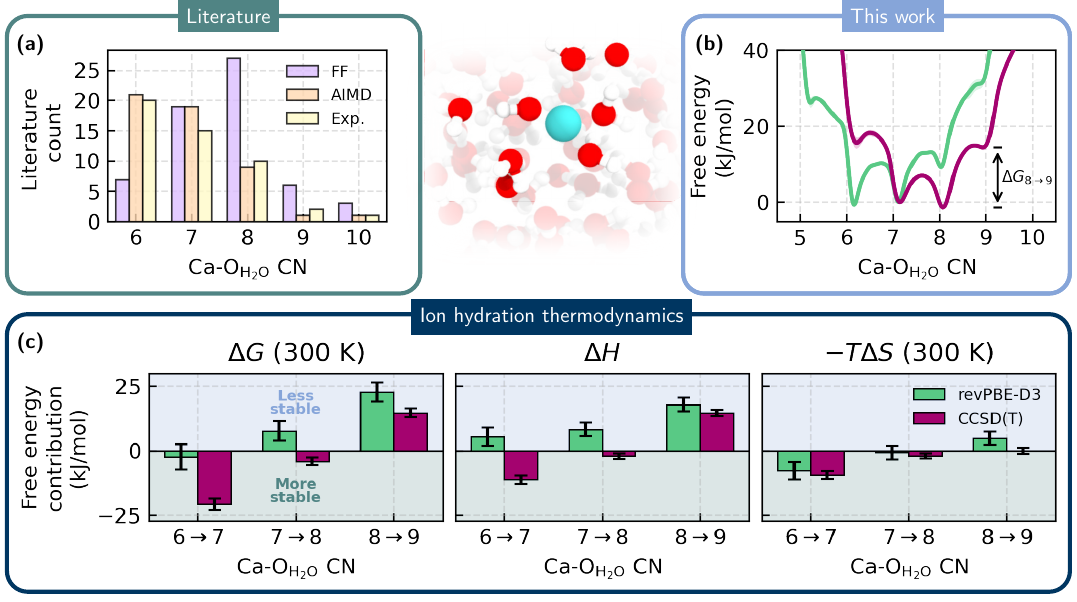}
    \caption{\textbf{Resolving the calcium hydration mechanism:} (a) Summary of previous literature reporting the coordination number of calcium in water. Results are further broken down into experiment and simulations using classical force field models (FF) and first principles methods (AIMD). Full details of the systems, experimental and simulation setups are tabulated in Tables \ref{si-tab:aimd_mlip_calcium_cn}, \ref{si-tab:force_field_calcium_cn} and \ref{si-tab:experimental_calcium_cn} of the SI. (b) Calcium-water coordination number free energy from OPES simulations for revPBE-D3 and CCSD(T) MLPs. The profiles are set to 0 at CN=7. Other levels of theory studied in this work are included in the SI,  Fig.~\ref{si-fig:coordination}, as well as a comparison to the classical FF from \citet{armstrongSolubilityconsistentForceField2023} (c) Breakdown of thermodynamics of ion hydration. 7$\rightarrow$8 means the free energy to add a water, and is calculated as $\Delta G_{CN=8}-\Delta G_{CN=7}$.  From left to right are the free energy (at 300 K), the enthalpy and entropic free energy contribution. The full temperature dependence of CN transition free energies used to obtain the enthalpy and entropy are given in the SI Fig.~\ref{si-fig:temperature_cacn}.}
    \label{fig:ca-cn}
\end{figure*}
{\color{black}
The free energy cost to insert or remove a water and reorganise an ion’s hydration shell governs its (in)ability to access alternative coordination environments where dehydration from its bulk solvation structure is typically required.
Such processes span a remarkably broad range of chemistry, including ion pairing and mineral formation \cite{pianaAssistedDesolvationKey2006a}, binding at biological sites, transport through membranes and nanopores \cite{zhangCa2PermeationMechanism2020, moRegulatingIonAffinity2024}, and electrochemical phenomena \cite{rodellarIonSolvationKinetics2024,renRoleInnerSolvation2020} .
However, a survey of the literature in Figure~\ref{fig:ca-cn}(a) shows that even the seemingly simple quantity of the calcium--water (Ca--$\mathrm{O_{H_2O}}$) coordination number (CN) -- the number of water molecules in the ion's first hydration shell -- remains surprisingly poorly resolved, with substantial disagreement across both experiment and simulation.
Experimentally, inferring CNs from ensemble-averaged scattering or spectroscopic signals makes it difficult to distinguish between multiple, rapidly interconverting local hydration structures \cite{masonHydrationBiologicallyRelevant2024} and this results in a wide range of reported CNs (between 5-10).

However, even the atomistic resolution afforded by simulations does not resolve this ambiguity, with reported CNs spanning a similarly broad range.
The simulation literature can be further divided into studies based on classical force fields and those employing first-principles methods (mainly via \textit{ab initio} molecular dynamics (AIMD)).
Classical force fields parameterised primarily against macroscopic observables are not necessarily constrained to reproduce the microscopic hydration structure.
In fact, even if the classical force fields were to be explicitly parameterised against CN experimental data, the substantial spread among the available measurements would leave considerable freedom in the resulting microscopic hydration structure~\cite{masonHydrationBiologicallyRelevant2024}.
On the other hand, for AIMD results, statistical sampling presents a particular challenge: we show in the SI, Table~\ref{si-tab:ca_co3_water_structural_summary}, that the mean residence time of water within the first hydration shell of \ch{Ca^{2+}} is approximately 25~ps (regardless of functional choice), comparable to or longer than the total trajectory lengths accessible to many previous AIMD studies (see SI, Table \ref{si-tab:aimd_mlip_calcium_cn} for full details of literature AIMD simulation setups).
Furthermore, in the SI, Fig.~\ref{si-fig:finitesize_cn} we show significant finite size effects arising from the small box sizes typically associated with AIMD simulations, where lower CNs can be favoured by up to 13~\(\mathrm{kJ\,mol^{-1}}\).
In Figure~\ref{fig:ca-cn}(b), we resolve this question directly by computing the free energy as a function of the Ca--$\mathrm{O_{H_2O}}$ CN at 300~K using OPES simulations of an isolated Ca ion in water with the CCSD(T) model validated in the previous section.
{\color{black}To ensure convergence with respect to the aforementioned finite-size effects, the simulations contain 197 water molecules in a cubic box of length approximately $18$~\AA{}; further simulation details and convergence tests are respectively provided in Sec.~\ref{si-sec:cn_opes} and Fig.\ref{si-fig:finitesize_cn} of the SI.}
For comparison, we report here also results for revPBE-D3 since it shows the greatest dissimilarity to CCSD(T), with results for the other models developed in this work given in the SI, Fig. \ref{si-fig:coordination}.

Both CCSD(T) and revPBE-D3 predict a broad coordination landscape, with 5-, 6-, 7-, 8-, and 9-fold coordination states accessible, consistent with the wide range of CNs reported in the literature in panel~(a).
Crucially, however, the relative stability of these states changes substantially with the underlying model.
CCSD(T) stabilises the higher 7- and 8-fold coordination states, whereas revPBE-D3 preferentially stabilises the lower 6- and 7-fold states.
This trend towards higher coordination with increasing electronic structure accuracy is also evident in the radial distribution functions (RDFs) shown in SI, Fig.~\ref{si-fig:rdfs_ca}, where the cumulative coordination number distributions systematically shift toward larger values.

We next examine the origin of the increased preference for higher coordination at the CCSD(T) level by decomposing the free energy changes for each coordination transition (as illustrated in Fig.~\ref{fig:ca-cn}(b)) into enthalpic and entropic contributions for both CCSD(T) and revPBE-D3 in Fig.~\ref{fig:ca-cn} (c). 
Increasing coordination from 6 to 7 to 8 is actually entropically favoured for both CCSD(T) and revPBE-D3, and instead the enthalpy component is what determines its overall favourability, being positive (unfavourable) for revPBE-D3 and negative (favourable) for CCSD(T).
Since the dominant difference (in this case) between revPBE-D3 and CCSD(T) arises from the internal energy term, we further examine how the methods describe the underlying ion--water interactions.
In the SI, Fig.~\ref{si-fig:interaction_energy}, we compare the interaction energies of a set of clusters extracted at each coordination number. The relative energies show that MP2 and CCSD(T) have a preference for higher coordination numbers compared with revPBE-D3.
These results suggest that differences in the description of ion--water interactions provide an important explanation for the lower coordination numbers predicted by revPBE-D3.

Unlike the wide range of accessible states for the Ca--O$_{\mathrm{H_2O}}$ CN predicted by all methods, we show in the SI, Fig~\ref{si-fig:coordination} that the \ch{CO3}--O$_{\mathrm{H_2O}}$ profiles are much broader, with fewer discrete coordination states.
This observation is in line with carbonate's ability to act as hydrogen bond acceptor at 3 equivalent sites, allowing a more flexible hydration environment around carbonate.
Similarly as for \ch{Ca^{2+}}, we observe a shift towards lower coordination numbers of \ch{CO3} with water for lower levels of theory.
This may be explained by the greater localization of charge for MP2 and CCSD(T) enabling stronger binding with the waters in the first solvation shell, with lower levels of theory having more diffuse charge environments around the anions.}

These profiles again highlight the challenge of obtaining reliable atomistic insights to rationalise experiment from simulation when only considering 0 K properties, or alternatively when the free energy is computed at a lower level of theory.
The wide range of accessible states, coupled with the sensitive differences between levels of theory emphasise the importance of sampling the full free energy surface with a sufficiently accurate description of  the electronic structure.

\section{Discussion}
Years of effort in the parameterisation of classical force-field models for \ch{CaCO3} also achieve good experimental agreement for the ion pairing free energies, enthalpy, entropy and solvation structure \cite{armstrongSolubilityconsistentForceField2023} as shown in Figure \ref{si-fig:ff_comp} of the SI.
Such models have been directly fitted to match experimental hydration free energies of the ions, (which is correlated to a large extent to the ion pairing association free energy), and so transferability is not guaranteed for more complex situations.
\textcolor{black}{
The Ca$^{2+}$ hydration mechanism serves as a cautionary example. 
In this case, although the force field reproduces the targeted experimental free energies (SI, Fig.~\ref{si-fig:ff_comp}), its hydration structure differs markedly from CCSD(T) (SI, Fig.~\ref{si-fig:coordination}), showing that agreement with macroscopic thermodynamics instead reflects its parameterisation and does not guarantee a faithful description of the underlying microscopic hydration.}

On the other hand, our MLP-based framework provides a systematic path towards the correct solution, starting from a robust MP2 MLP, before adding the $\Delta$-MLP to reach CCSD(T).
\textcolor{black}{The close similarity in free energy predictions and mechanistic insights from both MP2 and CCSD(T) gives us confidence in the reliability of our predictions.}
Importantly, the models developed in this work require no empirical fitting parameters to achieve experimental agreement. 

The value of this approach becomes more apparent when the steps beyond ion pairing are considered, where an accurate model has been sorely needed to resolve open questions regarding the nucleation mechanism of \ch{CaCO3}.
In particular, there is intense debate over possible non-classical mechanisms including liquid-liquid phase separation \cite{wallaceMicroscopicEvidenceLiquidLiquid2013,zouPhaseDiagramCalcium2017, jinFormationChemicalEvolution2025}, and transformation to the solid phase via thermodynamically stable pre-nucleation clusters \cite{finneyProbingStructureStability2013,luRoleWaterCaCO32021, gebauerPrenucleationClustersSolute2014, demichelisStablePrenucleationMineral2011, gebauerStablePrenucleationCalcium2008}.
Alternative simulation work proposes that such clusters play no significant role in nucleation, and instead dissolve after extended simulation times, and that nucleation proceeds via aggregation of ion pairs that can be described from a purely classical nucleation theory based framework \cite{smeetsClassicalViewNonclassical2017, henzlerSupersaturatedCalciumCarbonate2018}.
Such differences manifest in the experimental range shown in Fig. \ref{fig:pmf}, which arises from fitting a speciation model to the linear binding profile from potentiometric titration measurements \cite{gebauerPrenucleationClustersNonclassical2011, plummerSolubilitiesCalciteAragonite1982}.
The methodological framework developed in this work is primed for extension to tackle the increased complexity of multiple ion aggregation.
In addition, obtaining reliable thermodynamics of the ions in solution relative to the solid-state will necessitate models that can faithfully reproduce the experimental solubility \cite{armstrongSolubilityconsistentForceField2023}, and so obtaining hydration free energies within our MLP framework will be an important next step \cite{harrymooreComputingSolvationFree2026} as well. 

    {\color{black} Accurately describing the ion hydration mechanism, as  shown here, also has implications beyond bulk solution.
In particular, ion transport through confined environments -- crucial for biological ion channels, desalination membranes and ion-conducting membranes in electrochemical technologies -- often requires (partial) dehydration of the solvation shell.
Crucially, relatively small changes in ion coordination can strongly affect permeation barriers and conductance.
For example, widely used non-polarisable classical forcefields have been shown to overbind Ca, leading to qualitatively differing transport mechanisms in biological \ch{Ca^{2+}} channels. \cite{zhangCa2PermeationMechanism2020,xueDecipheringCa2Permeation2025}
Quantitatively resolving the relative stability of individual hydration states therefore will provide an important foundation for predictive simulations of ion transport and selectivity in complex environments.}

The comparison between DFT and cWFT methods also highlights the importance of explicitly describing the solvent degrees of freedom at the same CCSD(T) level of theory as the main species of interest.
For example, implicit solvation models continue to be the workhorse for large scale studies of chemical reactivity in solution \cite{marenichUniversalSolvationModel2009}.
We show that entropic contributions to the ion pairing differ significantly from CCSD(T) to lower levels of theory, while previous work also highlighted the significant challenges in reproducing water properties with DFT \cite{oneillRoutineCondensedPhase2025, monterodehijesDensityIsobarWater2024}.
Therefore, the approach developed in this paper serves as a route forward for treating both the solvent and reactive components on an equal footing at a high level of theory.
\textcolor{black}{More broadly, the consistently high accuracy of CCSD(T) across both the solvent and ionic interactions considered here suggests a route towards transferable, high-accuracy models of chemical reactivity in solution. 
Although MP2 performs comparably to CCSD(T) for the present ion pairing energetics, it is known to substantially overstructure liquid water (see SI, Fig.~\ref{si-fig:mp2ccsdt_valid}) and produce overly sluggish dynamics~\cite{willowWhyMP2WaterCooler2016}.
Here we show that CCSD(T), by contrast, provides a quantitatively accurate description of both the solvent and the ion--water/ion--ion interactions, making it a more compelling foundation for transferable simulations of reactivity in solution.}
This approach would be particularly amenable to explore reactivity questions between carbonate, bicarbonate and carbonic acid \cite{piaggiInitioMachinelearningSimulation2025}, where the fixed topology of many classical force fields has prevented their application.

Finally, while still an order of magnitude or so slower than classical force fields, our MLP approach has allowed the accumulation of microseconds worth of cWFT-quality simulations.
This has enabled us to obtain uncertainties on the free energies, entropy and enthalpy that are of the same order as those from classical forcefield simulations. 
This has been made possible by coupling the Symmetrix library ( a C++, Kokkos optimized MACE MLP implementation\cite{wcwittWcwittSymmetrix2026, kovacsMACEOFFShortRangeTransferable2025}) and efficient statistical sampling from OPES. 
In terms of the cost investment to generate the data for training the MLP models, CCSD(T) is quite affordable with just 50k CPUh required, while the majority of the computational cost actually comes from the periodic MP2 calculations, which is between one and two orders of magnitude larger.

\section{Conclusions}

In this paper we have introduced an approach to study ions in solution at the `gold-standard' CCSD(T) level of accuracy.
This work generalizes previous $\Delta$-machine learning approaches for liquid water to now handle the increased complexity of solvated ions.
We combine this with state of the art enhanced sampling methods to converge both the accuracy of the PES and statistical sampling, while explicitly treating the solvent degrees of freedom.
We show such a level of accuracy on the PES is necessary to obtain reliable agreement to experiments on the free energy, enthalpy and entropy of ion pairing, as well as bulk water properties, with no other \textit{ab initio} methods capable of simultaneously getting all within experimental uncertainties.
This reliability has also enabled new atomistic understanding into ion pairing as well as the hydration structure around ions, highlighting salient differences with lower levels of theory.
Such detailed insights into the solvation structure of individual ions are challenging to obtain experimentally while its reliability can be difficult to ascertain when inferred from empirical force-field simulations.

The approach taken within this work can readily be adopted by other research groups.
All the codes used are freely available to academics and we provide the models, steps and code to generate the data for $\Delta$-learning.
We expect little modification is required to extend this towards other solution-based processes such as chemical reactions \cite{zhangModellingChemicalProcesses2024} as well as higher concentrations of ions.
We also foresee exciting new developments on the horizon on both the electronic structure theory and MLP fronts that will significantly benefit our approach.
For example, explicit treatment of long-range electrostatic interactions in the MLPs should allow for more diverse systems, including interfaces and highly charged systems, that are difficult to describe by current purely short-ranged models \cite{yueWhenShortrangeAtomistic2021, niblettLearningIntermolecularForces2021}, while ongoing developments in obtaining gradients of the CCSD(T) PES \cite{zhangDifferentiableQuantumChemistry2022} will further facilitate the training of such models for more complex and demanding systems.

Besides being able to derive new insights, the routine development of CCSD(T) models opens a new paradigm towards the development of newer, cost-efficient methods.
For example, classical force fields typically require parameterization against experimental data, which are not always readily available and can obscure mechanistic interpretation, whereas the CCSD(T) models developed here provide both quantitative accuracy and direct mechanistic insight.
Nevertheless, the next steps towards, for example, nucleation will still pose an arduous computational challenge even for the relative efficiency of the models trained here.
The idea of using these models to train cheaper, more efficient MLP architectures in the spirit of knowledge distillation \cite{gardnerDistillationAtomisticFoundation2025} while in principle retaining the CCSD(T) quality is an exciting one for further exploration.
Additionally these models will also be useful for benchmarking DFT functionals, where we show that commonly used functionals for aqueous systems, such as revPBE-D3 and revPBE0-D3, fail to achieve quantitative accuracy with experiment for the ion pair association free energy of \ch{CaCO3}, {\color{black} as well as predicting qualitatively differing ion hydration mechanisms.}
By bringing CCSD(T)-level accuracy to explicitly solvated systems, this work lays the foundation for predictive simulations of complex chemical processes in solution.

\section*{Methods}
\subsection{Machine learning potentials}
In this paper we follow two approaches to develop MLPs at various levels of electronic structure theory.
The key details are described here, with a more thorough description given in Section \ref{si-sec:caco3si_models} of the SI.
The ML framework used in this work is based on the MACE architecture \cite{batatiaMACEHigherOrder2022}, which has been shown to be particularly successful in faithfully reproducing the underlying PES with high accuracy \cite{kovacsEvaluationMACEForce2023} and data efficiency.

For all reference methods up to and including MP2 we use standard protocols to fit MLPs to periodic reference data. 
For CCSD(T) accuracy we correct this 'lower-level` cWFT PES using a $\Delta$-learning procedure.

\textbf{Periodic models:}
The models based on revPBE-D3, revPBE0-D3, RPA and MP2 were trained using periodic reference data labelled with energies and forces.
We exploit highly efficient implementations of both RPA \cite{delbenEnablingSimulationFifth2015} and MP2 \cite{delbenForcesStressSecond2015} within the CP2K code~\cite{kuhneCP2KElectronicStructure2020} which provides access to both energies and gradients for periodic boundary conditions (PBCs).

We use a 2-step process to obtain reliable models as illustrated in the central panel of Figure \ref{fig:schematic}, where first the PES is roughly sampled to generate an initial model that is then systematically improved in a second targeted sampling step.
A common initial dataset sampled from previous AIMD simulations \cite{raiteriIonPairingMultiple2020} containing roughly 300 structures was used to generate initial models for each level of theory.
We emphasise that AIMD simulations are not strictly necessary, and indeed sampling configurations from foundation models or classical force field simulations is sufficient to obtain a reliable initial model.

We and others have shown in particular that sampling from a range of (positive and negative) pressures, generates robust models for stable simulations in the isothermal-isobaric ensemble \cite{monterodehijesDensityIsobarWater2024, oneillRoutineCondensedPhase2025}.
Therefore, after generating an initial model, it is then refined to ensure that it is quantitatively accurate for the underlying reference level of theory, as well as suitable for NPT simulations, by sampling using the first generation model over a range of pressures, as well as from biased simulations to ensure adequate sampling over the various solvated and ion pairing states.
We find that simply one extra round of iterative training, as described, results in models that can then be immediately used in simulations to obtain thermodynamic properties.
Figure \ref{si-fig:dataset} in the SI illustrates the configurational space covered by each dataset.
The total number of configurations in each of the final models was between 700 - 900.
All periodic models were trained on energies and forces with 2 message passing layers with 128 channels and a 6~\AA~radial cutoff. 

\textbf{$\Delta$-learning workflow:}
To obtain the CCSD(T) level models, we employ a $\Delta$-learning strategy that we first introduce in Ref.~\cite{oneillRoutineCondensedPhase2025}.
We generalise our approach for bulk water to now tackle ions in solution.
The general strategy is that a baseline model trained on energies and forces at the MP2 level is corrected to CCSD(T) using clusters of the solvated ion pair cleaved out from bulk simulations at the MP2 level as depicted in the left-most panel of Figure \ref{fig:schematic}.
The CCSD(T) $\Delta$-MLP model was trained on the energy differences between MP2 and CCSD(T).
The close agreement between MP2 and CCSD(T), along with their absence of delocalization error, enables highly data-efficient $\Delta$-MLPs.
These models require far fewer data points to capture the more complex ion pairing problem (while also being able to model bulk liquid water) -- using only 1,133 configurations -- compared to a recent study~\cite{oneillRoutineCondensedPhase2025} that needed ${\sim}6,000$ configurations to model only bulk liquid water.

In practice, a similar 2-step procedure as the periodic models was used, where first, clusters are cut from the MP2 dataset to give an initial CCSD(T)-level $\Delta$-MLP (to sum with the final MP2 MLP).
The cluster dataset was then augmented by cutting out clusters from NPT simulations over a range of pressures and solvation states using the initial CCSD(T) model to give the final CCSD(T) $\Delta$-MLP.
The clusters were cut such that all waters within a 5.5~\AA~cutoff of each ion in the ion pair were included (as illustrated in Figure \ref{si-fig:cluster} of the SI).
All clusters contained an ion pair and were thus neutral.

In summary, this simple framework combines many advances in methods across the materials modeling field, namely electronic structure methods, machine learning architectures, data-efficient training procedures and enhanced sampling algorithms to develop models that can provide accurate thermodynamic insight at cWFT levels of theory.
Additional details on the models, including training errors and validation tests for both the periodic MP2 MLP and $\Delta$-MLP model are given in Section \ref{si-sec:caco3si_models} of the SI.
In particular we validate our $\Delta$-learning approach by demonstrating that the final CCSD(T) model can reliably correct the known deficiencies of MP2 for bulk water, where it overstructures the oxygen-oxygen radial distribution function \cite{delbenBulkLiquidWater2013}.
This CCSD(T) model also correctly reproduces the experimental water density.

\subsection{Coupled Cluster Theory}
For the (cluster-based) dataset used to train the $\Delta$-MLP, we calculated MP2 with local approximations~\cite{nagyIntegralDirectLinearScalingSecondOrder2016,szaboLinearScalingOpenShellMP22021} (LMP2) and CCSD(T) with the local natural orbital approximation (LNO)~\cite{nagyOptimizationLinearScalingLocal2018,gyevi-nagyIntegralDirectParallelImplementation2020}.
These calculations were both performed in MRCC~\cite{kallayMRCCProgramSystem2020}, with the LMP2 numbers coming naturally from the LNO-CCSD(T) calculation.
We used the default ``normal'' LNO thresholds -- recently validated for the properties of water in Ref.~\cite{oneillRoutineCondensedPhase2025} -- and use density-fitting (with the def2-QZVPP-RI-JK auxiliary basis set) to speed up the HF calculations.
We included the sub-valence correlation of the 3s and 3p electrons on the Ca atom with the corresponding cc-pwCV$X$Z~\cite{petersonAccurateCorrelationConsistent2002,balabanovSystematicallyConvergentBasis2005} basis set, where $X$ represents the cardinal number.
We use the aug-cc-pV$X$Z basis set on the O and cc-pV$X$Z basis set on the H and C atoms, while the cc-pwCV$X$Z basis set was used for the Ca atoms.
For all other atoms, we used the cc-pV$X$Z family of (valence) basis sets, in particular using the diffuse aug-cc-pV$X$Z basis sets on the O atom.
For the auxiliary basis sets required for the LNO-CCSD(T) calculations, we used those generated in Ref.~\citenum{weigendRIMP2OptimizedAuxiliary1998} and Ref.~\citenum{hellwegOptimizedAccurateAuxiliary2007} for each atom.
However, as none were available for Ca, we used the automatic auxiliary basis functions code of Stoychev \etal{}~\cite{stoychevAutomaticGenerationAuxiliary2017,lehtolaStraightforwardAccurateAutomatic2021} to generate these basis functions.
We performed complete basis set (CBS) extrapolations using parameters taken from Neese and Valeev~\cite{neeseRevisitingAtomicNatural2011}.
\subsection{Periodic DFT + MP2}
All reference calculations were done on a unit cell of length $\sim 14~\text{\AA}$ containing 126 waters and one ion pair.
revPBE-D3 calculations were performed with FHI-AIMs \cite{blumInitioMolecularSimulations2009}, using the \texttt{tight} family of basis sets and Grimme's D3 (zero-damping variant) \cite{grimmeConsistentAccurateInitio2010}. 
All periodic calculations were performed at the $\Gamma$-point.
revPBE0-D3, RPA and MP2 calculations were performed with the CP2K/Quickstep \cite{kuhneCP2KElectronicStructure2020} code.
The electronic density was partitioned into core and valence electrons, with core electrons described with the norm-conserving Goedecker, Teter and Hutter (GTH) pseudopotentials \cite{goedeckerSeparableDualspaceGaussian1996}.
For revPBE0-D3, hydrogen and oxygen atoms were described with the TZV2P family of basis sets, while we used the ccGRB family for carbon and calcium.
revPBE0-D3 calculations used the auxiliary density matrix methods (ADMM) approximation \cite{guidonAuxiliaryDensityMatrix2010} to improve the efficiency.
For RPA and MP2 we used the cc-TZ basis set for oxygen and hydrogen, with the ccGRB basis set for carbon and calcium.
PBE was used as a starting point for the RPA correlation energy
calculations.
The resolution-of-identity (RI) technique was used for these methods \cite{bussySparseTensorBased2023}.
We used triple-zeta (TZ) quality correlation-consistent basis sets for H and O (taken from CP2K’s \texttt{RI\_cc\_TZ set}) and correlation-consistent basis sets designed by Ye and Berkelbach \cite{yeCorrelationConsistentGaussianBasis2022} for
carbon and calcium.
Auxiliary basis sets for the RI integral operations were generated using the automatic auxiliary basis of Stoychev et al. \cite{stoychevAutomaticGenerationAuxiliary2017, lehtolaStraightforwardAccurateAutomatic2021} for calcium and carbon, with the defaults (from CP2K) used for H and O.
For consistency with previously developed models \cite{oneillPairNotPair2024} a planewave cutoff of 1200 Ry was used in the CP2K calculations (although in practice the forces are converged after around 400 Ry).
\subsection{Molecular Dynamics}
All simulations were performed using the Large-scale Atomic Molecular Massively Parallel Simulator (LAMMPS)~\cite{thompsonLAMMPSFlexibleSimulation2022} code combined with PLUMED \cite{bonomiPromotingTransparencyReproducibility2019}, coupled with the \texttt{Symmetrix} library \cite{wcwittWcwittSymmetrix2026, kovacsEvaluationMACEForce2023}, using the \texttt{symmetrix/}\texttt{mace} pair style in tandem with the \texttt{hybrid/overlay} pairstyle to sum the MP2 baseline and $\Delta$-MLPs. 

\subsection{Free energy calculations}
All MLP-based simulations to obtain the ion pairing free energies were performed in a simulation box of length $\sim18~\text{\AA}$ with 197 waters and 1 \ch{CaCO3} ion pair.
For each level of theory, the equilibrium density was first obtained for a given temperature from simulations in the NPT ensemble using the CSVR thermostat \cite{bussiCanonicalSamplingVelocity2007} with a temperature relaxation time of 0.1~ps and a 1.0~fs timestep using deuterium masses for hydrogen atoms. 
These trajectories were then used to generate initial configurations for the OPES simulations to obtain the free energy profiles from which the association free energies were computed \cite{invernizziRethinkingMetadynamicsBias2020}.
The efficient convergence behavior of OPES was used to enable multiple replicas to obtain sufficient statistics to robustly estimate the free energy of ion pairing.
For each level of theory, simulations were performed at 5 temperatures (300, 310, 320, 330, 340 K).
At each temperature 6 independent replicas were initialized from randomized velocities from the Maxwell-Boltzmann distribution for the given temperature, and run for between 20 - 30 ns each.
The bias update frequency was 500 steps, and the barrier for maximum free energy exploration 30 kJ/mol following Ref. \cite{piaggiInitioMachinelearningSimulation2025}, with the Ca -- \ch{CO3} separation being explicitly biased.
The Ca-O$_{\text{\ch{H2O}}}$ coordination number was computed on the OPES trajectory via a smooth switching function:
\begin{equation}
S(r)=
\frac{1-\left(\dfrac{r-d_0}{r_0}\right)^8}
{1-\left(\dfrac{r-d_0}{r_0}\right)^{16}} \,
\end{equation}
with $d_0$=1.8~\AA~and $r_0$=1.5~\AA~to overlap with the first peak in the \ch{Ca}-O RDF.
After $\sim$ 1 - 3 ns the bias during the simulation had stabilised, enabling reversible exploration of the free energy landscape.

The unbiased probability distribution ($P(d,n)$) was reconstructed from the OPES simulations by reweighting the OPES trajectory using the time-dependent bias potential $V_{\text{bias}}$, as a function of both collective variables, the \ch{Ca}--\ch{CO3} separation ($d(\mathbf{r})$) and the Ca-O$_{\text{\ch{H2O}}}$ coordination number ($n(\mathbf{r})$):
\begin{equation}
P(d,n) =
\frac{
\left\langle
\delta\!\left(d-d(\mathbf{r})\right)
\delta\!\left(n-n(\mathbf{r})\right)
e^{\beta V_{\text{bias}}(\mathbf{r})}
\right\rangle_V
}{
\left\langle
e^{\beta V_{\text{bias}}(\mathbf{r})}
\right\rangle_V
}
\end{equation}
where $\langle \cdot \rangle_V$ corresponds to the ensemble average taken over the biased simulation.
The final free energy surface is then given by:
\begin{equation}
F(d,n) = -k_B T \ln P(d,n)
\end{equation}

To obtain the binding free energies, the Ca-O coordination number first was integrated out from the 2D free energy to give the 1D PMF $\phi(\mathbf{r})$.
The tail of the PMF was then aligned with the analytic solution of two point charges $q_i$ and $q_j$ interacting via a screened dielectric potential with dielectric constant $\epsilon_r$:
\begin{equation}
    \Delta G = \frac{1}{4\pi \epsilon_0}\frac{q_iq_j}{\epsilon_rr}-k_BT\text{ln}(4\pi r^2).
\end{equation}
with the final standard ion pairing free energy $\Delta G_{\text{IP}}^{\ominus}$ then given by integrating the aligned PMF ($\phi'(r)$) up to the edge of the bound state $R_c$:
\begin{equation}
    \Delta G_{\text{IP}}^{\ominus} = -RT\ln\Bigg[ c^{\ominus} \int_0^{R_c}\exp\bigg(-\frac{\phi'(r)}{k_BT}\bigg)dr\Bigg].
\end{equation}
Here $c^{\ominus}$ is a concentration unit conversion from atoms/\AA$^3$ to mol/dm$^3$ required to correctly relate the simulated association free energy to that obtained from experiment. 
We used the experimental value of the dielectric constant $\epsilon_r$ (78), and show in Figure \ref{si-fig:dielectric} of the SI the sensitivity of our results to this choice.
The extent of finite size effects on the binding free energies and enthalpy/entropy calculations reported in Figure \ref{si-fig:finitesize} in the SI was tested using a classical force field \cite{armstrongSolubilityconsistentForceField2023} on a cubic box of length approximately $50~\text{\AA}$ containing 5319 waters and one ion pair.

\section*{Data availability}
The data required to reproduce this study is provided in the following Github repository: https://github.com/niamhon/caco3-ion-pairing

\section*{Supporting Information}
Details on model training, electronic structure convergence tests, Bader charge analysis, enhanced sampling simulations, protocol for computing free energies, system size convergence tests, comparison to previous simulation literature for ion pairing free energy and calcium-water coordination number and compilation of previous literature of calcium-water coordination number from simulation and experiment.

\section*{Code availability}
All simulations were performed with publicly available simulation software (\texttt{ACEsuit}, \texttt{LAMMPS}, \texttt{Symmetrix} and \texttt{PLUMED}).

\section*{Acknowledgments}
N.O.N acknowledges financial support from the Gates Cambridge Trust and is grateful for an International Exchange Grant funded by the Royal Society. The Flatiron Institute is a division of the Simons Foundation. W.C.W. acknowledges support from the EPSRC (Grant EP/V062654/1). A.M. acknowledges support from the European Union under the ``n-AQUA'' European Research Council project (Grant No. 101071937). J.D.G and B.I.A thank the Australian Research Council for funding under grant FL180100087, as well as the Pawsey Supercomputing Centre and National Computational Infrastructure for computing resources. C.S. acknowledges financial support from the Royal Society, grant number RGS/R2/242614 and the Isaac Newton Trust, award number G122390. We are grateful for computational support and resources from the UK national high-performance computing service, Advanced Research Computing High End Resource (ARCHER2). Access for ARCHER2 was obtained via the UK Car-Parrinello consortium, funded by EPSRC grant reference EP/P022561/1. We also acknowledge EuroHPC Joint Undertaking for awarding the project ID EHPC-REG-2024R02-130 access to Leonardo at CINECA, Italy. This work was also performed using resources provided by the Cambridge Service for Data Driven Discovery (CSD3) operated by the University of Cambridge Research Computing Service (www.csd3.cam.ac.uk), provided by Dell EMC and Intel using Tier-2 funding from the Engineering and Physical Sciences Research Council (capital grant EP/T022159/1), and DiRAC funding from the Science and Technology Facilities Council (www.dirac.ac.uk).

\bibliography{bibliography}

\newpage

\end{document}


\title{\mytitle}

\author{Niamh O'Neill}%
\affiliation{Yusuf Hamied Department of Chemistry, University of Cambridge, Lensfield Road, Cambridge, CB2 1EW, UK}
\affiliation{Cavendish Laboratory, Department of Physics, University of Cambridge, Cambridge, CB3 0HE, UK}
\affiliation{Lennard-Jones Centre, University of Cambridge, Trinity Ln, Cambridge, CB2 1TN, UK}
\affiliation{Max Planck Institute for Polymer Research, Mainz 55128, Germany}

\author{Benjamin X. Shi}%
\affiliation{Initiative for Computational Catalysis, Flatiron Institute, 160 5th Avenue, New York, NY 10010, USA}

\author{William C. Witt}%
\affiliation{Harvard John A. Paulson School of Engineering and Applied Sciences, Harvard University, Cambridge, MA 02134, USA}

\author{Blake I. Armstrong}%
\affiliation{School of Molecular and Life Sciences, Curtin University, PO Box U1987, Perth, Western Australia 6845, Australia}

\author{Asmita Niyogi}%
\affiliation{Yusuf Hamied Department of Chemistry, University of Cambridge, Lensfield Road, Cambridge, CB2 1EW, UK}
\affiliation{Lennard-Jones Centre, University of Cambridge, Trinity Ln, Cambridge, CB2 1TN, UK}

\author{William J. Baldwin}%
\affiliation{Lennard-Jones Centre, University of Cambridge, Trinity Ln, Cambridge, CB2 1TN, UK}
\affiliation{%
Engineering Department, University of Cambridge, Trumpington St, Cambridge CB2 1PZ, UK
}

\author{Paolo Raiteri}%
\affiliation{School of Molecular and Life Sciences, Curtin University, PO Box U1987, Perth, Western Australia 6845, Australia}

\author{Christoph Schran}%
\email{cs2121@cam.ac.uk}
\affiliation{Cavendish Laboratory, Department of Physics, University of Cambridge, Cambridge, CB3 0HE, UK}
\affiliation{Lennard-Jones Centre, University of Cambridge, Trinity Ln, Cambridge, CB2 1TN, UK}

\author{Angelos Michaelides}%
\email{am452@cam.ac.uk}
\affiliation{Yusuf Hamied Department of Chemistry, University of Cambridge, Lensfield Road, Cambridge, CB2 1EW, UK}
\affiliation{Lennard-Jones Centre, University of Cambridge, Trinity Ln, Cambridge, CB2 1TN, UK}

\author{Julian D. Gale}%
\email{j.gale@curtin.edu.au}
\affiliation{School of Molecular and Life Sciences, Curtin University, PO Box U1987, Perth, Western Australia 6845, Australia}

\date{\today}

\maketitle
\tableofcontents

\newpage

\section{Models}\label{sec:caco3si_models}
All of the MLPs in this work are based on the MACE framework \cite{batatiaMACEHigherOrder2022}.
Two approaches for obtaining MLPs were used depending on the reference method. 
For the DFT, RPA and MP2 based models, routine protocols based on fitting to \textit{periodic reference data} were used, as described in \ref{sec:periodic} below.
To obtain CCSD(T)-level models, a \textit{$\Delta$-learning approach} first introduced in Ref.\cite{oneillRoutineCondensedPhase2025} was used.
This has been extended within this work to enable routine application to ions in solution.
Rather than a baseline periodic DFT MLP model, we use a periodic MP2 model instead, which was `corrected' up to CCSD(T)-level of theory using a second $\Delta$-model trained on the difference between MP2 and CCSD(T) energies only.
This is described in greater detail in Section \ref{sec:delta} below.
\subsection{Periodic models}\label{sec:periodic}
All (MLP) models were trained over multiple generations, with the aim of faithfully reproducing the potential energy surface (PES) of the underlying level of theory.
An initial dataset of \ch{CaCO3}, sampled from previous \textit{ab initio} simulations from Ref.~\cite{raiteriIonPairingMultiple2020} was used to generate an initial `Generation-1' model at various levels of theory (revPBE-D3, revPBE0-D3, RPA, MP2).
As well as targeting the standard energy and force RMSEs with respect to a validation set, we found that directly benchmarking the models against desired thermodynamic properties was important.
Therefore we also `self-consistently' converged the models for the density and ion pairing free energy.
To ensure a robust dataset in the second generation, the following conditions were sampled using the Generation 1 models at each level of theory:
\begin{itemize}
    \item Configurations sampled from equilibrium molecular dynamics (MD) trajectories of 126-water boxes with one Ca--\ch{CO3} ion pair generated at a range of pressures (specifically -1500, -500, -300, 1, 2, 500, 1000, 4000, 8000 bar). Such density and pressure fluctuations have been shown to be important to obtain reliable densities \cite{oneillRoutineCondensedPhase2025, monterodehijesDensityIsobarWater2024}.
    \item Configurations obtained from OPES simulations to ensure complete sampling of both the ion pair separation and the coordination environments.
\end{itemize}
The total number of configurations in each of the final models was between 700 - 900.
All periodic models comprised two message passing layers, with 128 channels and a radial cutoff of 6 \AA. 
There was a 5 \% train/test split and training was run for 300 epochs, where the energy weight of the loss function was increased to 1000 for the last 50 epochs.
Figure \ref{fig:dataset} below summarises the sampling of configuration space for each model.
\begin{figure}[h!]
  \centering
   \includegraphics[width=1.0\textwidth]{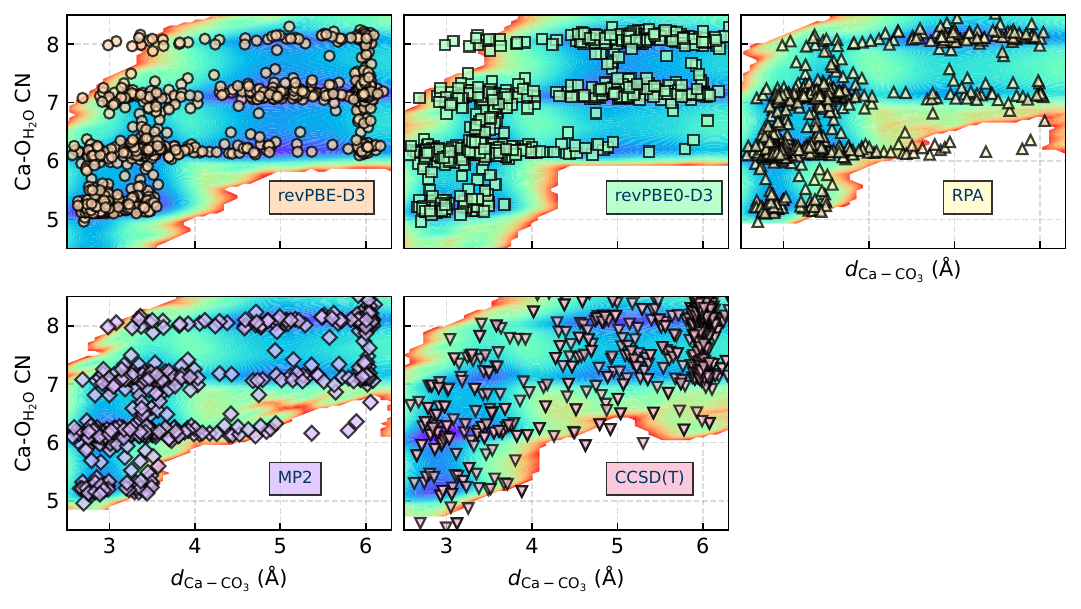}
    \caption[Periodic model validation]{\textbf{Dataset generation:} Summary of configurational space covered by each of the datasets for each level of theory, given in terms of the two collective variables important for ion pairing: $d_{Ca-CO_3}$ and Ca-O$_{\mathrm{H_2O}}$ CN. Each coloured point corresponds to a configuration, with the final 2D free energy profile underlaid for reference.}
   \label{fig:dataset}
\end{figure}

\subsubsection{Validation of periodic models}
\begin{figure}[h]
  \centering
   \includegraphics[width=1.0\textwidth]{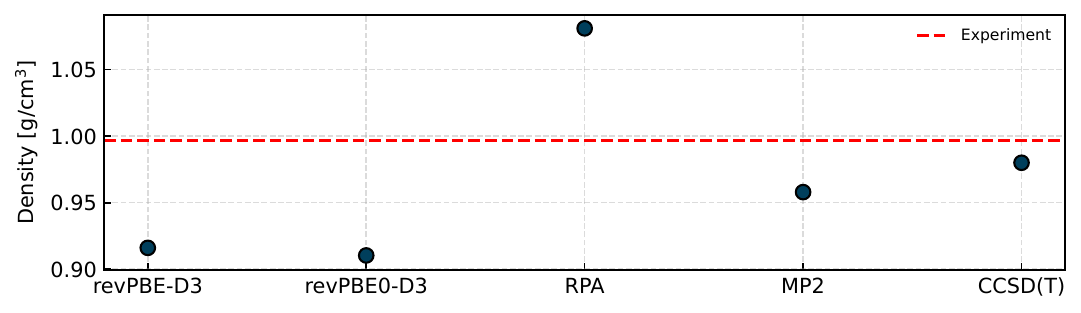}
    \caption[Periodic model validation]{\textbf{Validation of periodic models against bulk water density:} Density of bulk water for the levels of theory studied in this work compared to experiment at 298 K.}
   \label{fig:density_caco3}
\end{figure}

Training errors on the energies and forces are given in Table \ref{tab:mlp-errors}.
The final density prediction of all the models for bulk water is shown in Figure \ref{fig:density_caco3}.
Since we are in the end targeting a single Ca/\ch{CO3} ion pair in solution, this will have minimal effect on the density and so comparing to bulk water has more reliable experimental values as well as from previous literature.
\begin{table}[h]
\caption[MLP errors]{Training and validation errors for the final periodic models trained in this work.}
\centering
\begin{tabular}{@{}lcccc@{}}
\toprule
Model      & \begin{tabular}[c]{@{}c@{}}Train-Energy\\  {[}meV/atom{]}\end{tabular} & \begin{tabular}[c]{@{}c@{}}Validation-Energy \\ {[}meV/atom{]}\end{tabular} & \begin{tabular}[c]{@{}c@{}}Train-Forces \\ {[}meV/\AA{]}\end{tabular} & \begin{tabular}[c]{@{}c@{}}Validation-Forces \\ {[}meV/\AA{]}\end{tabular} \\ \midrule
revPBE-D3  & 0.3                                                                    & 0.3                                                                         & 25.1                                                                 & 26.1                                                                      \\
revPBE0-D3 & 0.6                                                                    & 0.6                                                                         & 26.9                                                                 & 27.1                                                                      \\
RPA        & 0.3                                                                    & 0.2                                                                         & 28.6                                                                 & 29.2                                                                      \\
MP2        & 0.3                                                                    & 0.3                                                                         & 25.6                                                                 & 26.3                                                                     
\end{tabular}
\label{tab:mlp-errors}
\end{table}
\subsection{CCSD(T) model}\label{sec:delta}
\begin{figure}[h]
    \centering
    \includegraphics[width=0.4\linewidth]{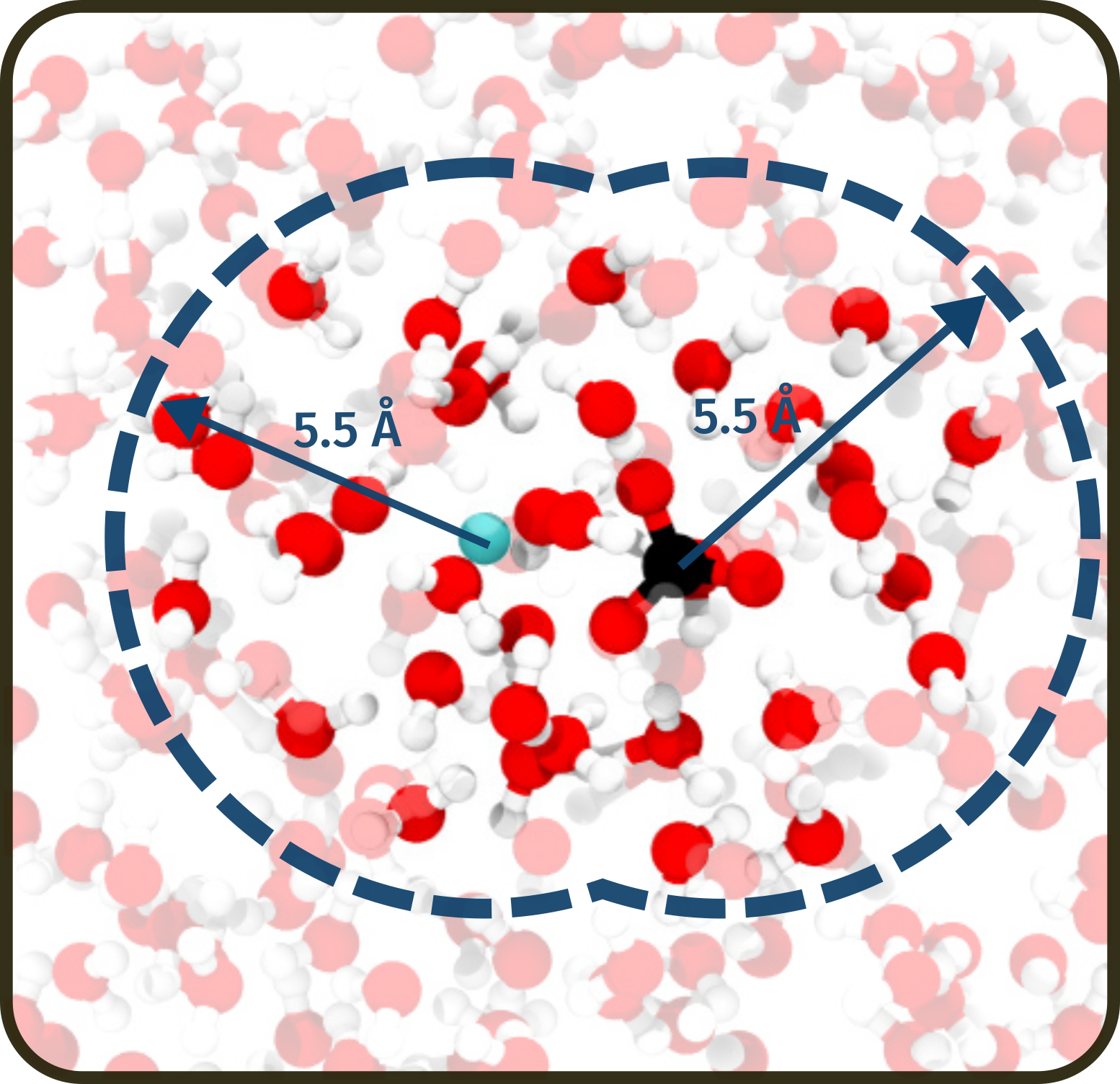}
    \caption[Example cluster cut from a periodic configuration]{\textbf{$\Delta-$model clusters: }Example cluster cut from periodic configuration. All waters within $5.5\,\text{\AA}$ of each of the Ca and \ch{CO3} are included (ensuring that the full molecule is included in the cluster i.e. no broken bonds).}
    \label{fig:cluster}
\end{figure}

The protocol to develop the CCSD(T) model is based on the framework we introduced in Ref. \cite{oneillRoutineCondensedPhase2025}.
The dataset for the $\Delta$ model was generated by cleaving clusters from periodic MP2 configurations, obtained from the pressure scan and OPES simulations described in the above section for generating the periodic dataset.
All waters within a $\sim 5.5\,\text{\AA}$ radius of each ion pair were included, where such a cluster size was chosen such that the ions were sufficiently solvated, while also allowing for affordable computational cost for the CCSD(T) calculations. 
Figure \ref{fig:cluster} shows a typical cluster.
The $\Delta$-model was then trained on the energy differences between MP2 and CCSD(T) for this dataset.
We used a smaller model architecture here, comprising 128 channels with 2 message passing layers and a 5~\AA~ radial cutoff, which we showed in Ref. \cite{oneillRoutineCondensedPhase2025} improves the efficiency (where there is otherwise a penalty of summing 2 MLPs) without any sacrifice on the accuracy or reliability.
The final $\Delta$-CCSD(T) model had energy errors of 0.1~meV/atom on both training and validation sets and the final dataset comprised 1133 structures.

\begin{figure}[h!]
    \centering
    \includegraphics[width=0.5\linewidth]{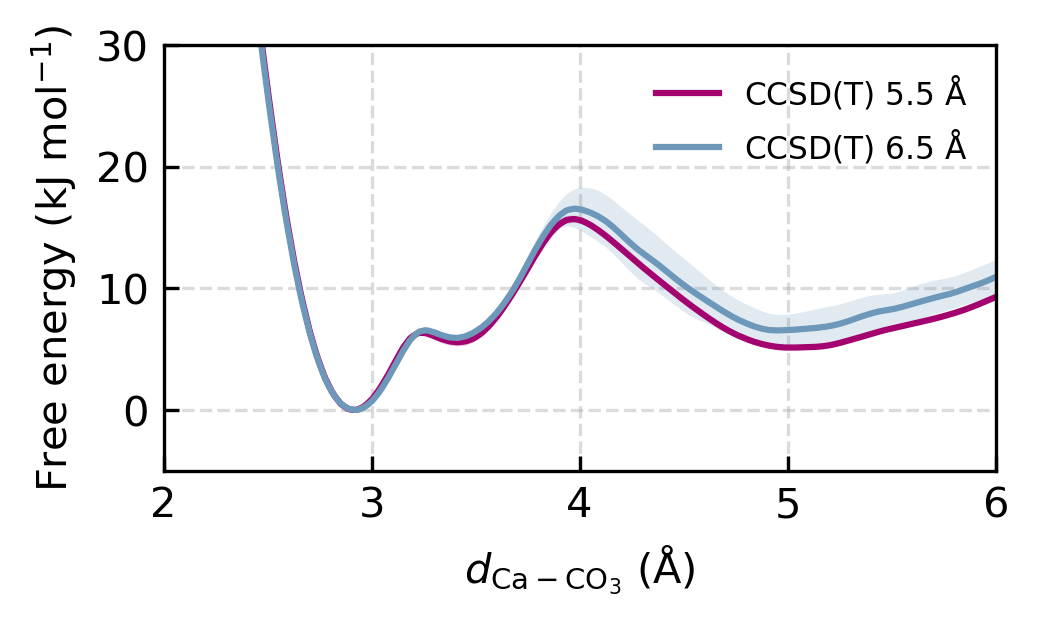}
    \caption{\textbf{Convergence of free energy with cluster cutoff in $\Delta$-model: }Comparison of free energy profiles for CCSD(T) models trained with datasets containing 6.5~\AA{} and 5.5~\AA{} radius clusters.}
    \label{fig:cluster_cutoff}
\end{figure}

In Fig.~\ref{fig:cluster_cutoff} we also examine the effect of cluster size on the convergence of the ion pairing thermodynamics.
To the dataset containing waters within a $\sim 5.5\,\text{\AA}$ radius of each ion, we added 300 $\sim 6.5\,\text{\AA}$ radius clusters.
We show that the free energy profile from the model with 5.5~\AA{} radius clusters is converged with respect to the larger cluster model.

\subsubsection{Validation of $\Delta$-CCSD(T) model}
\begin{figure}[h]
    \centering
    \includegraphics[width=1.0\linewidth]{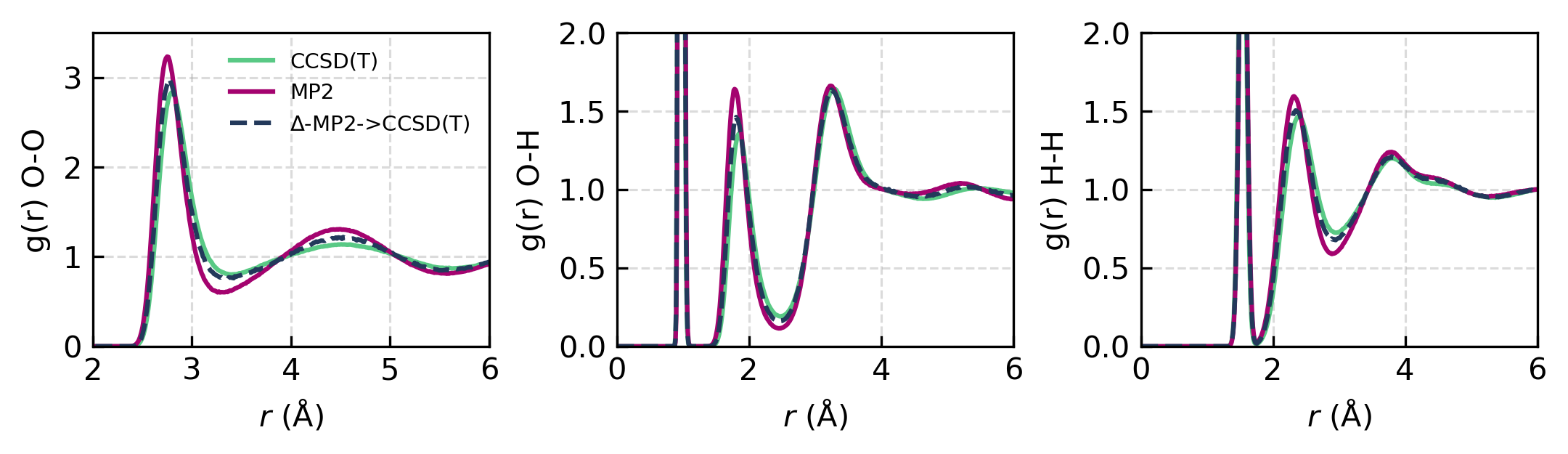}
    \caption{\textbf{Bulk water structure validation: }Comparison of $\Delta-$CCSD(T) model for bulk water structure, with periodic MP2 baseline (pink) and CCSD(T) model trained in Ref.~\cite{oneillRoutineCondensedPhase2025} (green). Note all simulations here have been performed with classical nuclei at 298 K.}
    \label{fig:mp2ccsdt_valid}
\end{figure}

Extensive validation tests on our $\Delta$-learning protocol using a DFT baseline are included in Ref.~\cite{oneillRoutineCondensedPhase2025}.
To show that our model can reliably correct the difference between MP2 and CCSD(T), in this Section, we validate it against a system with now well-established references for both MP2 and CCSD(T) -- liquid water.
In Figure \ref{fig:mp2ccsdt_valid}, we compare the RDFs predicted by the MP2 baseline model -- which is known to overstructure \cite{oneillPairNotPair2024} liquid water -- with the prediction of the $\Delta$-MLP model (summed on top of the MP2 baseline model).
We also include the CCSD(T) model developed in Ref.~\cite{oneillRoutineCondensedPhase2025} for comparison.
The $\Delta$-MLP model corrects the general overstructuring of the RDFs from MP2, especially the first peak of the O-O RDF and the hydrogen bonding peak of the O-H RDF.
Furthermore, it is in excellent agreement with the CCSD(T) model from Ref.~\cite{oneillRoutineCondensedPhase2025}.
We note that this model has been explicitly validated against experiment with full treatment of nuclear quantum effects, and thus serves as a stringent test.

\section{Electronic Structure}
Here the details of both the periodic and gas-phase calculations used for training the MLPs are described.
For completeness, some of the information given in the Methods Section of the main manuscript is repeated.
\subsection{Coupled Cluster Theory and MP2 for cluster calculations}
For the (cluster-based) dataset used to train the $\Delta$-MLP, we calculated MP2 with local approximations~\cite{nagyIntegralDirectLinearScalingSecondOrder2016,szaboLinearScalingOpenShellMP22021} (LMP2) and CCSD(T) with the local natural orbital approximation (LNO)~\cite{nagyOptimizationLinearScalingLocal2018,gyevi-nagyIntegralDirectParallelImplementation2020}.
These calculations were both performed in MRCC~\cite{kallayMRCCProgramSystem2020}, with the LMP2 numbers coming naturally from the LNO-CCSD(T) calculation.
We used the default ``normal'' LNO thresholds -- recently validated for the properties of water~\cite{oneillRoutineCondensedPhase2025} -- and use density-fitting (with the def2-QZVPP-RI-JK auxiliary basis set) to speed up the HF calculations.
We included the sub-valence correlation of the 3s and 3p electrons on the Ca atom with the corresponding cc-pwCV$X$Z~\cite{petersonAccurateCorrelationConsistent2002,balabanovSystematicallyConvergentBasis2005} basis set, where $X$ represents the cardinal number.
We use the aug-cc-pV$X$Z basis set on the O and cc-pV$X$Z basis set on the H and C atoms, while the cc-pwCV$X$Z basis set was used for the Ca atoms.
For all other atoms, we used the cc-pV$X$Z family (valence) basis sets, in particular using the diffuse aug-cc-pV$X$Z basis sets on the O atom.
This basis set combination we dub 'jul-'.
For the auxiliary basis sets required for the LNO-CCSD(T) calculations, we used those generated in Refs.~\cite{weigendRIMP2OptimizedAuxiliary1998,hellwegOptimizedAccurateAuxiliary2007} for each atom.
However, as none were available for Ca, we used the automatic auxiliary basis functions code of Stoychev \etal{}~\cite{stoychevAutomaticGenerationAuxiliary2017,lehtolaStraightforwardAccurateAutomatic2021} to generate these basis functions.
We performed complete basis set (CBS) extrapolations for the double (DZ) and triple-zeta (TZ) basis sets for LMP2 and TZ and quadruple-zeta (QZ) pair for LNO-CCSD(T), using parameters taken from Neese and Valeev.~\cite{neeseRevisitingAtomicNatural2011}.

\subsection{Periodic DFT + MP2}
All reference calculations were done on a unit cell of length $\sim 14~\text{\AA}$ containing 126 waters and one ion pair.
revPBE-D3 calculations were performed with FHI-AIMs \cite{blumInitioMolecularSimulations2009}, using the \texttt{tight} family of basis sets and the (in-built) Grimme's D3 \cite{grimmeConsistentAccurateInitio2010}. 
All periodic unit cell calculations were performed using the $\Gamma$-point.
revPBE0-D3, RPA and MP2 calculations were performed with the CP2K/Quickstep \cite{kuhneCP2KElectronicStructure2020} code.
The electronic density was partitioned into core and valence electrons, with core electrons and nuclei described with the norm-conserving Goedecker, Teter and Hutter (GTH) pseudopotentials \cite{goedeckerSeparableDualspaceGaussian1996}.
For revPBE0-D3, hydrogen and oxygen atoms were described with the TZV2P family of basis sets, while we used the ccGRB family for carbon and calcium.
revPBE0-D3 calculations used the auxiliary density matrix methods (ADMM) approximation \cite{guidonAuxiliaryDensityMatrix2010} to improve the efficiency.
For RPA and MP2 we used the cc-TZ basis set for oxygen and hydrogen, with the ccGRB basis set for carbon and calcium.
PBE was used as a starting point for the RPA correlation energy
calculations.
The resolution-of-identity (RI) technique was used for these methods. \cite{bussySparseTensorBased2023}
We used triple-zeta (TZ) quality correlation consistent basis sets for H and O (taken from CP2K’s \texttt{RI\_cc\_TZ set}) and correlation consistent basis sets designed by Ye and Berkelbach \cite{yeCorrelationConsistentGaussianBasis2022} for
carbon and calcium.
Auxiliary basis sets for the RI integral operations were generated using the automatic auxiliary basis of Stoychev et al. \cite{stoychevAutomaticGenerationAuxiliary2017, lehtolaStraightforwardAccurateAutomatic2021} for calcium and carbon, with the defaults (from CP2K) used for H and O.
For consistency with previously developed models \cite{oneillPairNotPair2024} a planewave cutoff of 1200 Ry was used (although in practice the forces are converged after around 400 Ry).

\subsection{LNO and Basis Set Convergence}
\subsubsection{Interaction energies}
\begin{figure}[h!]
    \centering
    \includegraphics[width=1.0\linewidth]{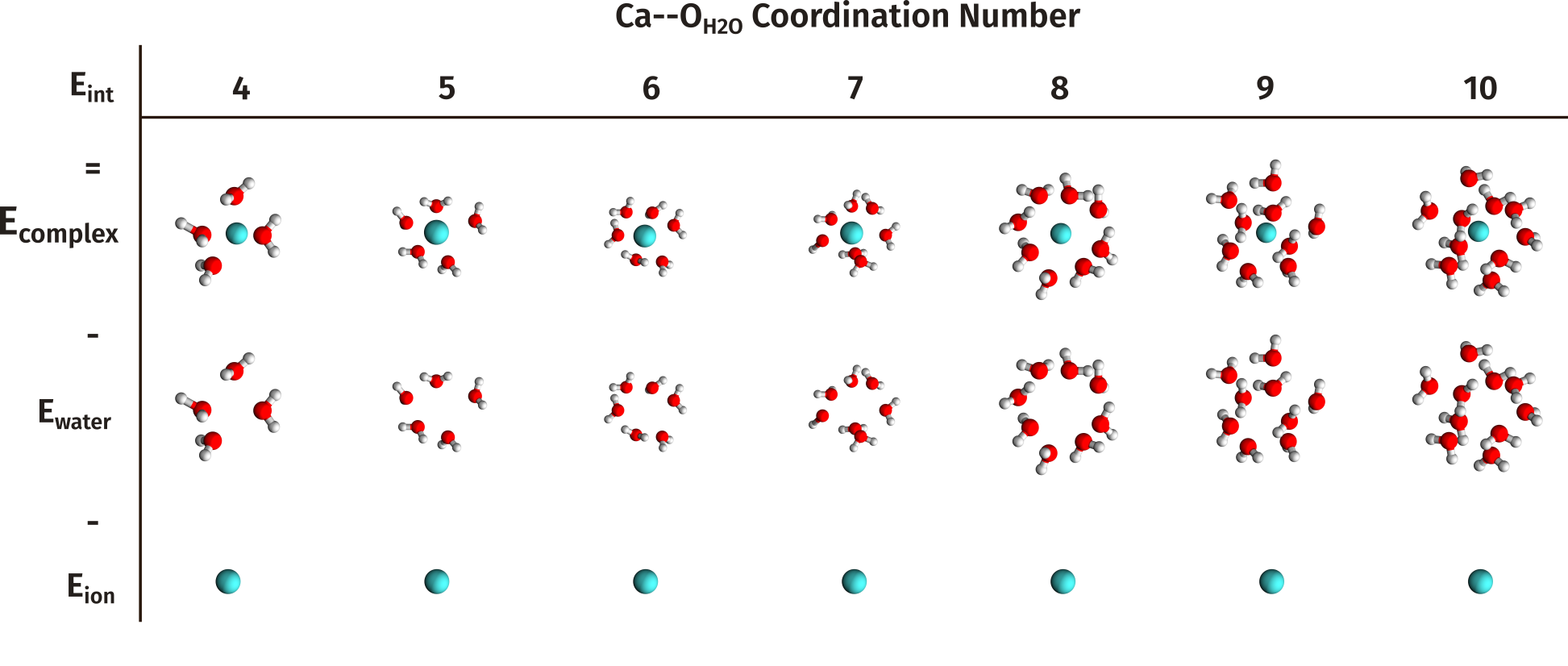}
    \caption{\textbf{Ion water cluster interaction energies: }Schematic of cluster calculations corresponding to Tables~\ref{tab:ccsdt_lno_convergence} and \ref{tab:ccsdt_basis_convergence}.}
    \label{fig:eint_schematic}
\end{figure}

{\color{black}
We compare the effect of different LNO thresholds as well as basis sets on the relative interaction energies between different coordination numbers.
We have selected 20 solvated Ca$^{2+}$ structures carved from periodic simulation cells at each coordination number from 5 to 10 water molecules.
As shown by Figure~\ref{fig:eint_schematic}, the interaction energy is defined as:
\[
E_{\mathrm{int}}=E_{\mathrm{complex}}-E_{\mathrm{water}}-E_{\mathrm{ion}}.
\]
We averaged the 20 solvated Ca$^{2+}$ structures for each coordination number to come up with an average $E_\text{int}$ which is then subtracted relative to the $E_\text{int}$ for CN=8.

In Table~\ref{tab:ccsdt_lno_convergence}, we have compared normal, tight and `very tight (vtight)' LNO thresholds against canonical calculations for the TZ basis set.
It can be seen that the relative interaction energy differences are very small, less than $3\,$meV already achieved by the use of the normal thresholds.

We also show the basis set convergence.
In Table~\ref{tab:ccsdt_basis_convergence}, we have made a comparison of the DZ, TZ, QZ, CBS(DZ/TZ) and CBS(TZ/QZ) for the same vtight LNO threshold.
It can be seen that the errors (given in parenthesis) with respect to the CBS(TZ/QZ) estimates become less than $1\,$kJ/mol by the TZ basis set and is much better with the CBS(TZ/QZ) approach taken within this work.
}
\begin{table}[h!]
\color{black}
\centering
\setlength{\tabcolsep}{5pt}
\caption{Mean CCSD(T) interaction energies relative to CN=8 for each coordination number, in meV. Differences from the canonical DF/TZ result are given in parentheses.}
\label{tab:ccsdt_lno_convergence}
\begin{tabular}{crrrr}
\toprule
CN & Canonical & LNO normal & LNO tight & LNO very tight \\
\midrule
5  & 2435 (0) & 2433 (-2) & 2431 (-4) & 2432 (-3) \\
6  & 1290 (0) & 1292 (+1) & 1288 (-2) & 1288 (-2) \\
7  &  478 (0) &  480 (+1) &  477 (-1) &  477 (-1) \\
8  &    0 (0) &    0 (0) &    0 (0) &    0 (0) \\
9  & -694 (0) & -695 (0) & -693 (+1) & -693 (+1) \\
10 & -705 (0) & -708 (-3) & -702 (+3) & -703 (+2) \\
\bottomrule
\end{tabular}
\end{table}
\begin{table}[h!]
\color{black}
\centering
\setlength{\tabcolsep}{6pt}
\caption{Mean CCSD(T) interaction energies relative to CN=8 for each coordination number, in meV. Differences from the CBS(TZ/QZ) result are given in parenthesis.}
\label{tab:ccsdt_basis_convergence}
\begin{tabular}{crrrrr}
\toprule
CN & DZ & TZ & QZ & CBS(DZ/TZ) & CBS(TZ/QZ) \\
\midrule
5 & 2432 (-11) & 2432 (-10) & 2438 (-5) & 2441 (-1) & 2443 (0) \\
6 & 1304 (+26) & 1288 (+11) & 1281 (+3) & 1285 (+8) & 1278 (0) \\
7 & 479 (+6) & 477 (+4) & 474 (+1) & 479 (+6) & 473 (0) \\
8 & 0 (0) & 0 (0) & 0 (0) & 0 (0) & 0 (0) \\
9 & -690 (+8) & -693 (+5) & -696 (+2) & -697 (+1) & -698 (0) \\
10 & -712 (-16) & -703 (-7) & -698 (-2) & -702 (-6) & -696 (0) \\
\bottomrule
\end{tabular}
\end{table}
\clearpage
\subsubsection{Ion pairing thermodynamics}
{\color{black}In order to directly test the effect of LNO thresholds and basis set combinations on the relevant observable of interest -- the ion pairing free energy -- we have trained several additional models.
Each model uses the same MP2 baseline.
In all cases, we compare to the specific settings used in this work, namely CCSD(T): jul-TZ/QZ, normal.
\begin{figure}[h!]
    \centering
    \includegraphics[width=1.0\linewidth]{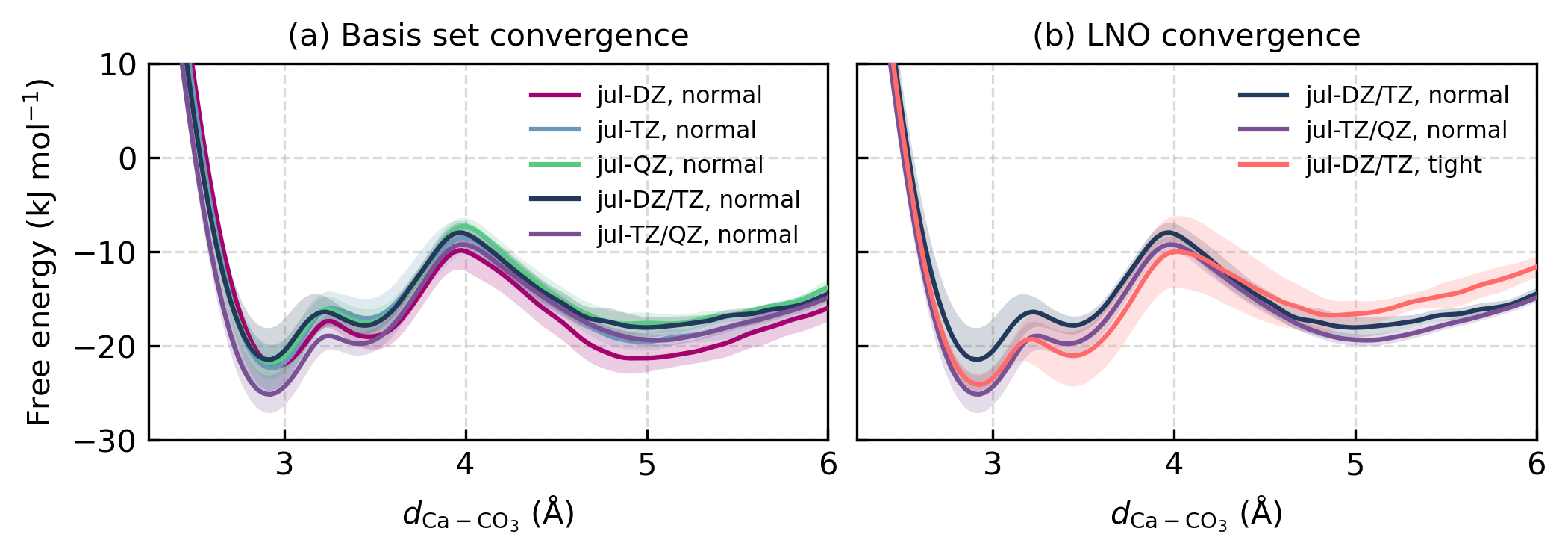}
    \caption{\textbf{Free energy convergence with CCSD(T) settings: }Free energy profiles at 300 K for different basis set and LNO threshold combinations. The jul-TZ/QZ, normal setup was used for the results reported in the main manuscript.}
    \label{fig:pmf_lno}
\end{figure}

Fig.~\ref{fig:pmf_lno} compares the computed free energy profiles between (a) different basis set combinations and (b) normal and tight LNO thresholds.
All of the free energy profiles agree within their respective error bars.

The association free energy for each model has also been computed and is shown in Fig.~\ref{fig:dg_lno}, confirming the convergence of the association free energy computed in the main manuscript using normal LNO thresholds, which are within \si{1}~{kJ mol$^{-1}$} of the stricter tight settings.
\begin{figure}[h!]
    \centering
    \includegraphics[width=1.0\linewidth]{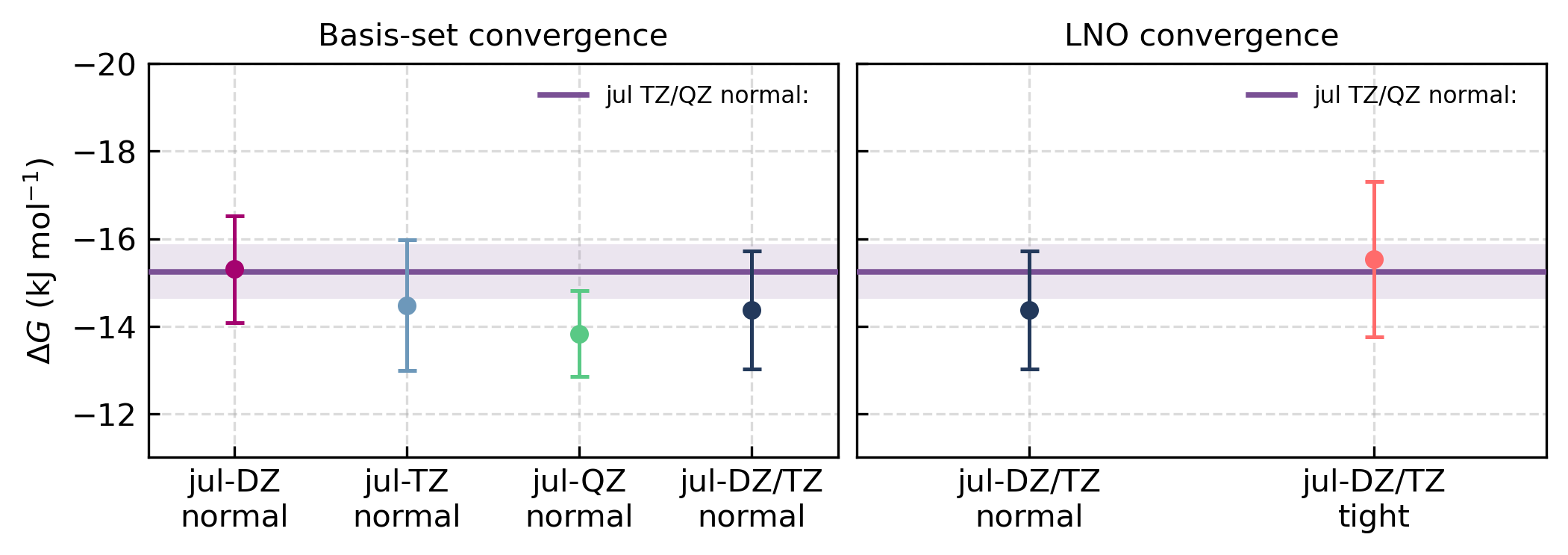}
    \caption{\textbf{Ion pairing free energy convergence with CCSD(T) settings: }Association free energies and corresponding standard error on the mean, obtained from the PMFs in Fig~\ref{fig:pmf_lno}. The free energy reported in the main manuscript using jul TZ/QZ, normal settings is shown as a horizontal purple line.}
    \label{fig:dg_lno}
\end{figure}
 
Again, all computed free energies lie within their respective error bars and well below chemical accuracy (\si{4}~{kJ mol$^{-1}$}).
Therefore, in summary, the tests in Tables \ref{tab:ccsdt_lno_convergence} and \ref{tab:ccsdt_basis_convergence}, as well as in Figs~\ref{fig:pmf_lno} and ~\ref{fig:dg_lno}, confirm the convergence of our electronic structure setup, from the fundamental interaction energies to the thermodynamic observable that is compared to experiment.}
\clearpage

\section{Bader charge analysis}\label{sec:bader}
\begin{figure*}[h!]
  \centering
   \includegraphics[width=0.75\textwidth]{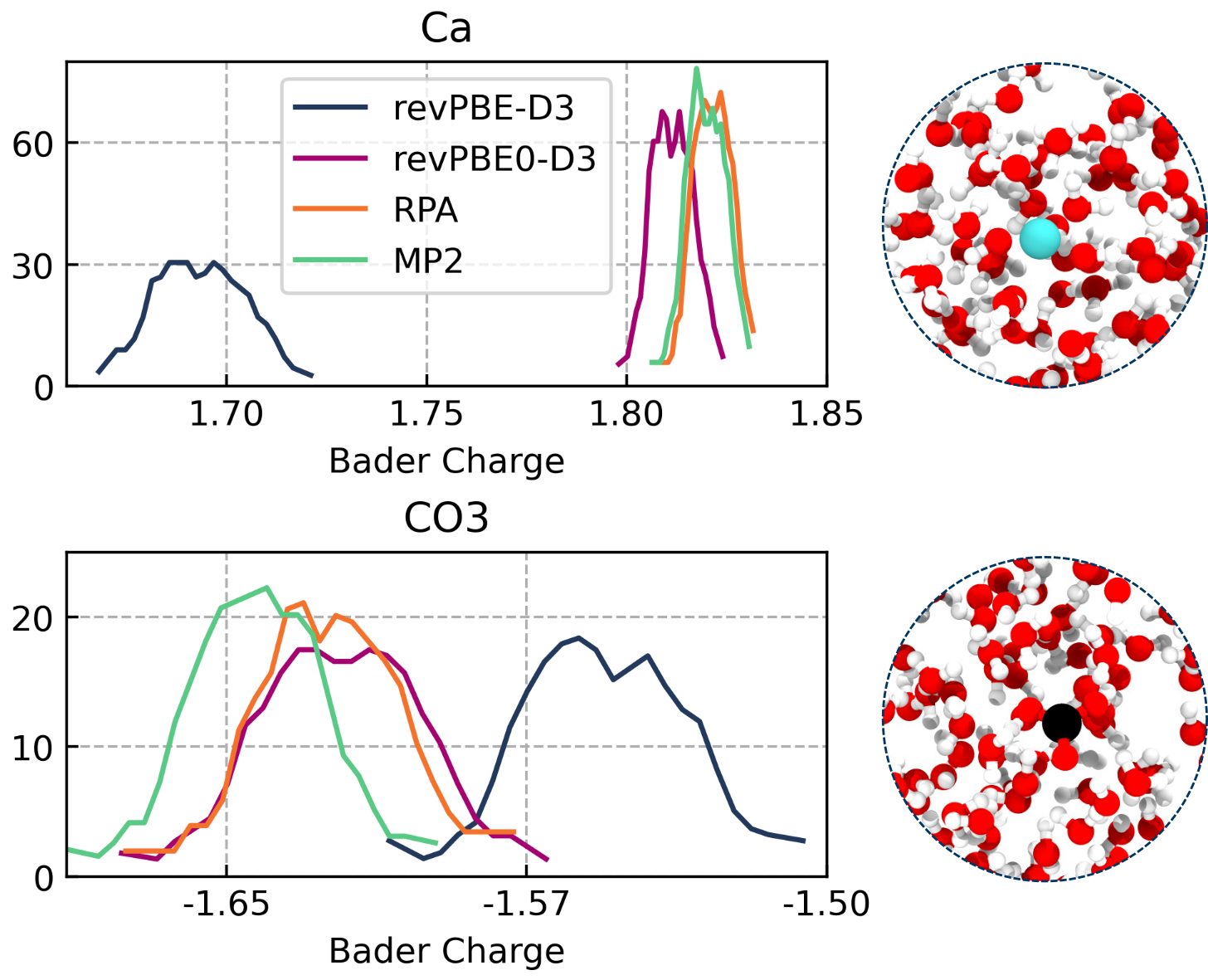}
    \caption{{\textbf{Bader charge distribution as a function of method} Bader charges on solvated \ch{Ca^{2+}} and \ch{{CO_3}^{2-}} ions computed at increasing levels of electronic structure theory. The snapshots show the box, of one \ch{Ca^{2+}} (\ch{CO3^{2-}}) ion surrounded by 63 waters.}}
   \label{fig:bader}
\end{figure*}

One of the well-documented shortcomings of DFT is its inability (to varying extents) to localise charge \cite{bryentonDelocalizationErrorGreatest2023, chengRedoxPotentialsAcidity2014}. 
This is in general evident in the 1-dimensional potential of mean force, where going from GGA (revPBE-D3) to hybrid (revPBE0-D3) to MP2 and CCSD(T) results in increasingly stable contact ion pair states.
To explore this further we used the Bader charge partitioning scheme to assign partial charges on the solvated ions  predicted by the studied levels of theory in this work.
Figure \ref{fig:bader} shows the Bader charges on the \ch{Ca^{2+}} and \ch{CO3^{2-}} ions.
For each ion, 100 snapshots were sampled from a trajectory containing the individual ion in a box of 63 waters.
The electronic density was then computed on these snapshots, and the Bader charge obtained using the Bader code from the Henkelman group \cite{yuAccurateEfficientAlgorithm2011}.
For both the  \ch{Ca^{2+}} and \ch{CO3^{2-}}, there is a trend of revPBE-D3 $<$ revPBE0-D3 $<$ RPA $<$ MP2 for the extent of charge localisation.
This is consistent with the expected degree of delocalisation error with DFT \cite{bryentonDelocalizationErrorGreatest2023}.
There is also a clear correlation between the extent of charge delocalisation and the depth of the contact ion pair minima in the 1-D PMFs in Figure 3 in the main text.

\section{Simulations}\label{sec:simulations}
For clarity, all MLP results shown in the SI are reported without finite size corrections (except Fig.~\ref{fig:ff_comp}), to enable future comparison with our setup.
The final results presented in the main paper have been corrected for finite size effects, which are discussed in detail below.
All simulations were performed using the Large-scale Atomic Molecular Massively Parallel Simulator (LAMMPS)~\cite{thompsonLAMMPSFlexibleSimulation2022} code, combined with PLUMED \cite{bonomiPromotingTransparencyReproducibility2019} and coupled with the \texttt{Symmetrix} library \cite{wcwittWcwittSymmetrix2026, kovacsEvaluationMACEForce2023}, using the \texttt{symmetrix/}\texttt{mace} pair style in tandem with the \texttt{hybrid/overlay} pairstyle to sum the MP2 baseline and $\Delta$-MLPs. 
All classical force field simulations used the parameters developed in Ref.~\cite{armstrongSolubilityconsistentForceField2023} for \ch{CaCO3} in water.

\section{Ion pairing free energy}\label{sec:ionpairing}
The calculation of ion pairing free energies has received substantial previous attention, and so here we summarise the main key steps and considerations, as well as showing convergence tests to reassure the reliability of our final results.
For a thorough description and discussion, the reader is referred to the supplementary information of Ref.~\cite{aufortComputationalInsightsMg22022}.
\subsection{OPES simulations}\label{sec:ip_opes}
The free energy profiles were obtained from on-the-fly probability enhanced (OPES) \cite{invernizziRethinkingMetadynamicsBias2020} simulations, where the Ca-\ch{CO3} ion pair separation was used as the collective variable to be biased.
It has been previously shown that sampling the orthogonal Ca-water coordination number is important to obtain reliable estimates of the free energy.
Therefore the Ca-O$_{\text{\ch{H2O}}}$ coordination number was computed on the OPES trajectory via a smooth switching function:
\begin{equation}
S(r)=
\frac{1-\left(\dfrac{r-d_0}{r_0}\right)^8}
{1-\left(\dfrac{r-d_0}{r_0}\right)^{16}} \,
\end{equation}
with $d_0$=1.8~\AA~and $r_0$=1.5~\AA~to overlap with the first peak in the \ch{Ca}-O RDF.

The unbiased probability distribution ($P(d,n)$) was reconstructed from the OPES simulations by reweighting the OPES trajectory using the time-dependent bias potential $V_{\mathrm{bias}}$, as a function of both collective variables, the \ch{Ca}--\ch{CO3} separation ($d(\mathbf{r})$) and the Ca-O$_{\ch{H2O}}$ coordination number ($n(\mathbf{r})$):
\begin{equation}
P(d,n) =
\frac{
\left\langle
\delta\!\left(d-d(\mathbf{r})\right)
\delta\!\left(n-n(\mathbf{r})\right)
e^{\beta V_{\mathrm{bias}}(\mathbf{r})}
\right\rangle_V
}{
\left\langle
e^{\beta V_{\mathrm{bias}}(\mathbf{r})}
\right\rangle_V
}
\end{equation}
where $\langle \cdot \rangle_V$ corresponds to the ensemble average taken over the biased simulation.
The final free energy surface is then given by:
\begin{equation}
F(d,n) = -k_B T \ln P(d,n)
\end{equation}
We note that it was important to first recover the 2D free energy, since there are some orthogonal steps to the Ca-\ch{CO3} separation.
To ensure the correct asymptotic behaviour, the tail of the PMF is aligned with the analytical solution for the pairing free energy of two point charges $q_i, q_j$ interacting via a screened electrostatic potential:
\begin{equation}
    \Delta G = \frac{1}{4\pi \epsilon_0}\frac{q_iq_j}{\epsilon_rr}-k_BT\mathrm{ln}(4\pi r^2).
\end{equation}
where $\epsilon_r$ is the dielectric constant of the medium.
The standard ion pairing free energy $\Delta G_{\mathrm{IP}}^{\ominus}$ can then be obtained by integrating the aligned PMF ($\phi'(r)$) up to $r_c$:
\begin{equation}\label{eq:dg}
    \Delta G_{\mathrm{IP}}^{\ominus} = -RT\ln\Bigg[ c^{\ominus} \int_0^{r_c}\exp\bigg(-\frac{\phi'(r)}{k_BT}\bigg)dr\Bigg].
\end{equation}
Here $c^{\ominus}$ is a concentration unit conversion from atoms/\AA$^3$ to mol/dm$^3$ required to correctly relate the simulated association free energy to that obtained from experiment. 
The choice of $r_c$ has been shown to only weakly affect the resulting association constant (provided that the free energy profile
is correctly aligned to the tail of the analytic solution) \cite{aufortComputationalInsightsMg22022}, and so here we use $r_c$ as 7~\AA.

Furthermore, since the MACE potential does not contain charge information, we make an approximation and use the experimental dielectric constant (78) to align the PMF to the analytic solution.
\begin{figure}[h!]
    \centering
    \includegraphics[width=0.5\linewidth]{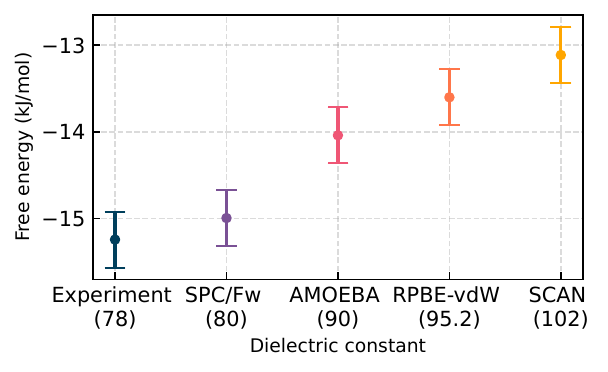}
    \caption{\textbf{Effect of dielectric constant on association free energy:} Comparison of ion pairing free energy for the CCSD(T) model at 300 K using different values for the dielectric constant (values for the dielectric constant given in brackets beneath the model name). RPBE-vdW value from Ref.~\cite{morawietzHowVanWaals2016} and SCAN from Ref.~\cite{caiSimulationsDielectricPermittivity2025}.}
    \label{fig:dielectric}
\end{figure}
Figure \ref{fig:dielectric} shows the dependence of the association free energy (for the CCSD(T) model) to the choice of dielectric constant, spanning experiment, non-polarisable (SPC/Fw) \cite{wuFlexibleSimplePointcharge2006} and polarisable (AMOEBA) \cite{ponderCurrentStatusAMOEBA2010} force field, as well as the value for some DFT functionals that have been published in the literature.
The dielectric constant for SCAN \cite{caiSimulationsDielectricPermittivity2025} and RPBE-vdW \cite{morawietzHowVanWaals2016} are both greater than experiment, which subsequently shifts the free energy systematically by up to 2 kJ/mol.
Nevertheless, given previous benchmarking on other properties of liquid water for CCSD(T) \cite{oneillRoutineCondensedPhase2025} including density and diffusion, we expect that the CCSD(T) values of the dielectric constant should be in close agreement with experiment, justifying our choice.
However, ongoing developments in the direction of including explicit electrostatics in MLPs, we anticipate should remove the requirement to make such a choice, since charge information will be readily available.
\clearpage
\subsection{Convergence of OPES simulations}
\begin{figure}[h!]
    \centering
    \includegraphics[width=1\linewidth]{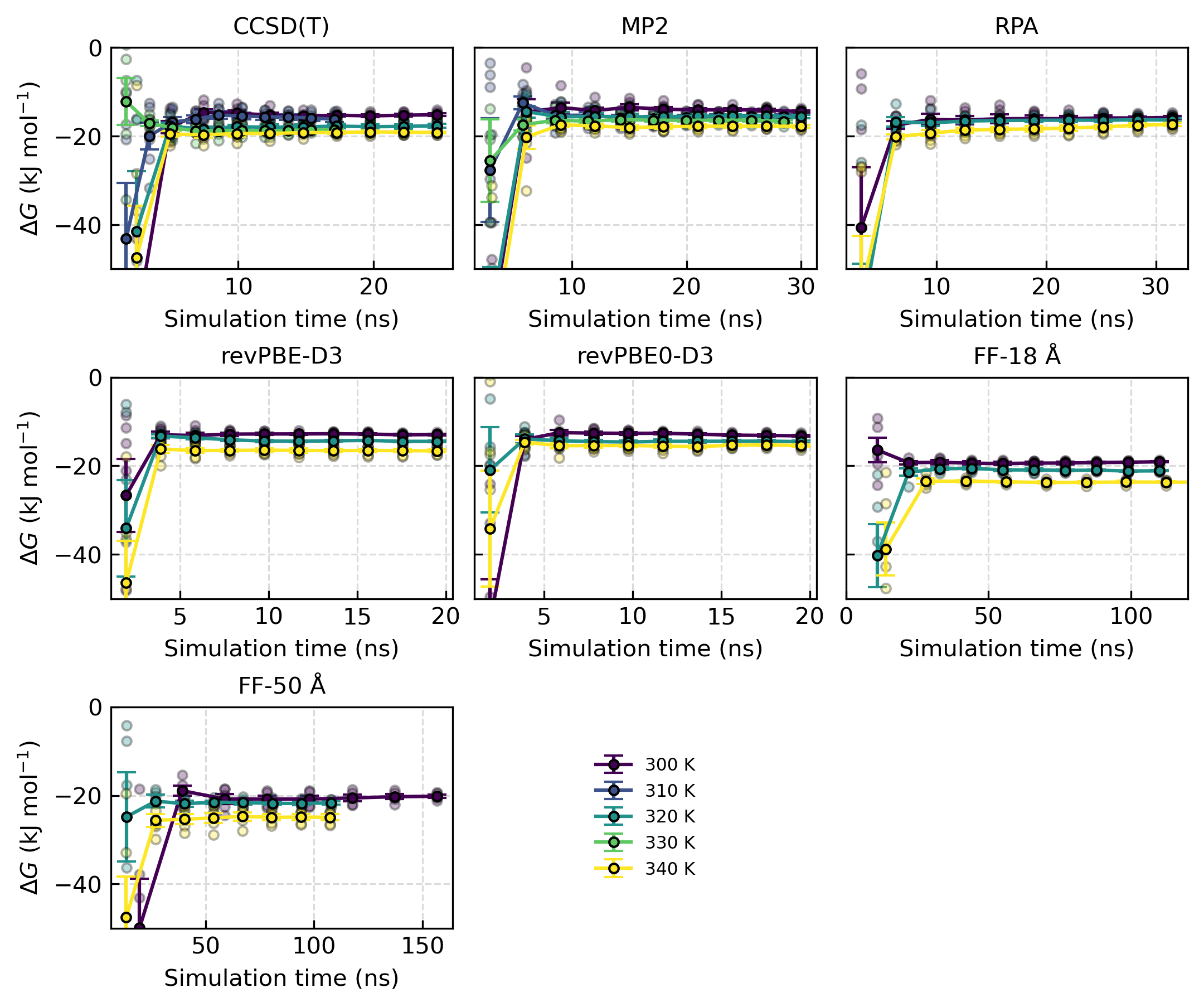}
    \caption{\textbf{OPES convergence:} Convergence of the association free energy computed from integration of the OPES PMF as a function of cumulative simulation time for temperatures from 300-340 K for all the models in this work. The transparent circles show the individual replicas at each temperature, while the translucent circles show the average and standard errors across the replicas as a function of cumulative simulation time.}
    \label{fig:opes_convergence}
\end{figure}

To obtain statistically converged estimates for all models, six independent replicas were performed at each temperature, each with a total simulation length of 20–30 ns (except for the classical force field results used for finite size tests, whose efficiency enabled simulations over 100 ns each. This enabled us to further confirm the convergence of our shorter MLP runs). 
To ensure statistical convergence of the final free energy, in Figure~\ref{fig:opes_convergence} we show the free energy of each replica at each temperature as a function of cumulated simulation time.
After approximately $\sim$3 ns, the OPES bias potential had stabilised, after which the simulations reversibly sampled the free energy surface.
This highlights a key advantage of OPES over other enhanced sampling methods, especially when used with MLPs where computational cost is still a concern: once the bias has stabilised, the majority of the simulation time can be devoted to sampling the flattened free energy landscape \cite{invernizziExplorationVsConvergence2022}.
\clearpage
\subsection{Computing enthalpy and entropy}
\begin{figure}[h!]
    \centering
    \includegraphics[width=0.5\linewidth]{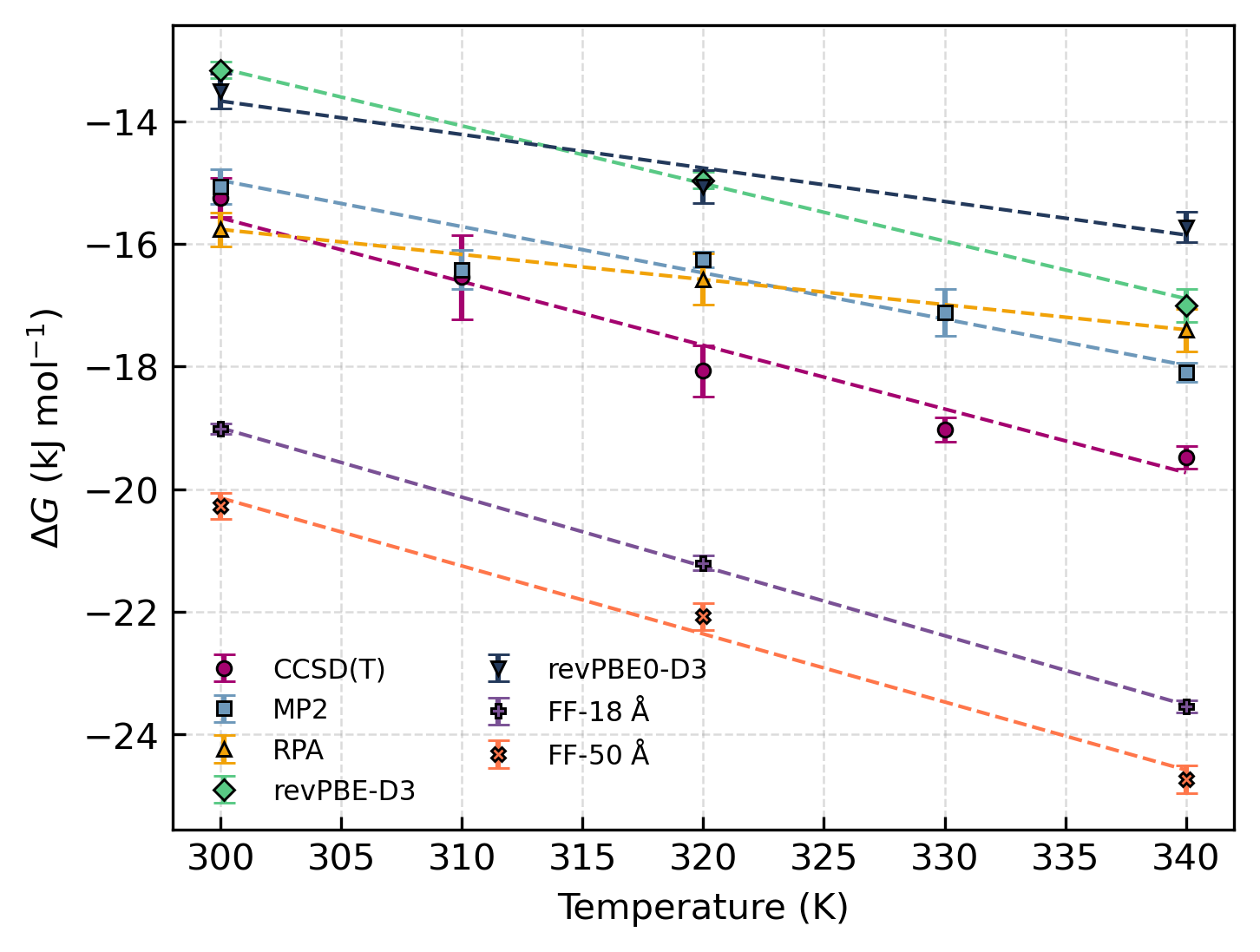}
    \caption{\textbf{$\Delta$G vs Temperature:} Comparison of $\Delta$G versus temperature from OPES simulations for all of the models developed in this work. Classical force field results are also shown for the 18~\AA~ and 50~\AA~ boxes used to compute finite size effects.}
    \label{fig:dgdt}
\end{figure}

Enthalpy $\Delta H$ and entropy $\Delta S$ contributions to the ion pairing free energy were computed assuming the standard thermodynamic relation:
\begin{equation}
    \Delta G=\Delta H-T\Delta S
\end{equation}
A weighted linear regression was performed to account for the uncertainty in the free energy values at each temperature.
Figure \ref{fig:dgdt} compares the temperature dependence of $\Delta$G and the associated linear fit for all of the models developed in this work, as well as classical force field results used for validation and computing finite size effects (see next Section).

\begin{figure}[h!]
    \centering
    \includegraphics[width=0.5\linewidth]{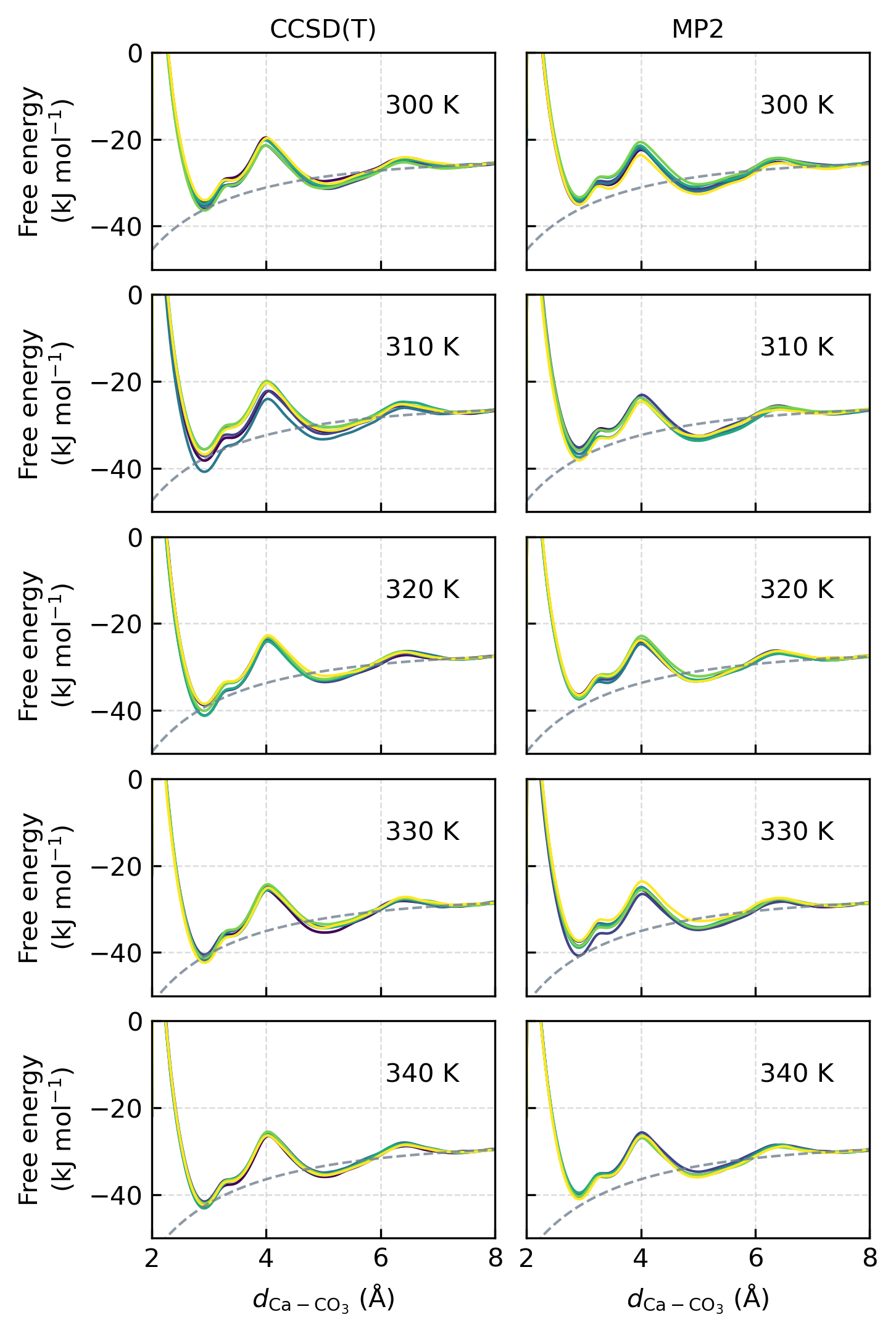}
    \caption{\textbf{1D PMFs -- cWFT:} Free energy as a function of Ca--\ch{CO3} separation for CCSD(T) and MP2 models at 300, 310, 320, 330 and 340 K. Each plot shows 6 profiles obtained from independent OPES simulations, as well as the analytical solution to which the tail is aligned.}
    \label{fig:all_pmfs_wf}
\end{figure}

\begin{figure}[h!]
    \centering
    \includegraphics[width=1\linewidth]{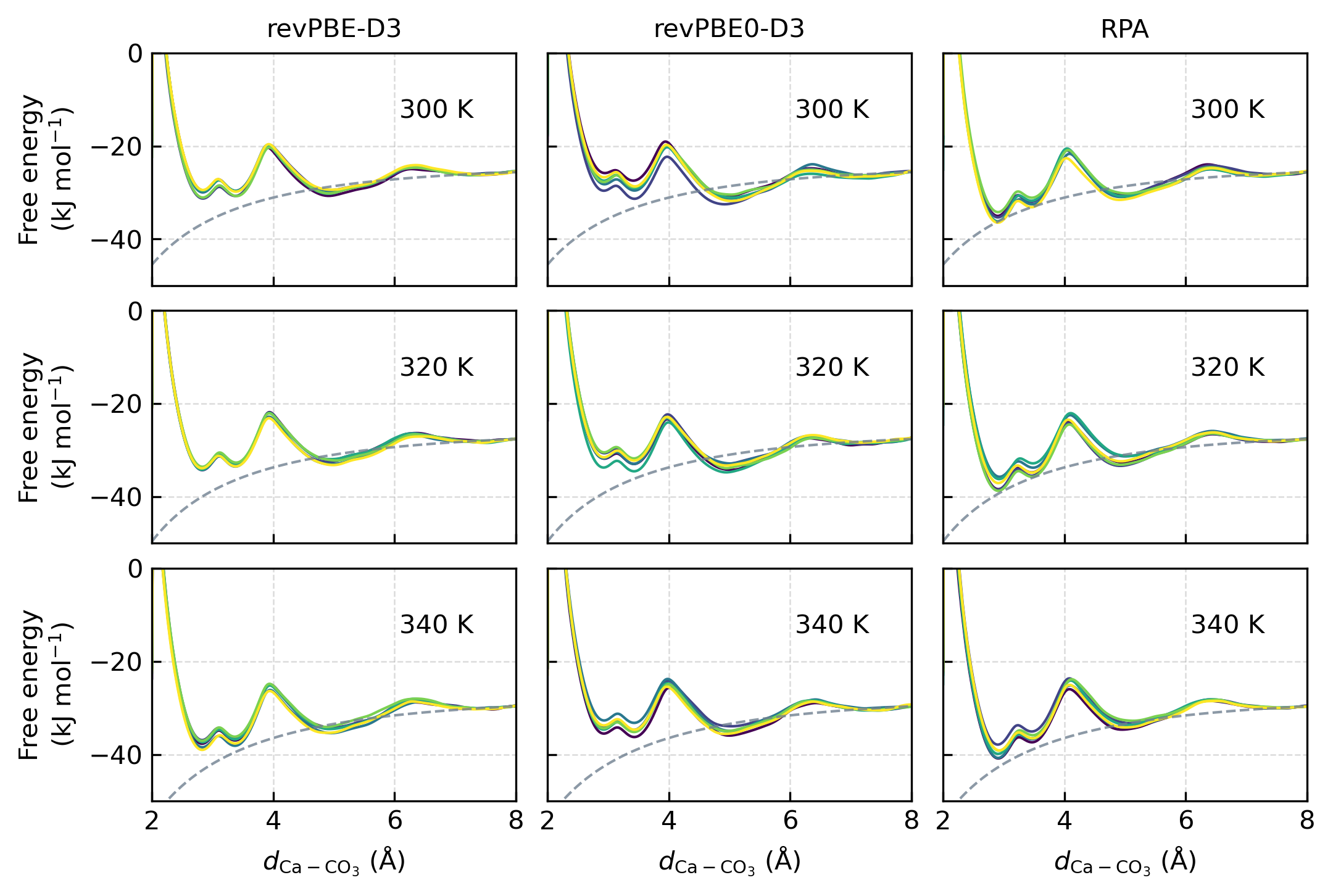}
    \caption{\textbf{1D PMFs -- DFT:} Free energy as a function of Ca--\ch{CO3} separation for the DFT models at 300, 320, and 340 K. Each plot shows 6 profiles obtained from independent OPES simulations, as well as the analytical solution to which the tail is aligned.}
    \label{fig:all_pmfs_dft}
\end{figure}

In Figures \ref{fig:all_pmfs_wf} and \ref{fig:all_pmfs_dft}, the 1D PMFs at each temperature for each method are also shown (from which $\Delta$G were computed), where there is a general trend of increasing CIP stability with increasing temperature for all methods (as has been previously observed for classical force field models \cite{armstrongSolubilityconsistentForceField2023}).

Finally to validate the OPES simulation set-up, in Table \ref{tab:ff_comparison} we compare the results obtained from previous well-tempered metadynamics simulations with the non-polarisable classical force field from Ref. \cite{armstrongSolubilityconsistentForceField2023} with the OPES setup in this work using the same force field parameters.

\begin{table}[h!]
\centering
\caption{Comparison of previous FF from well-tempered metadynamics in Ref. \cite{armstrongSolubilityconsistentForceField2023} and OPES from this work}
\begin{tabular}{lcc}
\hline
\textbf{Property} & \textbf{This Work} & \textbf{Literature} \\
\hline
$\Delta G$ (kJ/mol)
& $-20.3 \pm 0.2$
& $-19.5$ \\

$\Delta S$ (J/mol K)
& $+111 \pm 8$
& $+124 \pm 10$ \\

$\Delta H$ (kJ/mol)
& $+13.1 \pm 3$
& $+17.4 \pm 3$ \\\hline
\end{tabular}
\label{tab:ff_comparison}
\end{table}

\subsection{Finite size effects and error analysis}\label{sec:finitesize}
Despite the significant efficiency boost afforded by MACE over \textit{ab initio} methods, resolving the association free energy still requires substantial simulation effort.
While it is standard for classical force fields to resolve the PMF in a 50~\AA~ simulation box, ensuring the tail of the PMF can be aligned at the Bjerrum length, this would be impractical for the MACE models developed in this work. 
Therefore to quantify the finite size effects arising from the 18~\AA~box size used in the MLP simulations, we used the most recently developed classical force field parameterised for \ch{CaCO_3} in water \cite{armstrongSolubilityconsistentForceField2023} to compare the results between the 18~\AA~and 50~\AA~boxes.
This further allowed us to benchmark our OPES setup against the more established well-tempered metadynamics which has been widely used to study this system.
\begin{figure}[h!]
    \centering
    \includegraphics[width=0.5\linewidth]{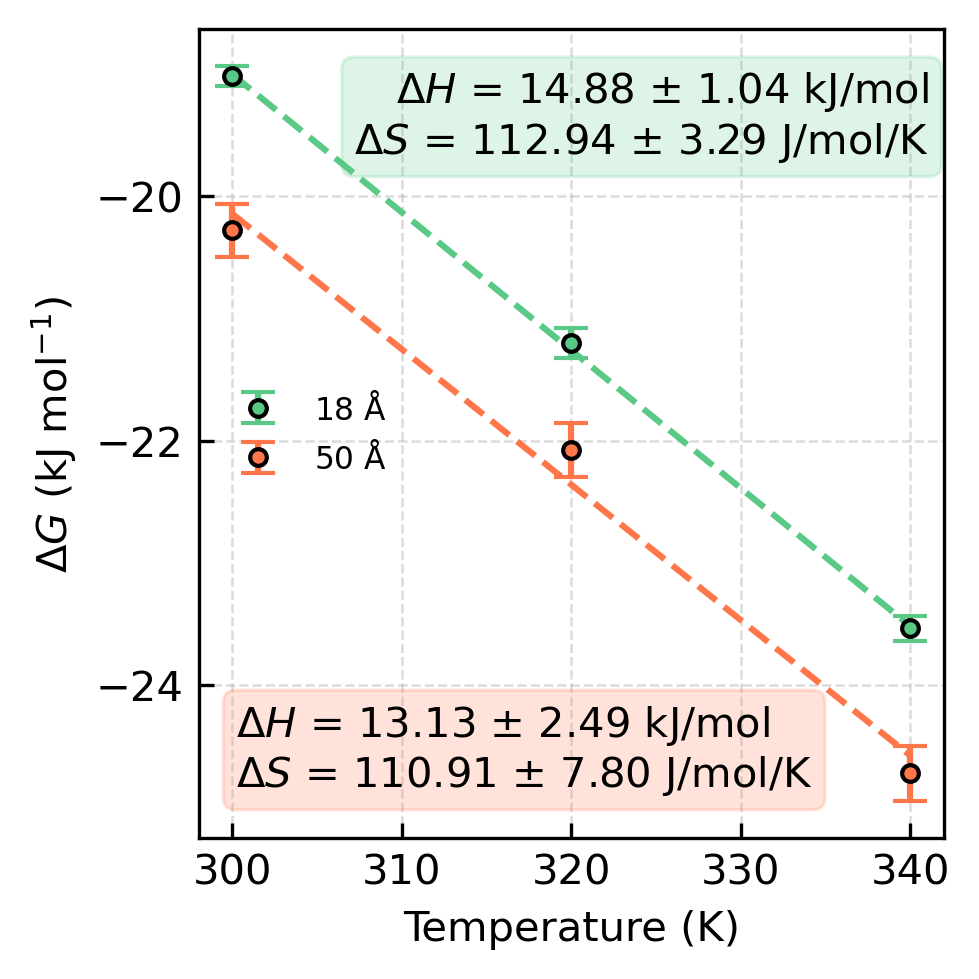}
    \caption{\textbf{Finite size effects:} Comparison of enthalpy and entropy computed from OPES simulations in simulation boxes of length 18~\AA~and 50~\AA~with a \ch{Ca}/\ch{CO3} ion pair and 197 and 5319 waters respectively. Standard errors were estimated from 6 independent simulations per temperature for the 18~\AA~system and 10 for the 50~\AA~system.}
    \label{fig:finitesize}
\end{figure}

Figure \ref{fig:finitesize} compares the association free energy computed from 6 and 10 independent replicas respectively for a 18~\AA~and 50~\AA~system as a function of temperature.
Both setups each contained a single \ch{CaCO3} ion pair and 197 and 5319 waters respectively, and were first equilibrated in the NPT ensemble at each temperature.
The finite size effect is roughly constant with temperature and has the effect of increasing the strength of the association energy by roughly 1.1 kJ/mol.
Table \ref{tab:finite_size} summarises the corrections.
All of the results shown in the main text have been corrected for this finite size effect, and the MLP and classical FF errors have been summed in quadrature. 
\begin{table}[t]
\centering
\begin{tabular}{lccc}
\toprule
 & 18~\AA~ & 50~\AA~ & finite size effect \\
\midrule
$\Delta G_{300 K}$ [kJ/mol] 
& $-19.0 \pm 0.1$ 
& $-20.3 \pm 0.2$
& $-1.3 \pm 0.2$ \\

$\Delta H$ [kJ/mol] 
& $14.9 \pm 1$ 
& $13.1 \pm 3$ 
& $-1.8 \pm 3$ \\

$\Delta S$ [J/mol\,K] 
& $112.9 \pm 3$ 
& $110.9 \pm 8$ 
& $-2 \pm 8$ \\
\bottomrule
\end{tabular}
\caption{\textbf{Finite size effects:} Comparison of enthalpy, entropy computed from OPES simulations in simulation boxes of length 18~\AA~and 50~\AA~with a \ch{Ca}/\ch{CO3} ion pair and 197 and 5319 waters respectively. Standard errors were estimated from 6 independent simulations per temperature for the 18~\AA~system and 10 for the 50~\AA~system.}
\label{tab:finite_size}
\end{table}
\subsection{Comparison to previous literature}\label{sec:lit_comp_ip}
\subsubsection{cWFT}
\begin{figure}[h]
    \centering
    \includegraphics[width=0.5\linewidth]{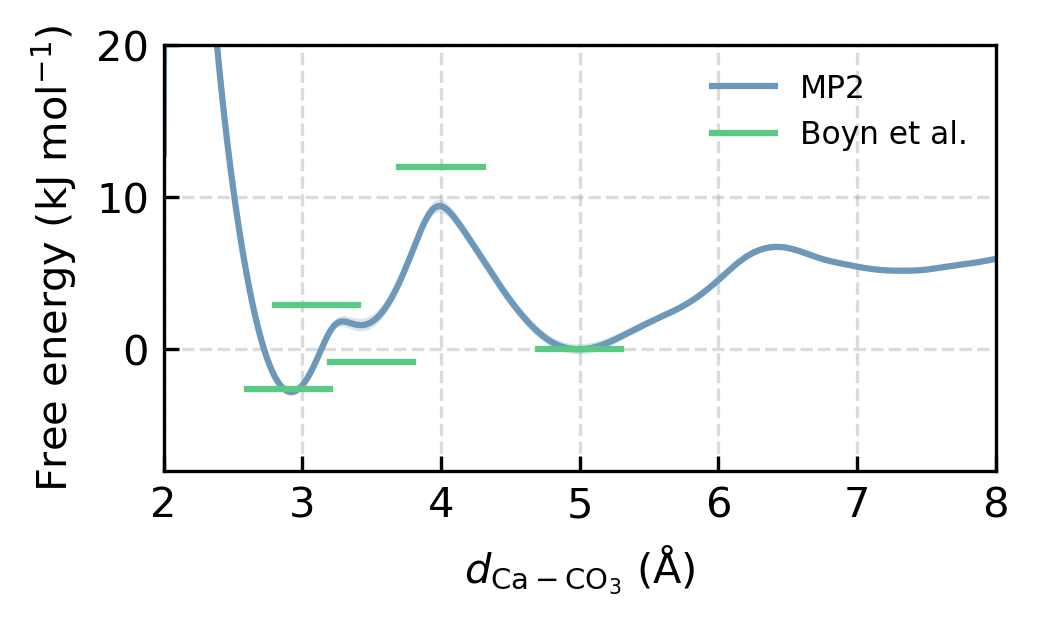}
    \caption{\textbf{Comparison of MP2 free energy profile to previous literature:} Comparison of MP2 free energy profile obtained from this work at 300 K (blue curve) compared to the corrected features of the DFT PMF to MP2 from~\citet{boynCharacterizingMechanismsCa2023} (green lines). The PMF and features are aligned such that the SShIP is at 0.}
    \label{fig:ec_comparison}
\end{figure}
 As described in the main text, previous studies have explored Ca--\ch{CO_3} ion pairing in water using quantum embedding approaches at the MP2 level \cite{boynCharacterizingMechanismsCa2023} and more recently at the CCSD(T) level~\cite{bianTransferLearningMeets2026}.
\color{black}
We now compare our results with these previous simulations and contextualise the differences between the studies.
Figure \ref{fig:ec_comparison} compares our full MP2 free energy profile to the free energies reported by \citet{boynCharacterizingMechanismsCa2023}. 
In the approach of \citet{boynCharacterizingMechanismsCa2023}, selected features of a DFT free energy profile are corrected to the MP2 level using an embedded correlated-wavefunction (ECW) multilevel scheme.
The relative stabilities of the bidentate and solvent
shared ion pair (SShIP) are in good quantitative agreement between both models.
Interestingly, the barrier between the CIP and SShIP is approximately 3~\(\mathrm{kJ\,mol^{-1}}\) higher in the multilevel embedding approach than in our simulations.
 Because \citet{boynCharacterizingMechanismsCa2023} correct only the internal energy differences along the underlying DFT PMF, this discrepancy suggests that explicitly sampling the full MP2-quality PES -- and thereby including entropic contributions -- can be important for obtaining reliable transition state barriers.
There are also important sampling limitations associated with the underlying DFT PES, which was generated from \textit{ab initio} blue moon ensemble (BME) simulations. We now discuss these in terms of both the accessible timescales and length scales:

\textbf{Timescale:}
As shown in Fig.~3 of the main manuscript, and in previous studies~\cite{piaggiInitioMachinelearningSimulation2025,raiteriIonPairingMultiple2020}, ion pairing is coupled to slow(er) solvent degrees of freedom that are orthogonal to the ion pair separation.
In particular, we show in Table~\ref{tab:ca_co3_water_structural_summary} that exchange around the calcium ion occurs on a timescale of roughly 25 ps.
If these solvent rearrangements are not explicitly included in the bias, they are therefore difficult to sample adequately within the accessible (\textit{ab initio}) simulation timescales, which can hinder convergence of the ion pairing free energy profile.
In particular, \citet{boynCharacterizingMechanismsCa2023} employed only 10 ps of sampling per BME window, biasing only the ion pair separation variable.
\begin{figure}[h!]
    \centering
    \includegraphics[width=0.5\linewidth]{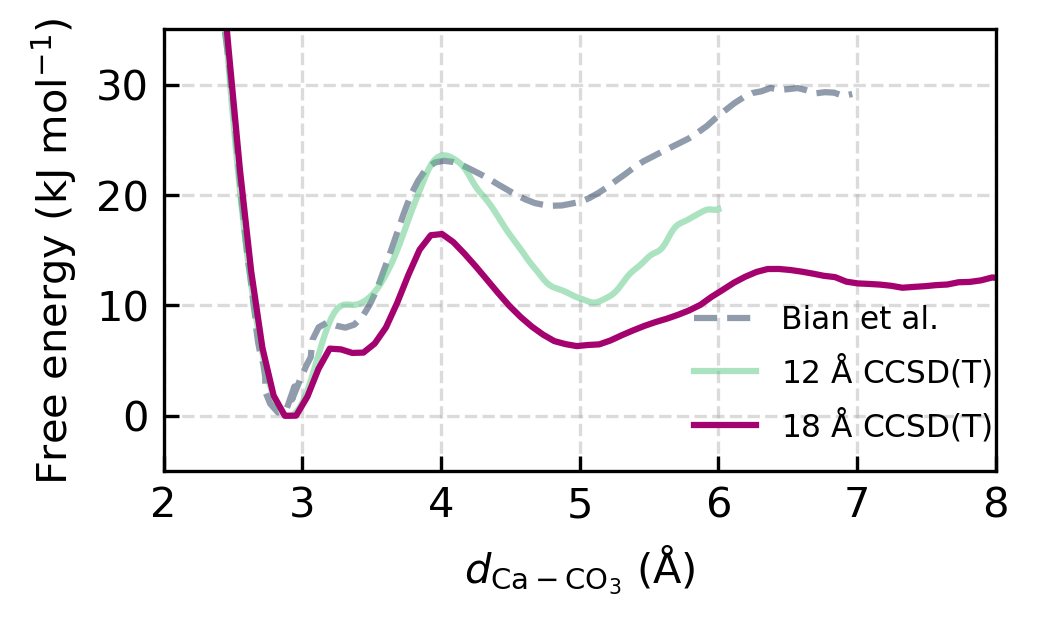}
\caption{\textbf{Comparison of CCSD(T) free energy profile to previous literature: }Comparison of the CCSD(T) result from this work with the transfer-learned emb-CCSD(T) free energy profile reported by ~\citet{bianTransferLearningMeets2026}, both evaluated at 330K. The magenta profile corresponds to our converged simulation setup, as reported in the main manuscript. The green profile was obtained using our CCSD(T) model in a 12~\AA{} simulation cell, matching the system size used in Ref.~\onlinecite{bianTransferLearningMeets2026}.}
    \label{fig:ccsdt_comparison}
\end{figure}

These sampling limitations are addressed more directly in the recent work of \citet{bianTransferLearningMeets2026}, which builds on \citet{boynCharacterizingMechanismsCa2023} by combining transfer learning of ECW data with an underlying DFT MLP.
Similar to our approach, the use of MLPs enables more extensive sampling at both the DFT and transfer-learned ECW levels. 
This represents a significant advance over the \textit{ab initio} MP2-based results shown in Fig.~\ref{fig:ec_comparison} from \citet{boynCharacterizingMechanismsCa2023}, and additionally allows \citet{bianTransferLearningMeets2026} to report a CCSD(T)-quality free energy profile.

In Fig.~\ref{fig:ccsdt_comparison}, we compare our CCSD(T) MLIP free energy profiles with that reported by~\citet{bianTransferLearningMeets2026}.
Because their approach now recovers the full ion pairing free energy landscape to a separation of 6~\AA{}, a direct comparison is possible.
For consistency, we show our results at 330 K, matching the temperature used in their simulations, and include both our converged 18~\AA{} simulation box and a smaller 12~\AA{} box to illustrate finite size effects.
We also point out that the requirement of \citet{bianTransferLearningMeets2026} to perform the simulations at 330~K is to overcome limitations in describing water structure at ambient conditions at GGA level (the level which their water remains at).
This is no longer necessary with our CCSD(T) model, which reproduces water structure at the correct experimental temperature (see Figs.~\ref{fig:density_caco3} and \ref{fig:mp2ccsdt_valid}).

Despite the same nominal level of electronic structure theory, substantial differences remain between the free energy profiles.
The relative stabilities of the bidentate and monodentate CIPs are comparable, but the CIP-to-SShIP barrier reported by~\citet{bianTransferLearningMeets2026} is approximately 5~\(\mathrm{kJ\,mol^{-1}}\) higher than in our 18\,\AA{} CCSD(T) simulations.
More strikingly, the SShIP is predicted to be approximately 15~\(\mathrm{kJ\,mol^{-1}}\) less stable than in our 18~\AA{} CCSD(T) profile.

\begin{figure}[h!]
    \centering
    \includegraphics[width=1.0\linewidth]{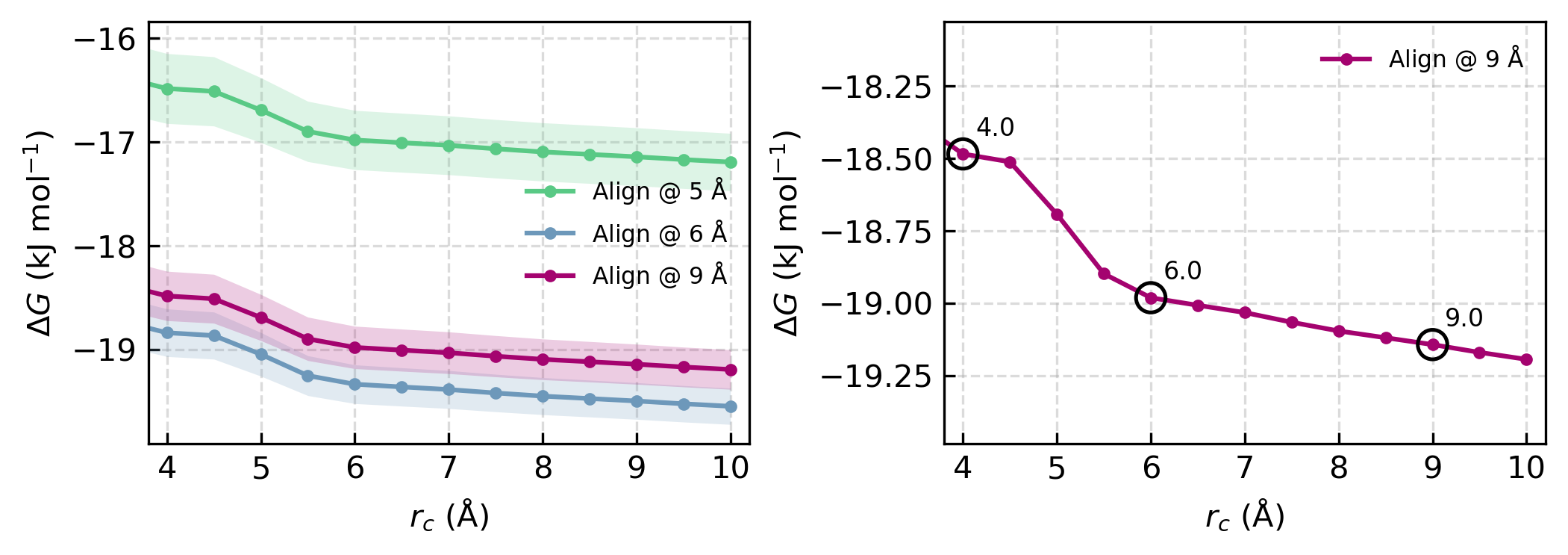}
    \caption{\textbf{Finite size effects on ion pairing free energy:} Left panel shows the computed $\Delta G$ for the CCSD(T) model at 330~K, as a function of the cut-off radius for the integral in Equation~5 and the choice of alignment distance of the PMF to the analytic solution of Equation~4. The right hand plot focuses on the result aligned at 9~\AA{}, showing the smaller effect on the choice of cut-off radius $r_c$.}
    \label{fig:alignment_cutoff}
\end{figure}

To further investigate the origin of these differences, we repeated our CCSD(T) simulations using the same $\sim$~12\,\AA{} length box containing 53 water molecules as~\citet{bianTransferLearningMeets2026}, thereby making the two setups directly comparable.
As shown also in Fig.~\ref{fig:ccsdt_comparison}, under these conditions, we recover excellent quantitative agreement not only in the CIP region, but also for the barrier separating the CIP and SShIP states.
This indicates that finite size effects associated with the smaller simulation box used by~\citet{bianTransferLearningMeets2026} account for a substantial part of the discrepancy between the two free energy profiles.
We next examine how these finite size effects propagate to the ion pair association free energy -- the key thermodynamic quantity that can be compared directly with experiment -- in Fig.~\ref{fig:alignment_cutoff}.

\textbf{Finite size effects on ion pairing free energy:}

As described previously in Sec.~\ref{sec:ionpairing}, extracting an ion pair association free energy from the calculated PMF requires careful treatment of both its long-range behaviour and the definition of the bound state.
First, the long-range tail of the simulated PMF must be aligned with the analytical solution in Eq.~4 for two point charges interacting through the dielectric response of the solvent.
The resulting aligned PMF, $\phi'(r)$ is then integrated up to a cutoff, $r_c$, defining the extent of the bound ion pair state to obtain the association constant and corresponding standard association free energy.

In Fig.~\ref{fig:alignment_cutoff}, we examine the sensitivity of the calculated ion pairing free energy to these two choices by varying both the PMF alignment range and the integration cutoff, $r_c$.
The left panel of Fig.~\ref{fig:alignment_cutoff} shows that the dominant source of uncertainty is the long-range alignment of the PMF.
In particular, a simulation box with a length of only 12 \AA{} necessarily limits the ion separations over which the PMF can be aligned reliably to the analytical point-charge result. 
Aligning the PMF at only 5 \AA, for example, introduces an error of almost 3~\(\mathrm{kJ\,mol^{-1}}\).
This is larger than many of the differences between electronic structure methods discussed in the main manuscript, highlighting the importance of the simulation setup itself.

The right panel of Fig.~\ref{fig:alignment_cutoff} shows that, once the PMF has been properly aligned in its long-range regime, the dependence on the precise choice of integration cutoff is considerably smaller.
Thus, the accuracy of the association free energy is controlled primarily by the quality of the long-range PMF alignment, provided that a physically reasonable cutoff is used to define the bound state \cite{aufortComputationalInsightsMg22022}.

Even after accounting for finite size effects using the smaller box simulation, the SShIP minimum reported by~\citet{bianTransferLearningMeets2026} remains approximately 10~\(\mathrm{kJ\,mol^{-1}}\) less stable than in our CCSD(T) MLP.
We suggest that this residual discrepancy may arise, at least in part, because the model of Bian and Carter does not represent a fully CCSD(T)-level PES.
As shown in Fig.~7 of~\citet{bianTransferLearningMeets2026}, the water description in their transfer-learned ECW CCSD(T) model remains at the underlying DFT level.

This distinction is important because we have shown that entropic contributions to ion pairing are significant and differ between DFT and CCSD(T).
At larger ion separations, where solvent reorganisation becomes increasingly important, the water degrees of freedom are expected to make a substantial contribution to the entropy.
We therefore emphasise that the model developed in our present work represents the full system at CCSD(T) accuracy, including both the ions and the surrounding water.
As further validation, in Figs~\ref{fig:density_caco3} and \ref{fig:mp2ccsdt_valid}, we show that the bulk-water properties of our CCSD(T) model, including the RDFs and density, agree closely with experiment and with previous CCSD(T)-based models of liquid water.

Finally, we summarise the key methodological features that, based on the comparisons above, make the protocol developed in this work more complete and reliable for computing reactive processes in aqueous solution:
\begin{enumerate}
    \item \textbf{Sampling the full free energy landscape at CCSD(T) accuracy:}
    Our MLP represents the full system at CCSD(T) accuracy and therefore enables direct sampling of both energetic and entropic contributions to the free energy landscape.
    This is important because we have shown that entropic contributions to ion pairing are significant and can differ appreciably between DFT and CCSD(T).
    \item \textbf{Converged sampling of solvent degrees of freedom:}
    Ion pairing is coupled to solvent rearrangements that occur orthogonally to the ion pair separation and on comparatively long timescales. The efficiency of the CCSD(T) MLP allows these degrees of freedom to be sampled explicitly through an additional collective variable and over sufficiently long trajectories to obtain a converged free energy profile.
    \item \textbf{A simulation cell large enough to recover the correct long-range PMF:}
    Accurate association free energies require the atomistic PMF to be aligned with its analytical long-range limit. We have shown that small simulation cells restrict the range over which this alignment can be performed and can introduce errors of several \(\mathrm{kJ\,mol^{-1}}\), which is comparable to or larger than the differences between electronic structure methods.
    \item \textbf{A robust determination of the association free energy:}
    Once the PMF is correctly aligned, the remaining dependence on the precise cutoff used to define the bound state is comparatively small. Our protocol therefore combines a converged long-range PMF with a physically appropriate integration range, allowing the resulting association free energy to be determined reliably and compared directly with experiment.
\end{enumerate}
Taken together, these results show that obtaining quantitative ion pairing thermodynamics requires not only an accurate electronic structure description, but also converged sampling in a sufficiently large simulation cell, and careful treatment of the long-range free energy profile. In contrast to previous studies, the approach developed here addresses all of these requirements.
\subsubsection{Comparison to classical force-field}\label{sec:ff_comp}
\begin{figure}[h!]
    \centering
    \includegraphics[width=0.75\linewidth]{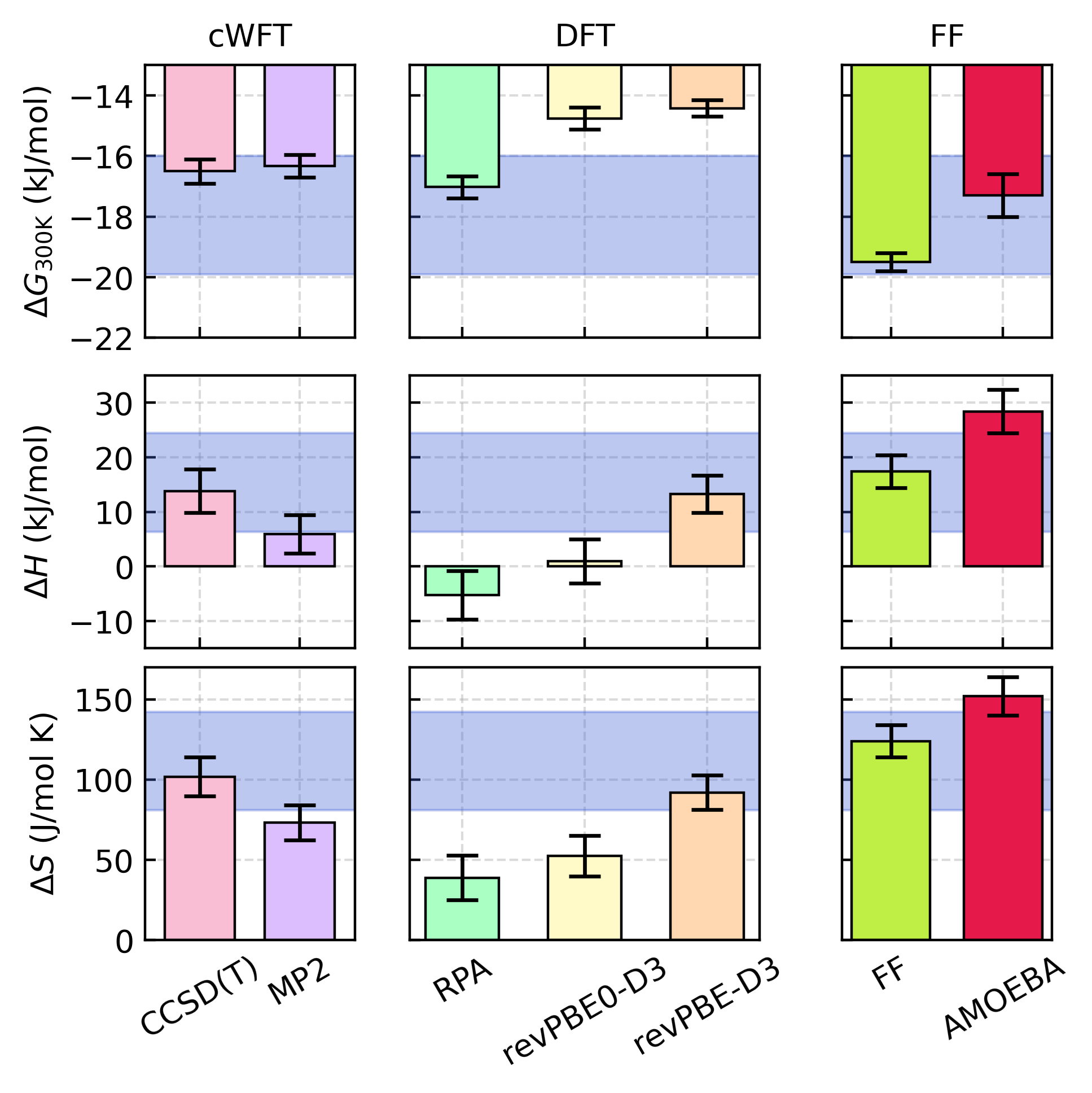}
    \caption{\textbf{Summary of simulation predictions for ion pairing thermodynamics:} Comparison of cWFT predictions of $\Delta G$, $\Delta H$ and $\Delta S$ with DFT and classical force field parameterisation from Ref. \cite{armstrongSolubilityconsistentForceField2023} as well as AMOEBA results~\cite{raiteriIonPairingMultiple2020}.}
    \label{fig:ff_comp}
\end{figure}
Finally, in Figure \ref{fig:ff_comp}  we compare the enthalpy and entropy predictions of CCSD(T), MP2 and DFT to the predictions of the classical force-field results from  Ref.~\cite{armstrongSolubilityconsistentForceField2023}, as well as previous results from the AMOEBA polarisable force field \cite{raiteriIonPairingMultiple2020}.
\clearpage
\subsection{Additional analysis}
\begin{figure}[h!]
    \centering
    \includegraphics[width=1.0\linewidth]{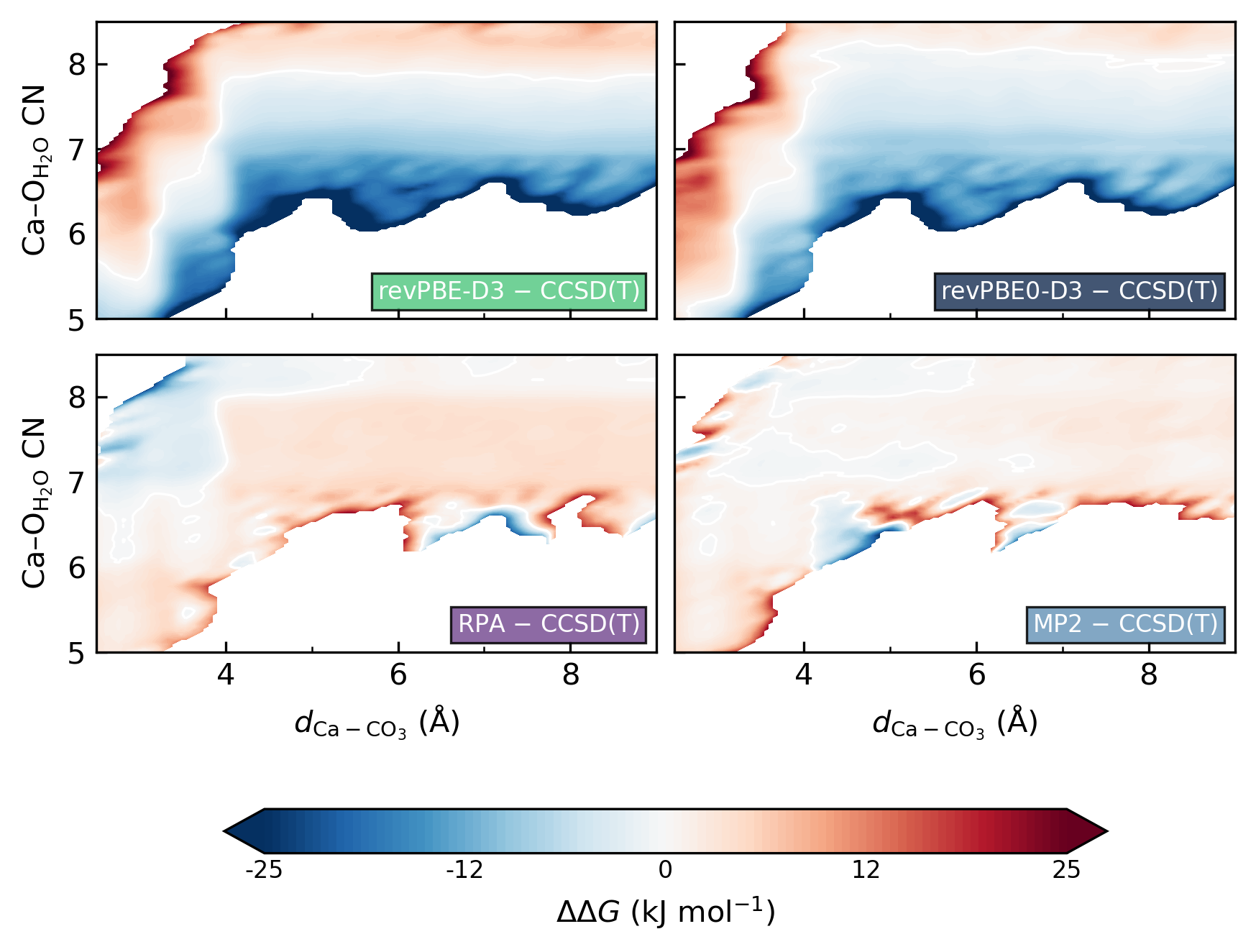}
    \caption{\textbf{2D-free energy differences w.r.t. CCSD(T):} Free energy difference as defined by Equation 7, between the levels of theory studied in this work and the CCSD(T) result. Blue is more stable w.r.t. CCSD(T) and red is less stable.}
    \label{fig:difference_fes}
\end{figure}
\color{black}{
Figure \ref{fig:difference_fes} compares the differences between free energy surfaces predicted by lower levels of theory with respect to CCSD(T).
Since the free energy is defined up to an arbitrary constant, for this analysis all free energy surfaces were first normalised to zero at their own global minimum.
Thus the final plotted surface $\Delta\Delta G(q)
$ corresponds to (using revPBE-D3 as an example):
\begin{equation}
\Delta\Delta G(q)
=
\left[
G_{\mathrm{revPBE-D3}}(q)
-
G_{\mathrm{revPBE-D3}}^{\min}
\right]
-
\left[
G_{\mathrm{CCSD(T)}}(q)
-
G_{\mathrm{CCSD(T)}}^{\min}
\right], 
\mathrm{where~} q =
\left(
d_{\mathrm{Ca-CO_3}},
\mathrm{CN}
\right).
\end{equation}
The most striking differences are observed for the lower levels of theory; revPBE-D3 and revPBE0-D3.
The differing calcium ion hydration mechanisms discussed in the main manuscript are again manifest in this analysis.
Clearly both revPBE-D3 and revPBE0-D3 overstabilise lower coordination number configurations with respect to CCSD(T), while understabilising configurations with larger coordination numbers.
While this effect is a combination of ion-ion, ion-solvent and solvent-solvent interactions, the fact that lower coordination numbers are overstabilised for almost all ion separations suggests that reliably describing ion-water interactions is a clear deficiency of lower levels of theory with respect to CCSD(T).
Furthermore in Table~\ref{tab:interaction_energy_comparison}, we show that indeed revPBE-D3 exhibits significant differences from CCSD(T), over(under)-stabilising ion-water interactions for small (large) ion-water CNs by up to almost 300 meV.

Previous literature has also shown such issues with using GGAs to describe ions, where \citet{eganNatureAlkaliIon2020} show revPBE-D3 has an error of over 1 kcal/mol on the ion-water dimer energy with respect to CCSD(T).
While these errors are somewhat improved with revPBE0-D3, the differences are still significant.
It has also been shown that using D3 dispersion, which takes atomic rather than ionic polarisabilities can introduce significant errors in the structure of solvated ions \cite{kostalCommonCationsAre2023} (although this is less pronounced for Ca$^{2+}$ than for K$^{+}$ or Na$^{+}$, which exhibit weaker electrostatic interactions.)
In contrast to revPBE-D3 and revPBE0-D3, RPA exhibits the opposite tendency, understabilising lower coordination states relative to CCSD(T) across most ion separations. 
RPA without additional corrections underbinds non-covalent interactions because of the neglect of single-excitation contributions and in particular it has been shown to underbind water clusters~\cite{tahirEfficientStructuralRelaxation2024, chedidEnergiesStructuresHarmonic2021} as well as being a poorer description of ion-water interaction energies compared to MP2 \cite{duignanQuantifyingHydrationStructure2020}.
MP2 on the other hand for the ion pairing (and also the ion-water system) performs comparably to CCSD(T), showing only very subtle differences on the free energy profile as well as having differences of less than 2 meV compared to LNO-CCSD(T)/tight for both standard and tight local-MP2 settings across all coordination numbers in Table~\ref{tab:interaction_energy_comparison}.

Nevertheless, in the quest for fully general models for complex aqueous systems, MP2 still has remaining deficiencies.
For example, it significantly overstructures liquid water, resulting in a substantially lower diffusion coefficient~\cite{beckMoreConvergedLess2026,willowWhyMP2WaterCooler2016, delbenBulkLiquidWater2013}.}
\section{Ion hydration mechanism}
\subsection{Simulation setups}\label{sec:cn_opes}
This section describes both the enhanced sampling and equilibrium molecular dynamics simulations performed to obtain the ion hydration properties.
In all cases, simulations were performed in a box of 197 waters with either a Ca or \ch{CO_3} ion corresponding to a cubic box length of roughly 18~\AA{}.
\subsubsection{Equilibrium molecular dynamics}
Using the simulation setup just described the system was first equilibrated for at least 1~ns in the NPT ensemble using the CSVR thermostat \cite{bussiCanonicalSamplingVelocity2007} with a temperature relaxation time of 0.1~ps and a 1.0~fs timestep using deuterium masses for hydrogen atoms. 
The resulting equilibrium density was then used for NVT simulations at 300~K, from which the radial distribution function and water residence times were computed.
All simulations were at least 1~ns long.
\subsubsection{OPES}\label{sec:opes_cn}
Similar to Section~\ref{sec:ionpairing} for the ion pairing free energies, the free energy profiles for the ion hydration were obtained from OPES \cite{invernizziRethinkingMetadynamicsBias2020} simulations, where the Ca-O$_{\ch{H2O}}$ (\ch{CO_3}-O$_{\ch{H2O}}$) coordination number was used as the collective variable to be biased, with an initial barrier of 50~kJ/mol.
This was computed via a smooth switching function $S(r)$:
\begin{equation}
S(r)=
\frac{1-\left(\dfrac{r-d_0}{r_0}\right)^8}
{1-\left(\dfrac{r-d_0}{r_0}\right)^{16}} \,
\end{equation}
with $d_0$=1.8~\AA~(2.6~\AA) and $r_0$=1.5~\AA~(1.5~\AA~) to overlap with the first peak in the \ch{Ca}-O (\ch{CO_3}-O) RDF.
The unbiased probability distribution ($P(n)$) was reconstructed from the OPES simulations by reweighting the OPES trajectory using the time-dependent bias potential $V_{\mathrm{bias}}$, as a function of the Ca-O$_{\ch{H2O}}$ (\ch{CO_3}-O$_{\ch{H2O}}$) coordination number ($n(\mathbf{r})$):
\begin{equation}
P(n) =
\frac{
\left\langle
\delta\!\left(n-n(\mathbf{r})\right)
e^{\beta V_{\mathrm{bias}}(\mathbf{r})}
\right\rangle_V
}{
\left\langle
e^{\beta V_{\mathrm{bias}}(\mathbf{r})}
\right\rangle_V
}
\end{equation}
where $\langle \cdot \rangle_V$ corresponds to the ensemble average taken over the biased simulation.
The final free energy surface is then given by:
\begin{equation}
F(n) = -k_B T \ln P(n)
\end{equation}

\subsection{Properties of ions in water for all models}
{\color{black}Figs~\ref{fig:rdfs_ca} and \ref{fig:rdfs_co3} compare the structure of the calcium and carbonate ions in water.
Both the radial distribution function, $g_{\mathrm{Ca(CO_3)-O_{H_2O}}}(r)$, and the cumulative coordination number, $N(r)$, are plotted.
The Ca$^{2+}$ plot again demonstrates clearly the change in coordination with increasing levels of
theory favouring larger coordination numbers, with a clear partition between revPBE-D3 and revPBE0-D3
and higher level methods.
Meanwhile, differences between levels of theory are more subtle for \ch{CO_3^{2-}}, reflecting the shallow and broad range of coordination states accessible.
\begin{figure}[h!]
    \centering
    \includegraphics[width=0.5\linewidth]{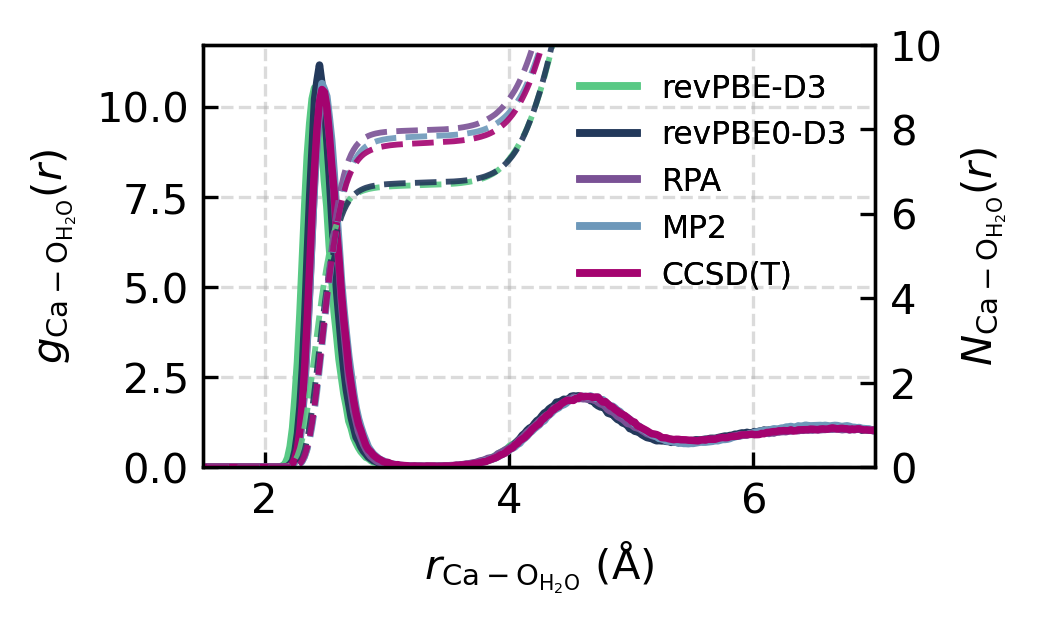}
    \caption{\textbf{Structure of Ca$^{2+}$ in water: } Radial distribution function (g$_\mathrm{{Ca-O_{H_2O}}}(r)$) and cumulative CN ($\mathrm{N_{Ca-O_{H_2O}}}$(r)) for the models trained in this work.}
    \label{fig:rdfs_ca}
\end{figure}
\begin{figure}[h!]
    \centering
    \includegraphics[width=1.0\linewidth]{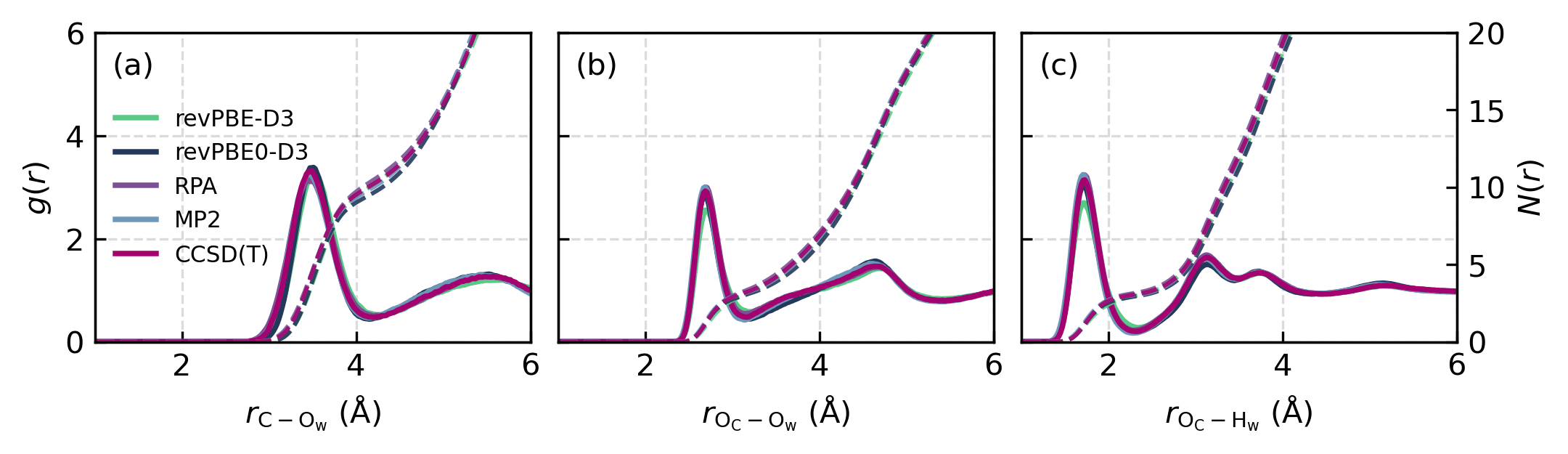}
    \caption{\textbf{Structure of CO$_3^{2-}$ in water: }Radial distribution function ($g(r)$) and cumulative CN ($N(r)$) for the models trained in this work for (a) $\mathrm{C-O_w}$, (b) $\mathrm{O_c-O_w}$ and (c) $\mathrm{O_C-H_w}$, where $\mathrm{O_w}$, $\mathrm{H_w}$ and $\mathrm{O_c}$ correspond to water oxygen, hydrogen and carbonate oxygen atoms respectively.}
    \label{fig:rdfs_co3}
\end{figure}

We have also computed the lifetimes of the first solvation shell waters, using the direct approach by \citet{hoferCharacterizationDynamicsReactivities2004}.
In this approach, the mean residence time is defined as;
\begin{equation}
\tau
=
\frac{t_{\mathrm{sim}}\langle \mathrm{CN}\rangle}
{N_{\mathrm{ex}}},
\label{eq}
\end{equation}
where $t_{\mathrm{sim}}$ is the total analysed simulation time, $\langle \mathrm{CN}\rangle$ is the mean number of water molecules in the first solvation shell, and $N_{\mathrm{ex}}$ is the number of accepted crossings of the first-shell boundary.
A crossing was counted only when the new coordination state persisted for at least 0.5~ps, thereby excluding short-lived recrossings. 
We note that comparison of the mean residence time with experiment is similarly challenging to that of the coordination number, with \citet{rampalTestingHypothesisThat2024} recently showing that experimental estimates for aqueous Ca$^{2+}$ span more than an order of magnitude, from below $100$~ps to several nanoseconds.}

\begin{table}[h!]
\color{black}
\centering
\caption{Hydration properties of Ca$^{2+}$ and CO$_3^{2-}$ from simulations at different electronic structure levels. The distance is Ca--O$_\mathrm{w}$ for Ca$^{2+}$ and C$_{\mathrm{CO_3}}$--O$_\mathrm{w}$ for CO$_3^{2-}$. Coordination-number uncertainties are equilibrium distribution standard deviations. Distance and residence time uncertainties are standard errors from five contiguous trajectory blocks.}
\label{tab:ca_co3_water_structural_summary}
\begin{tabular}{lcccccc}
\toprule
& \multicolumn{3}{c}{Ca$^{2+}$} & \multicolumn{3}{c}{CO$_3^{2-}$} \\
\cmidrule(lr){2-4}\cmidrule(lr){5-7}
Method & CN & $\langle r_{\mathrm{Ca-O_w}}\rangle$ / $\mathrm{\AA}$ & $\tau$ / ps & CN & $\langle r_{\mathrm{C_{CO_3}-O_w}}\rangle$ / $\mathrm{\AA}$ & $\tau$ / ps \\
\midrule
revPBE-D3 & $6.7 \pm 0.5$ & $2.454 \pm 0.003$ & $19.83 \pm 0.43$ & $9.5 \pm 0.7$ & $3.581 \pm 0.002$ & $2.01 \pm 0.06$ \\
revPBE0-D3 & $7.3 \pm 0.4$ & $2.517 \pm 0.001$ & $22.70 \pm 0.79$ & $9.1 \pm 0.7$ & $3.558 \pm 0.004$ & $2.68 \pm 0.10$ \\
RPA & $8.1 \pm 0.3$ & $2.532 \pm 0.002$ & $29.17 \pm 1.35$ & $10.6 \pm 0.7$ & $3.544 \pm 0.003$ & $2.28 \pm 0.05$ \\
MP2 & $7.6 \pm 0.5$ & $2.527 \pm 0.001$ & $25.24 \pm 0.63$ & $9.8 \pm 0.7$ & $3.532 \pm 0.006$ & $3.01 \pm 0.15$ \\
CCSD(T) & $7.7 \pm 0.4$ & $2.537 \pm 0.002$ & $25.97 \pm 3.44$ & $9.9 \pm 0.7$ & $3.544 \pm 0.003$ & $2.65 \pm 0.08$ \\
\bottomrule
\end{tabular}
\end{table}

\subsection{Ion hydration}
\begin{figure}[h!]
  \centering
   \includegraphics[width=1.0\textwidth]{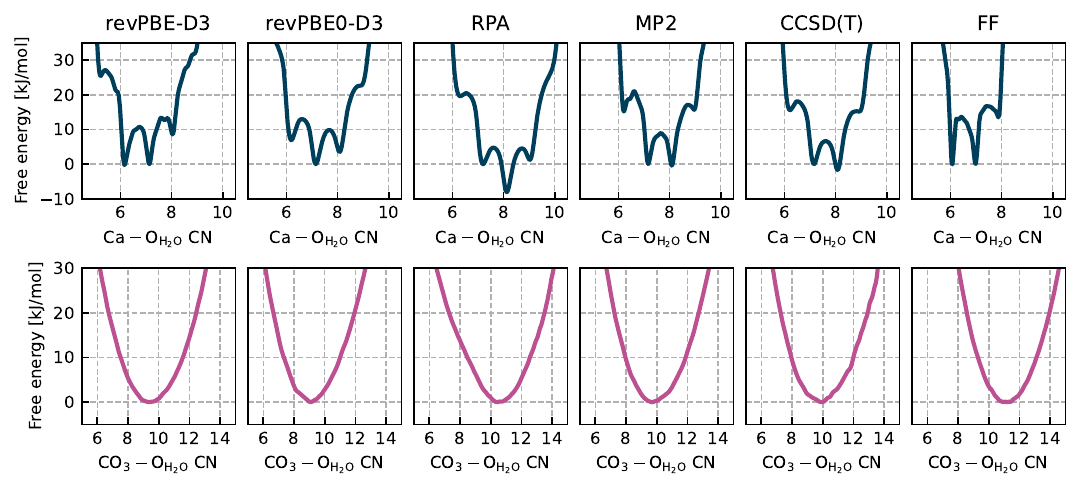}
    \caption{\textbf{Ion-water coordination free energies:} Free energy of \ch{Ca^{2+}} and \ch{CO3^{2-}} ions in water as a function of water coordination number for all of the levels of theory studied in this work. The \ch{Ca^{2+}} profiles are aligned to zero at CN=7, while the \ch{CO3^{2-}} profiles are aligned to zero at their own global minimum.}
   \label{fig:coordination}
\end{figure}
Figure \ref{fig:coordination} shows the free energy as a function of water coordination number for the Ca and \ch{CO_3} ions obtained from the OPES simulations described in Section~\ref{sec:cn_opes}.
The \ch{Ca} profile has many free energy minima.
Going to increasing levels of theory shifts the profiles towards higher coordination numbers; RPA, MP2 and CCSD(T) all giving 8 waters as the lowest free energy, while revPBE0-D3 gives a clear minimum at 7 and revPBE-D3 gives 6 and 7 as almost degenerate and the lowest in free energy.
The \ch{CO3} profiles are much broader, with fewer discrete coordination states. 
Here there is a shift towards higher coordination with water with increasing levels of theory.
Comparison of our results with force field models \cite{armstrongSolubilityconsistentForceField2023,kohagenAccurateDescriptionCalcium2014} shows reasonably good agreement, with an average calcium (carbonate)-water coordination number of 7.2 (10.7), compared to 7.7 (9.9) predicted by MP2 and CCSD(T) - however, the overall profiles are quite different.

\subsection{Free energy simulations}
\begin{figure}[h!]
    \centering
    \includegraphics[width=1.0\linewidth]{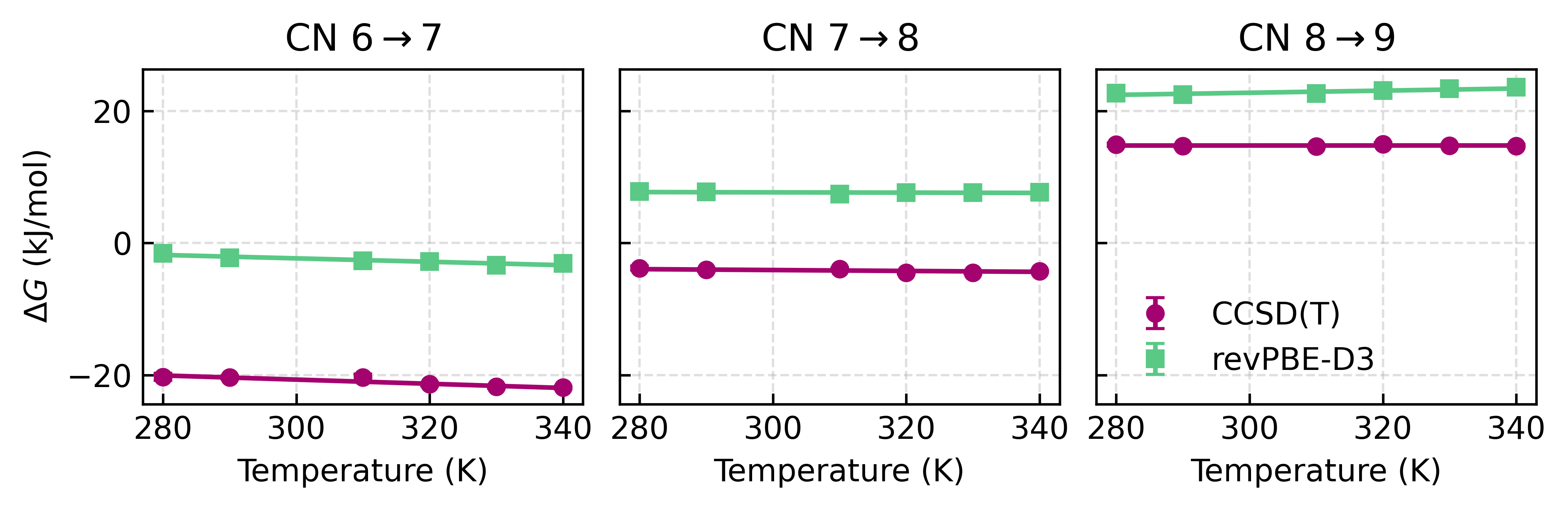}
    \caption{\textbf{Temperature dependence of free energy for coordination number transitions:} $\Delta$G as a function of temperature for transition A$\rightarrow$B where $\Delta G_{A\rightarrow B}$ = $\Delta G_{CN=B}-\Delta G_{CN=A}$. Computed for revPBE-D3 and CCSD(T) from OPES simulations shown in Fig.~\ref{fig:coordination}.}
    \label{fig:temperature_cacn}
\end{figure}

    Fig.\ref{fig:temperature_cacn} shows the computed free energy for coordination number transitions as a function of temperature.
    These were obtained from temperature-dependent OPES simulations of the calcium ion in water for both revPBE-D3 and CCSD(T) models.
    Each point is the average and standard error of 12 independent replicas, run for 1-2~ns, resulting in cumulatively over 100 ns of simulation.
    Simulation details are given in Sec.~\ref{sec:cn_opes}.
    The entropy and enthalpy were computed from a weighted linear regression.

\subsection{Finite size effects}\label{sec:finitesize_cn}
\begin{figure}[h!]
    \centering
    \includegraphics[width=0.5\linewidth]{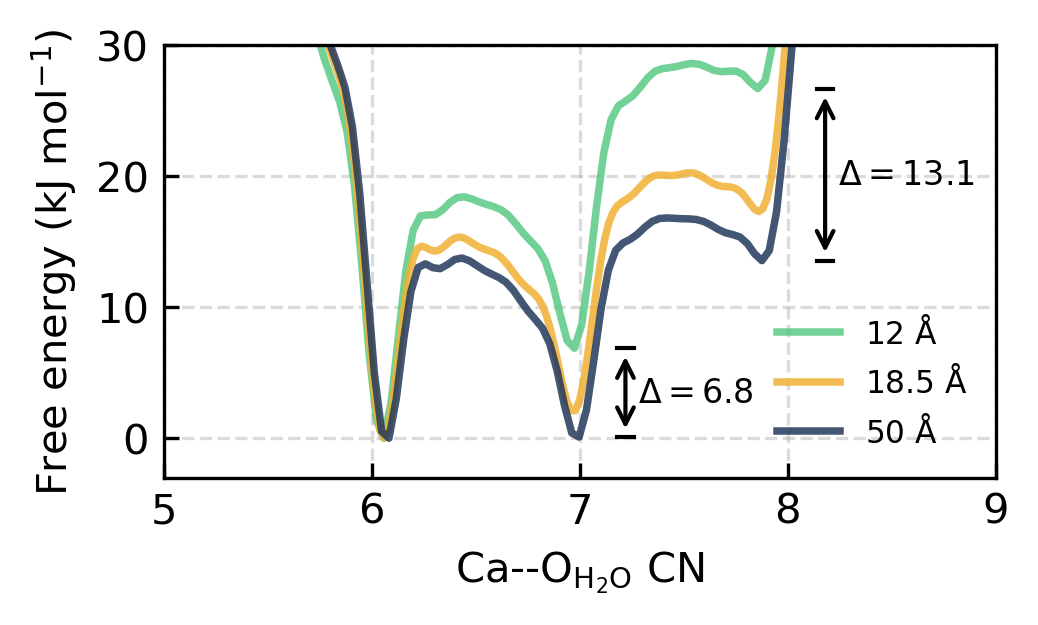}
    \caption{\textbf{Finite size effects on \ch{Ca-H2O} CN simulations:} Free energy profiles for \ch{Ca-H2O} simulations performed in simulation boxes of length 12, 18.5 and 50~\AA{}. Profiles are aligned to 0 at CN=6.}
    \label{fig:finitesize_cn}
\end{figure}

Well-tempered metadynamics simulations were performed using the force field of \citet{armstrongSolubilityconsistentForceField2023} to ensure convergence of the MLP simulation set-up, as well as to quantify errors arising from previous studies being performed in smaller simulation boxes.
Fig.~\ref{fig:finitesize_cn} shows the significant finite size effects resulting from the smaller 12~\AA{} box, with 7- and 8- fold coordinated states significantly understabilised by $\approx$7 and 13~\(\mathrm{kJ\,mol^{-1}}\) respectively.

\subsection{Gas phase interaction energies}
\begin{figure}[h!]
    \centering
    \includegraphics[width=1.0\linewidth]{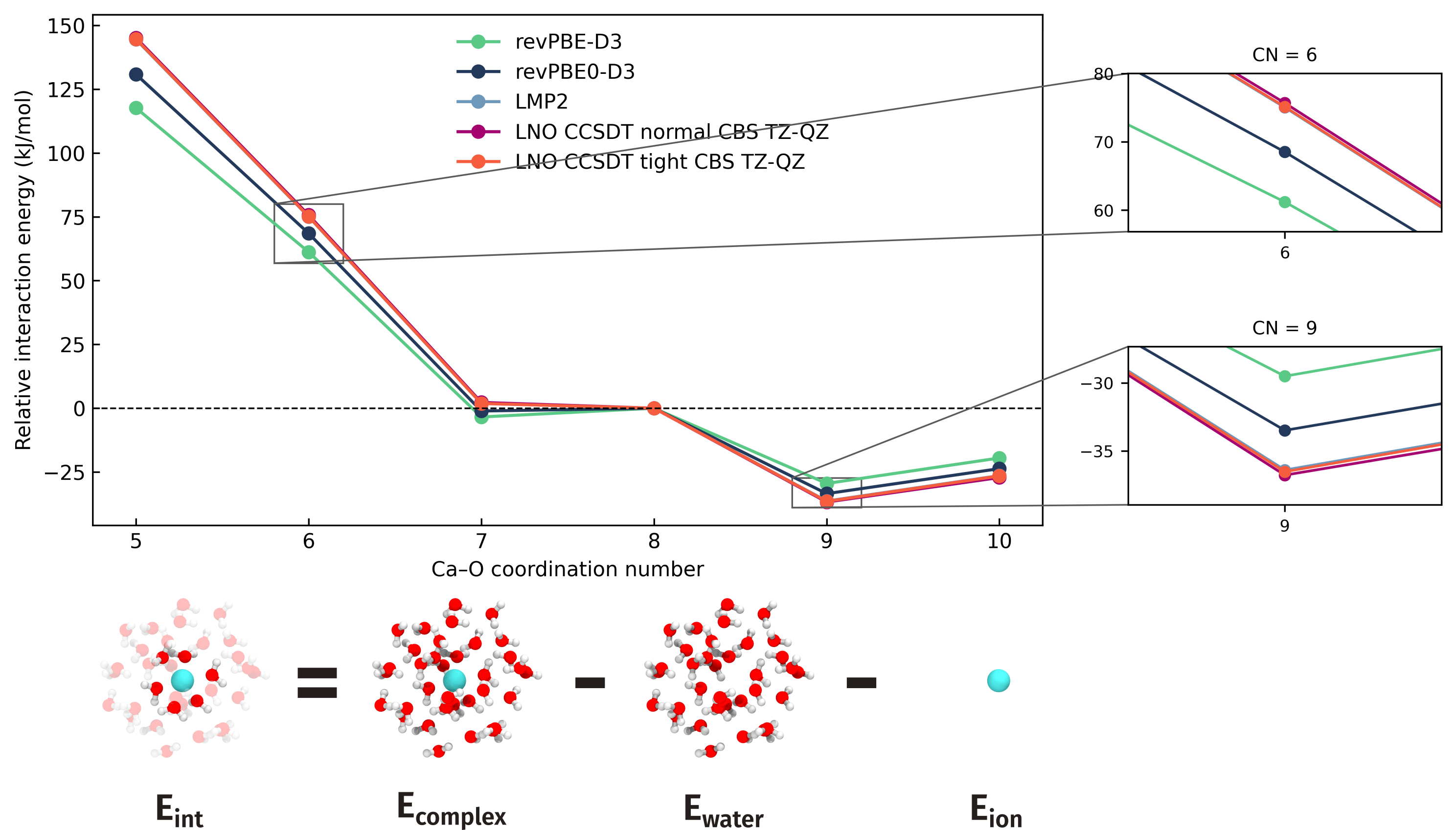}
    \caption{\textbf{Calcium-water interaction energies:} Calcium-water interaction energy as a function of number of coordinating waters in the first hydration shell. All interaction energies are given relative to CN=8. Representative snapshots are shown underneath.}
    \label{fig:interaction_energy}
\end{figure}
\begin{table}[h!]
\color{black}
\centering
\setlength{\tabcolsep}{6pt}
\caption{Mean interaction energies relative to CN=8 for each coordination number, in meV. Differences from the LNO-CCSD(T) tight CBS(TZ/QZ) result are given in parentheses.}
\label{tab:interaction_energy_comparison}
\begin{tabular}{crrrrrr}
\toprule
CN & revPBE-D3 & revPBE0-D3 & LMP2 & LMP2 (tight) & LNO-CCSD(T) & LNO-CCSD(T) (tight) \\
\midrule
5 & 1220 (-280) & 1356 (-143) & 1498 (-1) & 1498 (-2) & 1504 (+5) & 1500 (0) \\
6 & 634 (-144) & 710 (-69) & 779 (0) & 778 (-1) & 784 (+6) & 779 (0) \\
7 & -36 (-55) & -11 (-31) & 21 (+2) & 19 (0) & 23 (+4) & 19 (0) \\
8 & 0 (0) & 0 (0) & 0 (0) & 0 (0) & 0 (0) & 0 (0) \\
9 & -306 (+73) & -347 (+31) & -378 (+1) & -377 (+1) & -381 (-3) & -378 (0) \\
10 & -202 (+73) & -246 (+30) & -274 (+1) & -274 (+1) & -282 (-7) & -275 (0) \\
\bottomrule
\end{tabular}
\end{table}
{\color{black}
    To further understand and validate our observed calcium-water hydration mechanism on an energetic level, we have computed the calcium-water interaction energy for several of the levels of theory studied in this work.
    For a given hydrated ion cluster, we define the interaction energy $E_{\mathrm{int}}$ as:
    \begin{equation}
        E_{\mathrm{int}}=E_{\mathrm{complex}}-E_{\mathrm{water}}-E_{\mathrm{ion}}.
    \end{equation}
    where $E_{\mathrm{complex}}$ is the energy of the full hydrated ion cluster, $E_{\mathrm{water}}$ is the energy of the hydrated cluster with the ion removed, and $E_{\mathrm{ion}}$ is the energy of the bare ion.
    We compute this for clusters extracted from periodic simulations that have different numbers (5, 6, 7, 8, 9 and 10) of coordinating waters in the first hydration shell.
    These clusters are selected to be $6\,$\AA{} in radius for the (O atoms) of the water molecule around the Ca ion, leading to clusters with roughly 27 waters surrounding the ion.
    As a reference, we use LNO-CCSD(T) with `tight' LNO thresholds using the CBS(TZ/QZ) complete basis set extrapolation with the triple-$\zeta$ (TZ) and quadruple-$\zeta$ basis set pairs~\cite{neeseRevisitingAtomicNatural2011} discussed in the Methods of the main text.
    We show in a benchmark in Table~\ref{tab:ccsdt_lno_convergence} that this gives relative interaction energies to within $4\,$meV of the canonical CCSD(T) results.
    For each level of theory, the relative interaction energy with respect to CN=8 is computed, and the result is plotted in Fig.~\ref{fig:interaction_energy}, and tabulated in Table~\ref{tab:interaction_energy_comparison}.
    The trends observed here reinforce the trend we see from the free energies reported in the main manuscript, where revPBE-D3 significantly stabilises the 6 fold coordination state (relative to 8) by roughly 15~\(\mathrm{kJ\,mol^{-1}}\) compared to CCSD(T) and MP2.
    On the other-hand, this trend is flipped for the 9-fold coordination, with CCSD(T) stabilizing it (w.r.t.\ CN=8) by 7~\(\mathrm{kJ\,mol^{-1}}\) relative to revPBE-D3.
Additionally, these results further validate the use of `normal' thresholds in the LNO approximation for CCSD(T), with a maximum of 7~meV error with respect to the tight CBS(TZ/QZ) CCSD(T) result.}
\subsection{Literature Review}
\subsubsection{Compilation of experimental and simulation literature}
Tables~\ref{tab:aimd_mlip_calcium_cn}, \ref{tab:experimental_calcium_cn} and \ref{tab:force_field_calcium_cn} below summarise the main details of the reported calcium-water coordination number from experiment and simulation, plotted in Fig.4(a) in the main manuscript.
\input{si/figures/CN/literature_summary/aimd_mlip_cn_table}
\input{si/figures/CN/literature_summary/force_field_cn_table}
\input{si/figures/CN/literature_summary/experiment_cn_table}
\begin{figure}[h!]
    \centering
    \includegraphics[width=0.5\linewidth]{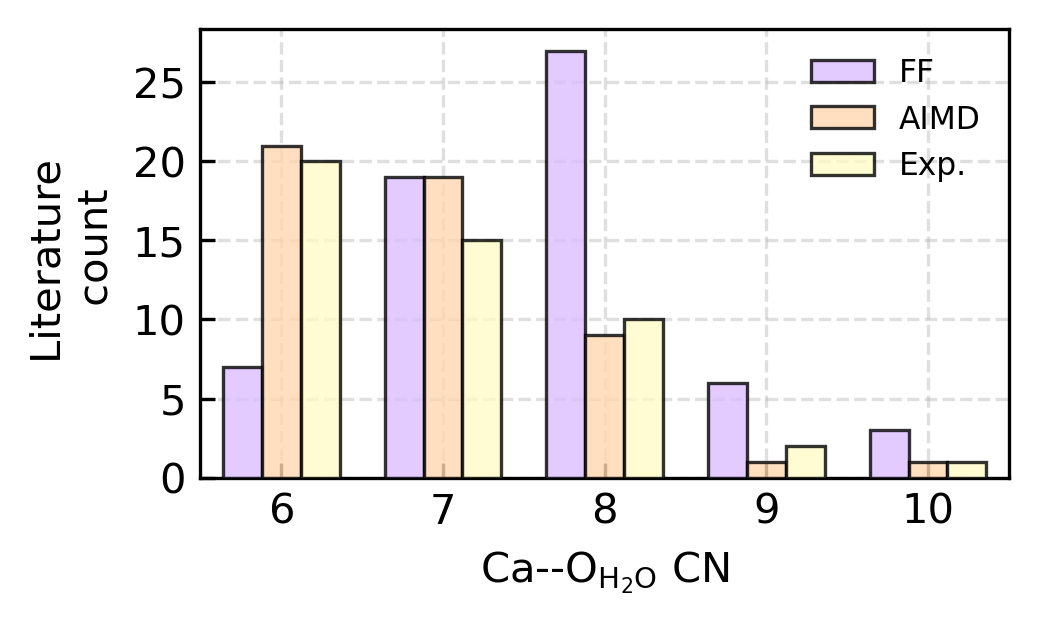}
    \caption{Summary of previously reported Ca-O$_{\mathrm{H_2O}}$ coordination numbers (CN) from the literature broken down by their source into FF, AIMD and experiment.}
    \label{fig:literature}
\end{figure}
\color{black}

\subsubsection{Comparison to previous literature}
{\color{black}
As a validation of our MLIP's ability to reproduce the underlying DFT description, we compare the revPBE-D3 results with previous AIMD studies.
Although most AIMD studies compiled in Table~\ref{tab:aimd_mlip_calcium_cn} do not report a full free energy profile, direct comparison is possible with the works of \citet{boynProbingPHDependentDehydration2023} and \citet{baerLocalAqueousSolvation2016}.
We note that both studies are subject to several of the limitations discussed above, including restricted sampling, relatively small simulation cells, and differences in the underlying electronic structure treatment (\citet{baerLocalAqueousSolvation2016} use BLYP-D instead of revPBE-D3).
Therefore while such comparison should be interpreted with considerable caution, they provide useful independent benchmarks for assessing the consistency of our revPBE-D3 MLP results with previous first-principles simulations.
\begin{figure}[h!]
    \centering
    \includegraphics[width=0.5\linewidth]{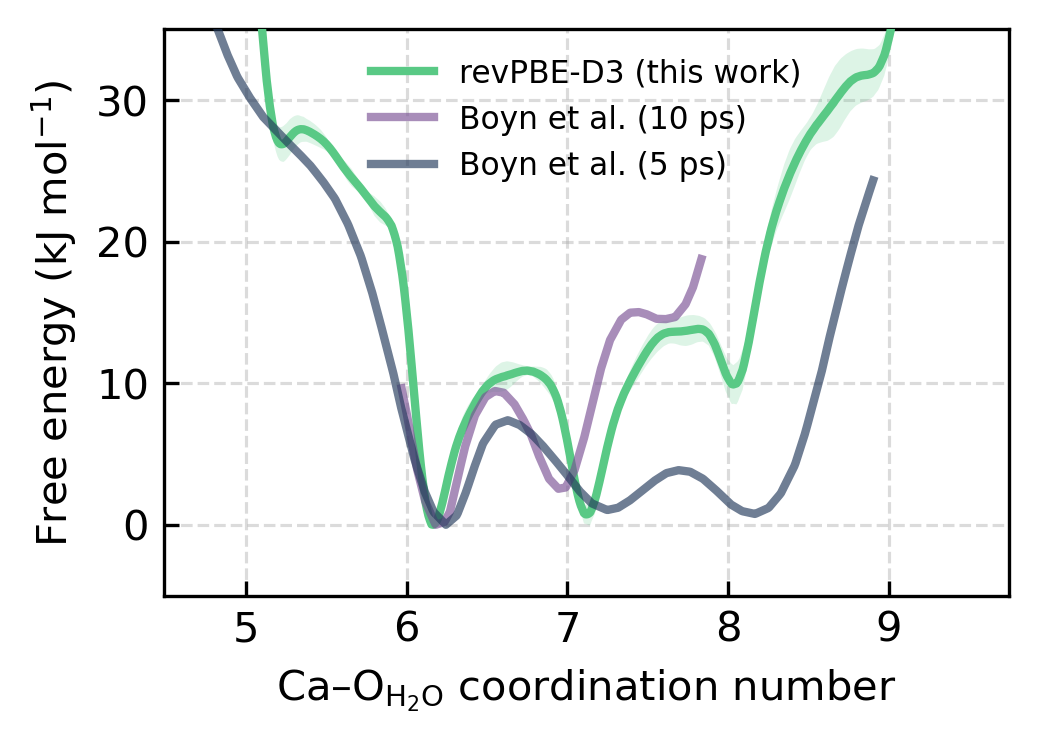}
    \caption{Comparison between revPBE-D3 MLP result from this work and revPBE-D3 AIMD result from \citet{boynProbingPHDependentDehydration2023}. All profiles are aligned so that CN=6 is set to 0 and the profiles of Boyn and Carter have also been shifted such that the reported CNs are integers. }
    \label{fig:boyn_comparison}
\end{figure}

Fig.~\ref{fig:boyn_comparison} compares our results to those of blue moon ensemble simulations from \citet{boynProbingPHDependentDehydration2023}.
While the BME is intended to accelerate exploration of the relevant configurations, each window was still restricted to 5 or 10~ps sampling time, and the simulations were performed in an approximately 11~\AA{} cell and remain subject to the sampling and finite size limitations discussed above in Sec.~\ref{sec:finitesize_cn}. 
They report 2 profiles, one spanning the full relevant CN space with 5~ps of sampling per BME window, and another one constrained with 10~ps sampling per BME window.
The 2 profiles of \citet{boynProbingPHDependentDehydration2023} differ significantly with each other, with the 5~ps one showing near-degeneracy between 6-, 7-, and 8- coordinated states.
The picture changes substantially when the sampling time is doubled, with the 10~ps profile placing the 8-coordinated state approximately 15~\(\mathrm{kJ\,mol^{-1}}\) higher in free energy.
This longer simulation begins to approach our revPBE-D3 free energy profile, suggesting that the limited sampling in \citet{boynProbingPHDependentDehydration2023} may preclude a fully converged description of the relative stability of the different coordination states.
We also point out that the definition of the ion-water coordination number used by Boyn and Carter does not yield integer minima of the CNs.
Therefore here we have rounded and shifted their reported free energy profiles such that the minima correspond to integer coordination numbers to enable comparison with our results.%

\begin{figure}[h!]
    \centering
    \includegraphics[width=0.5\linewidth]{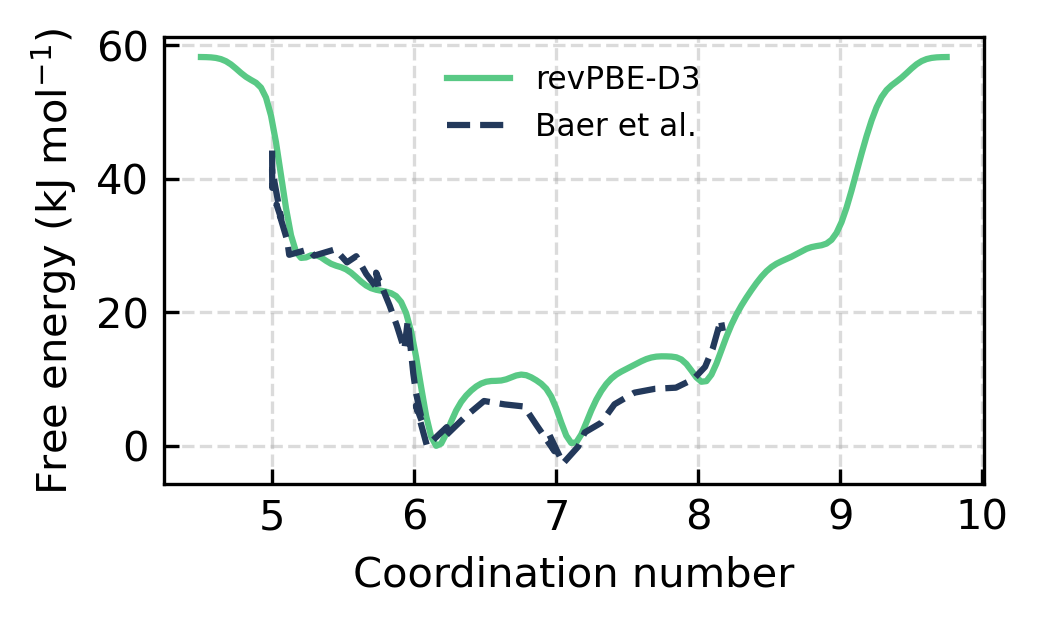}
    \caption{\textbf{Comparison of revPBE-D3 free energy profile to previous GGA literature: }Comparison between revPBE-D3 MLP result from this work and BLYP-D umbrella sampling AIMD result from \citet{baerLocalAqueousSolvation2016}. All profiles are aligned so that CN=6 is set to 0.}
    \label{fig:mundy_comparison}
\end{figure}
Fig.~\ref{fig:mundy_comparison} compares our revPBE-D3 results with the umbrella-sampling AIMD simulations of \citet{baerLocalAqueousSolvation2016}.
In this case, finite size and sampling effects are expected to be substantially reduced, with simulations performed using 96 water molecules in a $\sim$14~\AA{} box and approximately 50~ps of sampling per umbrella window.
The principal methodological difference is therefore the choice of DFT functional, with \citet{baerLocalAqueousSolvation2016} employing BLYP-D compared with the revPBE-D3 reference used to train our MLIP.
Overall, the two free energy profiles are in good agreement.
This suggests that, once statistical sampling and finite size effects are adequately controlled, differences between GGA functionals may be smaller than implied by the broad spread of coordination numbers reported in the AIMD literature compiled in Table~\ref{tab:aimd_mlip_calcium_cn}, where the predominantly GGA-level simulations have predicted coordination numbers ranging from 6 to 10.}

\bibliography{bibliography}

%% file: si/figures/CN/literature_summary/aimd_mlip_cn_table.tex
\begin{longtable}{@{}l r l l l c r@{}}
\caption{AIMD and MLIP calcium--water coordination numbers.}\label{tab:aimd_mlip_calcium_cn}\\
\toprule
Ref. & Year & Technique & Functional & System & \shortstack{Production time\\ / ps} & CN \\
\midrule
\endfirsthead
\multicolumn{7}{c}{\tablename\ \thetable{} -- continued}\\
\toprule
Ref. & Year & Technique & Functional & System & \shortstack{Production time\\ / ps} & CN \\
\midrule
\endhead
\midrule
\multicolumn{7}{r}{Continued on next page}\\
\endfoot
\bottomrule
\endlastfoot
\cite{tongraarSolvationCa2Water1997} & 1997 & QM/MM & HF (STO-3G) & \parbox[t]{2cm}{\raggedright 1Ca$^{2+}$ + 199H$_2$O} & 4 & 10 \\ 
\cite{tongraarSolvationCa2Water1997} & 1997 & QM/MM & HF (LANL2DZ) & \parbox[t]{2cm}{\raggedright 1Ca$^{2+}$ + 199H$_2$O} & 1.6 & 8.3 \\ 
\cite{schwenkMolecularDynamicsSimulations2001} & 2001 & QM/MM & HF & \parbox[t]{2cm}{\raggedright 1Ca$^{2+}$ + 199H$_2$O} & 7.6 & 7.1 \\ 
\cite{schwenkMolecularDynamicsSimulations2001} & 2001 & QM/MM & B3LYP & \parbox[t]{2cm}{\raggedright 1Ca$^{2+}$ + 199H$_2$O} & 8.1 & 7.6 \\ 
\cite{schwenkDynamicsSolvationProcess2001} & 2001 & QM/MM & HF & \parbox[t]{2cm}{\raggedright 1Ca$^{2+}$ + 199H$_2$O} & 10 & 8 \\ 
\cite{schwenkDynamicsSolvationProcess2001} & 2001 & QM/MM & B3LYP & \parbox[t]{2cm}{\raggedright 1Ca$^{2+}$ + 199H$_2$O} & 9 & 8 \\ 
\cite{bakoCarParrinelloMolecular2002} & 2002 & CPMD & BLYP & \parbox[t]{2cm}{\raggedright 1Ca$^{2+}$ + 54H$_2$O} & 7 & 6 \\ 
\cite{naorCarParrinelloMolecular2003} & 2003 & CPMD & BLYP & \parbox[t]{2cm}{\raggedright 1Ca$^{2+}$ + 31H$_2$O} & 14.2 & 7.5 \\ 
\cite{schwenkInitioQMMM2004} & 2004 & QM/MM & HF & \parbox[t]{2cm}{\raggedright 1Ca$^{2+}$ + 199H$_2$O} & 28 & 7.9 \\ 
\cite{schwenkInitioQMMM2004} & 2004 & QM/MM & B3LYP & \parbox[t]{2cm}{\raggedright 1Ca$^{2+}$ + 199H$_2$O} & 20 & 8 \\ 
\cite{lightstoneFirstPrinciplesMolecularDynamics2005} & 2005 & AIMD & PBE & \parbox[t]{2cm}{\raggedright 1Ca$^{2+}$ + 60H$_2$O} & 40 & 6.2--7.0 \\ 
\cite{ikedaHydrationPropertiesMagnesium2007} & 2007 & AIMD (constrained) & HCTH & \parbox[t]{2cm}{\raggedright 1Ca$^{2+}$ + 63H$_2$O} & 50 & 6.1 \\ 
\cite{todorovaCarParrinelloMolecular2008} & 2008 & CPMD & BLYP & \parbox[t]{2cm}{\raggedright 2Ca$^{2+}$ + 62H$_2$O} & 7.2 & 6 \\ 
\cite{todorovaCarParrinelloMolecular2008} & 2008 & CPMD & BLYP & \parbox[t]{2cm}{\raggedright 2CaCl$_2$ + 58H$_2$O} & 7.2 & 6 \\ 
\cite{todorovaCarParrinelloMolecular2008} & 2008 & CPMD & BLYP & \parbox[t]{2cm}{\raggedright 1CaCl$_2$ + 58H$_2$O} & 19.2 & 6.5 \\ 
\cite{ditommasoFirstPrinciplesSimulations2010} & 2010 & CPMD & PBE & \parbox[t]{2cm}{\raggedright 1Ca$^{2+}$ + 53H$_2$O} & 19.3 & 5.9 \\ 
\cite{tongraarStructureHydratedCa22010} & 2010 & QM/MM & HF & \parbox[t]{2cm}{\raggedright 1Ca$^{2+}$ + 199H$_2$O} & 55 & 7.4 \\ 
\cite{timkoInitioCalculationPotential2011} & 2011 & AIMD & BLYP & \parbox[t]{2cm}{\raggedright 1Ca$^{2+}$ + 1Cl$^-$ + 62H$_2$O} & 18 & 5--6 \\ 
\cite{wanprakhonHydrationStructureDynamics2011} & 2011 & QM/MM & HF & \parbox[t]{2cm}{\raggedright 1Ca$^{2+}$ + 199H$_2$O} & -- & 7.8 \\ 
\cite{wanprakhonHydrationStructureDynamics2011} & 2011 & ONIOM-XS & HF & \parbox[t]{2cm}{\raggedright 1Ca$^{2+}$ + 199H$_2$O} & -- & 7.6 \\ 
\cite{adeagboCa2SolvationFunction2012} & 2012 & CPMD & BLYP & \parbox[t]{2cm}{\raggedright 1Ca$^{2+}$ + 63H$_2$O} & 20 & 6.2 \\ 
\cite{bogatkoAqueousCa2System2013} & 2013 & CPMD & PBE96 & \parbox[t]{2cm}{\raggedright 1Ca$^{2+}$ + 64H$_2$O} & 33 & 6.7 \\ 
\cite{chizhikStructureHydrationShell2016} & 2016 & AIMD & BLYP & \parbox[t]{2cm}{\raggedright 1Ca$^{2+}$ + 48H$_2$O} & 50 & 6 \\ 
\cite{baerLocalAqueousSolvation2016} & 2016 & AIMD (constrained) & BLYP & \parbox[t]{2cm}{\raggedright 1Ca$^{2+}$ + 64H$_2$O} & -- & 6.3 \\ 
\cite{baerLocalAqueousSolvation2016} & 2016 & AIMD (constrained) & BLYP-D2 & \parbox[t]{2cm}{\raggedright 1Ca$^{2+}$ + 64H$_2$O} & -- & 6.7 \\ 
\cite{koskampReconsideringCalciumDehydration2019} & 2019 & AIMD & BLYP-D3 & \parbox[t]{2cm}{\raggedright 1Ca$^{2+}$ + 63H$_2$O} & 500 & 6.8 \\ 
\cite{koskampReconsideringCalciumDehydration2019} & 2019 & AIMD & revPBE-D3 & \parbox[t]{2cm}{\raggedright 1Ca$^{2+}$ + 63H$_2$O} & 40 & 6.1 \\ 
\cite{wangDensityFunctionalTheory2020} & 2020 & AIMD & PBE & \parbox[t]{2cm}{\raggedright 1Ca$^{2+}$ + 63H$_2$O} & 50 & 6 \\ 
\cite{wangDensityFunctionalTheory2020} & 2020 & AIMD & PBE-D3 & \parbox[t]{2cm}{\raggedright 1Ca$^{2+}$ + 63H$_2$O} & 50 & 6 \\ 
\cite{wangDensityFunctionalTheory2020} & 2020 & AIMD & PBE-D3 & \parbox[t]{2cm}{\raggedright 1Ca$^{2+}$ + 63H$_2$O} & 100 & 6 \\ 
\cite{wangDensityFunctionalTheory2020} & 2020 & AIMD & PBE-D3 & \parbox[t]{2cm}{\raggedright 1Ca$^{2+}$ + 63H$_2$O} & 500 & 6 \\ 
\cite{wangDensityFunctionalTheory2020} & 2020 & AIMD & PBE-D3 & \parbox[t]{2cm}{\raggedright 1Ca$^{2+}$ + 124H$_2$O} & 200 & 6.5 \\ 
\cite{wangDensityFunctionalTheory2020} & 2020 & AIMD & revPBE & \parbox[t]{2cm}{\raggedright 1Ca$^{2+}$ + 63H$_2$O} & 50 & 6.6 \\ 
\cite{wangDensityFunctionalTheory2020} & 2020 & AIMD & revPBE-D3 & \parbox[t]{2cm}{\raggedright 1Ca$^{2+}$ + 63H$_2$O} & 50 & 6.1 \\ 
\cite{wangDensityFunctionalTheory2020} & 2020 & AIMD & BLYP & \parbox[t]{2cm}{\raggedright 1Ca$^{2+}$ + 63H$_2$O} & 50 & 6.3 \\ 
\cite{wangDensityFunctionalTheory2020} & 2020 & AIMD & BLYP-D3 & \parbox[t]{2cm}{\raggedright 1Ca$^{2+}$ + 63H$_2$O} & 500 & 6.9 \\ 
\cite{liuSolvationStructuresCalcium2023} & 2023 & MLP MD & SCAN & \parbox[t]{2cm}{\raggedright 1Ca$^{2+}$ +2OH$^{-}$ 64H$_2$O} & 1000 & 6.91 \\ 
\cite{kostalCommonCationsAre2023} & 2023 & AIMD & revPBE-D3 & \parbox[t]{2cm}{\raggedright 1Ca$^{2+}$ + 128H$_2$O} & 100 & 6.68 \\ 
\cite{kostalCommonCationsAre2023} & 2023 & AIMD & revPBE-D3 (no D3 Ca) & \parbox[t]{2cm}{\raggedright 1Ca$^{2+}$ + 128H$_2$O} & 100 & 6.33 \\ 
\cite{boynProbingPHDependentDehydration2023} & 2023 & AIMD & revPBE-D3 & \parbox[t]{2cm}{\raggedright 1CaCl$_{2}$ + 54H$_2$O} & 60 & 6.81 \\ 
\cite{boynProbingPHDependentDehydration2023} & 2023 & AIMD (blue-moon ensemble) & revPBE-D3 & \parbox[t]{2cm}{\raggedright 1CaCl$_{2}$ + 54H$_2$O} & 60 & 7.22 \\ 
\cite{moisonLocalStructureEquilibration2024} & 2024 & AIMD & PBE & \parbox[t]{2cm}{\raggedright 1Ca$^{2+}$ + 62H$_2$O} & 30 & 6.14 \\ 
\cite{moisonLocalStructureEquilibration2024} & 2024 & AIMD & BLYP & \parbox[t]{2cm}{\raggedright 1Ca$^{2+}$ + 62H$_2$O} & 30 & 6.03 \\ 
\cite{moisonLocalStructureEquilibration2024} & 2024 & AIMD & BLYP-D2 & \parbox[t]{2cm}{\raggedright 1Ca$^{2+}$ + 62H$_2$O} & 100 & 6.5 \\ 
\cite{moisonLocalStructureEquilibration2024} & 2024 & AIMD & vdW-DF2 & \parbox[t]{2cm}{\raggedright 1Ca$^{2+}$ + 62H$_2$O} & 100 & 6.54 \\ 
\cite{kalvodaHydrationNumbersBiologically2024} & 2024 & AIMD & PBE-D3 & \parbox[t]{2cm}{\raggedright 1Ca$^{2+}$ + 64H$_2$O} & 80 & 6 \\ 
\cite{liangIonspecificHydrationStructures2025} & 2025 & AIMD & PBE & \parbox[t]{2cm}{\raggedright 1Ca$^{2+}$ + 63H$_2$O} & -- & 6 \\ 
\cite{liangIonspecificHydrationStructures2025} & 2025 & AIMD & SCAN & \parbox[t]{2cm}{\raggedright 1Ca$^{2+}$ + 63H$_2$O} & -- & 7.16 \\ 
\cite{yuSolvationStructuresIon2025} & 2025 & MLIP MD (metadynamics) & revPBE-D3 & \parbox[t]{2cm}{\raggedright 1CaCl$_{2}$ 64H$_2$O} & 10000 & 9 \\ 
\cite{piaggiInitioMachinelearningSimulation2025} & 2025 & MLIP MD (metadynamics) & SCAN & \parbox[t]{2cm}{\raggedright 1CaCO$_{3}$ +15763H$_2$O} & -- & 6.85 \\ 
\cite{youCalciumComplexationAqueous2026} & 2026 & AIMD & B3LYP-D3 & \parbox[t]{2cm}{\raggedright 1Ca$^{2+}$ + 40H$_2$O} & -- & 7 \\ 
\cite{salhaBenchmarkingOntheFlyMachine2026} & 2026 & AIMD & PBE-D3 & \parbox[t]{2cm}{\raggedright 1CaCl$_{2}$ + 62H$_2$O} & 50 & 6.2 \\ 
\end{longtable}

%% file: si/figures/CN/literature_summary/force_field_cn_table.tex
\begin{longtable}{@{}l r l l r r@{}}
\caption{Classical force-field calcium--water coordination numbers.}\label{tab:force_field_calcium_cn}\\
\toprule
Ref. & Year & Water model & Ca model & \shortstack{Concentration\\ / mol kg$^{-1}$} & CN \\
\midrule
\endfirsthead
\multicolumn{6}{c}{\tablename\ \thetable{} -- continued}\\
\toprule
Ref. & Year & Water model & Ca model & \shortstack{Concentration\\ / mol kg$^{-1}$} & CN \\
\midrule
\endhead
\midrule
\multicolumn{6}{r}{Continued on next page}\\
\endfoot
\bottomrule
\endlastfoot
\cite{boundsMolecularDynamicsStudy1985} & 1985 & TIP4P~\cite{jorgensenComparisonSimplePotential1983} & \cite{boundsMolecularDynamicsStudy1985} & 0.87 & 9.3 \\ 
\cite{boundsMolecularDynamicsStudy1985} & 1985 & TIP4P~\cite{jorgensenComparisonSimplePotential1983} & \cite{boundsMolecularDynamicsStudy1985} & 0.11 & 9.6 \\ 
\cite{probstMolecularDynamicsXray1985} & 1985 & CF~\cite{stillingerRevisedCentralForce1978} & \cite{probstMolecularDynamicsXray1985} & 1.1 & 9.2 \\ 
\cite{palinkasHydrationShellStructure1986} & 1986 & CF~\cite{stillingerRevisedCentralForce1978} & \cite{probstMolecularDynamicsXray1985} & 1.11 & 9--10 \\ 
\cite{straatsmaFreeEnergyIonic1988} & 1988 & SPC~\cite{berendsenInteractionModelsWater1981} & \cite{telemanMolecularDynamicsSimulation1986} & 0.11 & 9 \\ 
\cite{florisHydrationShellStructure1994} & 1994 & SPC/E~\cite{berendsenMissingTermEffective1987} & \cite{florisHydrationShellStructure1994} & 0.16 & 9.1 \\ 
\cite{florisHydrationShellStructure1994} & 1994 & SPC/E~\cite{berendsenMissingTermEffective1987} & \cite{florisHydrationShellStructure1994,florisInitioEffectivePair1992} & 0.16 & 8.6 \\ 
\cite{dangCommentMeanForce1995} & 1995 & SPC/E~\cite{berendsenMissingTermEffective1987} & \cite{dangCommentMeanForce1995} & 0.26 & 8 \\ 
\cite{bernal-uruchurtuRefinedMonteCarlo1995} & 1995 & \cite{saint-martinNonadditivityAnalyticalIntermolecular1990} & \cite{bernal-uruchurtuRefinedMonteCarlo1995} & 0.116 & 7.01 \\ 
\cite{kalkoEffectsTruncatingElectrostatic1996} & 1996 & SPC/E~\cite{berendsenMissingTermEffective1987} & \cite{straatsmaFreeEnergyIonic1988} & 0.26 & 7.9 \\ 
\cite{obstMolecularDynamicsStudy1996} & 1996 & TIP3P~\cite{jorgensenComparisonSimplePotential1983} & \cite{brooksCHARMMProgramMacromolecular1983} & 0.26 & 8 \\ 
\cite{perioleSimpleTwoBodyCationWater1997} & 1997 & TIP3P~\cite{jorgensenComparisonSimplePotential1983} & \cite{perioleSimpleTwoBodyCationWater1997} & $\sim$0.12 & 8 \\ 
\cite{koneshanSolventStructureDynamics1998} & 1998 & SPC/E~\cite{berendsenMissingTermEffective1987} & \cite{dangCommentMeanForce1995} & 0.26 & 7.9 \\ 
\cite{guardiaMolecularDynamicsSimulation1999} & 1999 & SPC/E~\cite{berendsenMissingTermEffective1987} & \cite{aqvistIonwaterInteractionPotentials1990} & 0.26 & 7.4 \\ 
\cite{guardiaMolecularDynamicsSimulation1999} & 1999 & SPC/E~\cite{berendsenMissingTermEffective1987} & \cite{dangCommentMeanForce1995} & 0.26 & 7.9 \\ 
\cite{guardiaMolecularDynamicsSimulation1999} & 1999 & SPC/E~\cite{berendsenMissingTermEffective1987} & \cite{scottGROMOSBiomolecularSimulation1999} & 0.26 & 8 \\ 
\cite{guardiaMolecularDynamicsSimulation1999} & 1999 & SPC/E~\cite{berendsenMissingTermEffective1987} & \cite{florisHydrationShellStructure1994} & 0.26 & 8.5 \\ 
\cite{babuTheoryIonicHydration1999} & 1999 & TIP3P~\cite{jorgensenComparisonSimplePotential1983} & \cite{brooksCHARMMProgramMacromolecular1983} & $\sim$0.05 & 8 \\ 
\cite{spangbergModelExtendedXray2000} & 2000 & TIP4P~\cite{jorgensenComparisonSimplePotential1983} & \cite{scottGROMOSBiomolecularSimulation1999} & 0.11 & 8 \\ 
\cite{spangbergModelExtendedXray2000} & 2000 & TIP4P~\cite{jorgensenComparisonSimplePotential1983} & \cite{boundsMolecularDynamicsStudy1985} & 0.11 & 9.5 \\ 
\cite{schwenkMolecularDynamicsSimulations2001} & 2001 & CF2~\cite{stillingerRevisedCentralForce1978} & \cite{schwenkMolecularDynamicsSimulations2001} & 0.28 & 7.1 \\ 
\cite{chialvoStructureCaCl2Aqueous2003} & 2003 & SPC/E~\cite{berendsenMissingTermEffective1987} & \cite{dangCommentMeanForce1995} & -- & 8.05 \\ 
\cite{chialvoStructureCaCl2Aqueous2003} & 2003 & SPC/E~\cite{berendsenMissingTermEffective1987} & \cite{dangCommentMeanForce1995} & 0.2 & 8 \\ 
\cite{chialvoStructureCaCl2Aqueous2003} & 2003 & SPC/E~\cite{berendsenMissingTermEffective1987} & \cite{dangCommentMeanForce1995} & 1.0 & 7.52 \\ 
\cite{chialvoStructureCaCl2Aqueous2003} & 2003 & SPC/E~\cite{berendsenMissingTermEffective1987} & \cite{dangCommentMeanForce1995} & 2.5 & 6.8 \\ 
\cite{chialvoStructureCaCl2Aqueous2003} & 2003 & SPC/E~\cite{berendsenMissingTermEffective1987} & \cite{dangCommentMeanForce1995} & 4.0 & 5.9 \\ 
\cite{chialvoStructureCaCl2Aqueous2003} & 2003 & SPC/E~\cite{berendsenMissingTermEffective1987} & \cite{dangCommentMeanForce1995} & 4.5 & 5.5 \\ 
\cite{chialvoStructureCaCl2Aqueous2003} & 2003 & SPC/E~\cite{berendsenMissingTermEffective1987} & \cite{dangCommentMeanForce1995} & 5.32 & 5.4 \\ 
\cite{chialvoStructureCaCl2Aqueous2003} & 2003 & SPC/E~\cite{berendsenMissingTermEffective1987} & \cite{dangCommentMeanForce1995} & 6.4 & 5.3 \\ 
\cite{chialvoStructureCaCl2Aqueous2003} & 2003 & SPC/E~\cite{berendsenMissingTermEffective1987} & \cite{dangCommentMeanForce1995} & 9.26 & 3.9 \\ 
\cite{megyesIonPairingAqueous2006} & 2006 & SPC/E~\cite{berendsenMissingTermEffective1987} & \cite{dangCommentMeanForce1995} & 4.62 & 6.2 \\ 
\cite{megyesIonPairingAqueous2006} & 2006 & SPC/E~\cite{berendsenMissingTermEffective1987} & \cite{dangCommentMeanForce1995} & 2.74 & 6.5 \\ 
\cite{dangMolecularSimulationAnalysis2006} & 2006 & \cite{dangMolecularDynamicsStudy1997} & \cite{glezakouElectronicStructureStatistical2006} & - & 6.5 \\ 
\cite{piquemalAccurateSolvationDynamics2006} & 2006 & AMOEBA~\cite{renPolarizableAtomicMultipole2003} & \cite{piquemalAccurateSolvationDynamics2006} & 0.93 & 7.2 \\ 
\cite{piquemalAccurateSolvationDynamics2006} & 2006 & AMOEBA~\cite{renPolarizableAtomicMultipole2003} & \cite{piquemalAccurateSolvationDynamics2006} & 0.26 & 7.7 \\ 
\cite{piquemalAccurateSolvationDynamics2006} & 2006 & AMOEBA~\cite{renPolarizableAtomicMultipole2003} & \cite{piquemalAccurateSolvationDynamics2006} & 0.11 & 7.7 \\ 
\cite{jiaoSimulationCa2Mg22006} & 2006 & AMOEBA~\cite{renPolarizableAtomicMultipole2003} & \cite{jiaoSimulationCa2Mg22006} & 0.108 & 7.3 \\ 
\cite{larentzosMolecularDynamicsStudy2008} & 2008 & SPC~\cite{berendsenInteractionModelsWater1981} & \cite{aqvistIonwaterInteractionPotentials1990} & 0.108 & 7.9 \\ 
\cite{liStructureDynamicsSolvation2008} & 2008 & SPC/E~\cite{berendsenMissingTermEffective1987} & \cite{dangCommentMeanForce1995} & 1.00 & 7.9 \\ 
\cite{timkoInitioCalculationPotential2011} & 2011 & TIP3P & \cite{mackerellAllAtomEmpiricalPotential1998} & 0.91 & 7.2 \\ 
\cite{timkoInitioCalculationPotential2011} & 2011 & SPC/E~\cite{berendsenMissingTermEffective1987} & \cite{dangCommentMeanForce1995} & 0.91 & 8 \\ 
\cite{deubleinSetMolecularModels2012} & 2012 & SPC/E~\cite{berendsenMissingTermEffective1987} & \cite{deubleinSetMolecularModels2012} & 0.34 & 7.8 \\ 
\cite{ditommasoModellingEffectsSalt2014} & 2014 & SPC/E~\cite{berendsenMissingTermEffective1987} & \cite{raiteriWaterKeyNonclassical2010} & 0.06 & 7 \\ 
\cite{ditommasoModellingEffectsSalt2014} & 2014 & SPC/E~\cite{berendsenMissingTermEffective1987} & \cite{deleeuwModelingAbsorptionSegregation2000} & 0.06 & 7.1 \\ 
\cite{ditommasoModellingEffectsSalt2014} & 2014 & SPC/E~\cite{berendsenMissingTermEffective1987} & \cite{dangCommentMeanForce1995} & 0.06 & 8 \\ 
\cite{hartkampStructureTransportAqueous2014} & 2014 & TIP4P/2005~\cite{abascalGeneralPurposeModel2005} & \cite{mamatkulovForceFieldsDivalent2013} & 0.11 & 7.9 \\ 
\cite{kohagenAccurateDescriptionCalcium2014} & 2014 & TIP4P/2005~\cite{abascalGeneralPurposeModel2005} & \cite{oostenbrinkBiomolecularForceField2004} & 4.00 & 5.84 \\ 
\cite{kohagenAccurateDescriptionCalcium2014} & 2014 & SPC/E~\cite{berendsenMissingTermEffective1987} & \cite{oostenbrinkBiomolecularForceField2004} & 4.00 & 7.76 \\ 
\cite{duarteForceFieldIndependent2014} & 2014 & \cite{duarteForceFieldIndependent2014} & \cite{duarteForceFieldIndependent2014} & 0.07 & 7 \\ 
\cite{salanneCa2+ClAssociationWater2017} & 2017 & \cite{dangMolecularDynamicsStudy1997} & \cite{taziTransferableInitioBased2012} & 0.079 & 7--8 \\ 
\cite{naleemKirkwoodBuffDerivedForce2018} & 2018 & SPC/E~\cite{berendsenMissingTermEffective1987} & \cite{naleemKirkwoodBuffDerivedForce2018} & - & 7.78 \\ 
\cite{mamatkulovForceFieldsMonovalent2018} & 2018 & TIP3P~\cite{jorgensenComparisonSimplePotential1983} & \cite{mamatkulovForceFieldsMonovalent2018} & 0.5 & 7.79 \\ 
\cite{koskampReconsideringCalciumDehydration2019} & 2019 & SM-lj~\cite{deleeuwMoleculardynamicsSimulationMgO1998} & \cite{deleeuwMoleculardynamicsSimulationMgO1998} & 0.07 & 8.4 \\ 
\cite{koskampReconsideringCalciumDehydration2019} & 2019 & SPC/fw-buck~\cite{wuFlexibleSimplePointcharge2006} & \cite{deleeuwMoleculardynamicsSimulationMgO1998} & 0.07 & 6.9 \\ 
\cite{koskampReconsideringCalciumDehydration2019} & 2019 & SPC/fw-tail~\cite{wuFlexibleSimplePointcharge2006} & \cite{raiteriThermodynamicallyConsistentForce2015} & 0.07 & 7 \\ 
\cite{zeronForceFieldLi2019} & 2019 & TIP4P/2005~\cite{abascalGeneralPurposeModel2005} & \cite{zeronForceFieldLi2019} & 6 & 7.1 \\ 
\cite{dopkeTransferabilityIonParameters2020} & 2020 & SPC/E~\cite{berendsenMissingTermEffective1987} & \cite{mamatkulovForceFieldsDivalent2013} & 0.11 & 7.21 \\ 
\cite{dopkeTransferabilityIonParameters2020} & 2020 & TIP4P/2005~\cite{abascalGeneralPurposeModel2005} & \cite{mamatkulovForceFieldsDivalent2013} & 0.11 & 7.61 \\ 
\cite{dopkeTransferabilityIonParameters2020} & 2020 & TIP4P/2005~\cite{abascalGeneralPurposeModel2005} & \cite{zeronForceFieldLi2019} & 0.11 & 7.44 \\ 
\cite{liSystematicParametrizationDivalent2020} & 2020 & OPC3~\cite{izadiAccuracyLimitRigid2016} & \cite{liSystematicParametrizationDivalent2020} & - & 7.1 \\ 
\cite{liSystematicParametrizationDivalent2020} & 2020 & OPC~\cite{izadiBuildingWaterModels2014} & \cite{liSystematicParametrizationDivalent2020} & - & 6.4 \\ 
\cite{liSystematicParametrizationDivalent2020} & 2020 & TIP3P-FB~\cite{wangBuildingForceFields2014} & \cite{liSystematicParametrizationDivalent2020} & - & 6.9 \\ 
\cite{liSystematicParametrizationDivalent2020} & 2020 & TIP4P-FB~\cite{wangBuildingForceFields2014} & \cite{liSystematicParametrizationDivalent2020} & - & 6.4 \\ 
\cite{yuABEEMMMCalcium2025} & 2025 & ABEEM~\cite{yangAtombondElectronegativityEqualization2004} & \cite{yuABEEMMMCalcium2025} & 0.085 & 8 \\ 
\cite{cancadeAdvancingCLPolPolarizable2026} & 2026 & SWM4-NDP~\cite{lamoureuxPolarizableModelWater2006} & \cite{cancadeAdvancingCLPolPolarizable2026} & 0.01 & 5.6 \\ 
-- & -- & -- & -- & - & -- \\ 
\end{longtable}

%% file: si/figures/CN/literature_summary/experiment_cn_table.tex
\begin{longtable}{@{}l r r l r@{}}
\caption{Experimental calcium--water coordination numbers.}\label{tab:experimental_calcium_cn}\\
\toprule
Ref. & Year & \shortstack{Concentration\\ / mol kg$^{-1}$} & Method & CN \\
\midrule
\endfirsthead
\multicolumn{5}{c}{\tablename\ \thetable{} -- continued}\\
\toprule
Ref. & Year & \shortstack{Concentration\\ / mol kg$^{-1}$} & Method & CN \\
\midrule
\endhead
\midrule
\multicolumn{5}{r}{Continued on next page}\\
\endfoot
\bottomrule
\endlastfoot
\cite{vanpanthaleonvaneckXrayDiffractionAqueous1957} & 1957 & 1.5 & XRD & 6 \\ 
\cite{bolHydrationDivalentCations1970} & 1970 & 1 & XRD & 6 \\ 
\cite{albrightXRayDiffractionStudies1972} & 1972 & 3.26 & XRD & 8.2 \\ 
\cite{albrightXRayDiffractionStudies1972} & 1972 & 5.22 & XRD & 8 \\ 
\cite{licheriXrayDiffractionStudy1975} & 1975 & 1.26 & XRD & 6 \\ 
\cite{licheriXrayDiffractionStudy1975} & 1975 & 2.14 & XRD & 8 \\ 
\cite{licheriXrayDiffractionStudy1976} & 1976 & 0.99 & XRD & 6 \\ 
\cite{licheriXrayDiffractionStudy1976} & 1976 & 2.09 & XRD & 6 \\ 
\cite{licheriXrayDiffractionStudy1976} & 1976 & 4.52 & XRD & 6 \\ 
\cite{cummingsIonHydrationAqueous1980} & 1980 & 4.49 & neutron diffraction & 5.5 \\ 
\cite{hewishEnvironmentCa2Ions1982} & 1982 & 1 & neutron diffraction & 10 \\ 
\cite{hewishEnvironmentCa2Ions1982} & 1982 & 2.8 & neutron diffraction & 7.2 \\ 
\cite{hewishEnvironmentCa2Ions1982} & 1982 & 4.5 & neutron diffraction & 6.4 \\ 
\cite{probstMolecularDynamicsXray1985} & 1985 & 1.1 & XRD & 6.9 \\ 
\cite{yamaguchiXrayDiffractionStudy1989} & 1989 & 6.45 & XRD & 6 \\ 
\cite{yamaguchiXrayDiffractionStudy1989} & 1989 & 9.25 & XRD & 5.9 \\ 
\cite{smirnovXrayDiffractionStudy1997} & 1997 & 8.98 & XRD & 7.3 \\ 
\cite{smirnovXrayDiffractionStudy1997} & 1997 & 4.59 & XRD & 7.7 \\ 
\cite{spangbergModelExtendedXray2000} & 2000 & 0.11 & EXAFS & 8 \\ 
\cite{jalilehvandHydrationCalciumIon2001} & 2001 & 0.94 & EXAFS/LAXS & 8 \\ 
\cite{jalilehvandHydrationCalciumIon2001} & 2001 & 1.88 & EXAFS/LAXS & 8 \\ 
\cite{jalilehvandHydrationCalciumIon2001} & 2001 & 1.5 & EXAFS/LAXS & 8 \\ 
\cite{fultonUnderstandingEffectsConcentration2003} & 2003 & 0.2 & EXAFS & 6.8 \\ 
\cite{fultonUnderstandingEffectsConcentration2003} & 2003 & 6 & EXAFS & 7.2 \\ 
\cite{megyesSolvationCalciumIon2004} & 2004 & 1.04 & XRD & 8 \\ 
\cite{megyesSolvationCalciumIon2004} & 2004 & 2.74 & XRD & 6.9 \\ 
\cite{megyesSolvationCalciumIon2004} & 2004 & 4.62 & XRD & 5.8 \\ 
\cite{megyesSolvationCalciumIon2004} & 2004 & 8.64 & XRD & 5.1 \\ 
\cite{badyalUnderstandingEffectsConcentration2004} & 2004 & 4 & neutron diffraction & 7.3 \\ 
\cite{badyalUnderstandingEffectsConcentration2004} & 2004 & 6.4 & neutron diffraction & 6.95 \\ 
\cite{zavitsasAqueousSolutionsCalcium2005} & 2005 & - & freezing-point depression & 12 \\ 
\cite{zavitsasAqueousSolutionsCalcium2005} & 2005 & - & boiling-point elevation & 6.7 \\ 
\cite{megyesIonPairingAqueous2006} & 2006 & 2.74 & XRD & 6.2 \\ 
\cite{megyesIonPairingAqueous2006} & 2006 & 2.74 & neutron diffraction & 6.5 \\ 
\cite{megyesIonPairingAqueous2006} & 2006 & 4.62 & XRD & 5.9 \\ 
\cite{tongraarStructureHydratedCa22010} & 2010 & 2.12 & EXAFS & 6.9 \\ 
\cite{rudolphHydrationCalciumiiIon2013} & 2013 & 0.1157 & Raman spectroscopy & 6 \\ 
\cite{yamaguchiIonHydrationAssociation2019} & 2019 & 2 & neutron diffraction + EPSR & 7 \\ 
\cite{renDecipheringSolventEffect2020} & 2020 & 0.5 & EXAFS/XANES & 9.04 \\ 
\cite{renDecipheringSolventEffect2020} & 2020 & 3 & EXAFS/XANES & 8.08 \\ 
\cite{renDecipheringSolventEffect2020} & 2020 & 10 & EXAFS/XANES & 5.15 \\ 
\cite{yangProbingCalciumSolvation2020} & 2020 & 0.5 & Ca L-edge XAS & 9 \\ 
\cite{smirnovXRayDiffractionData2023} & 2023 & 6.94 & XRD & 5.2 \\ 
\cite{smirnovXRayDiffractionData2023} & 2023 & 3.7 & XRD & 5.3 \\ 
\cite{smirnovXRayDiffractionData2023} & 2023 & 2.78 & XRD & 5.6 \\ 
\cite{smirnovXRayDiffractionData2023} & 2023 & 1.39 & XRD & 5.8 \\ 
\cite{smirnovXRayDiffractionData2023} & 2023 & 0.694 & XRD & 6 \\ 
\cite{yamaguchiVisualization3DStructure2025} & 2024 & 0.985 & XRD + Raman + EPSR & 7 \\ 
\cite{bajnocziStructureConcentratedAqueous2026} & 2026 & 5.98 & LAXS & 8 \\ 
\cite{youCalciumComplexationAqueous2026} & 2026 & 0.01 & Ca K-edge XANES/EXAFS & 7 \\ 
\end{longtable}